\documentclass[11pt]{article}
\usepackage{epsfig} 
\graphicspath{ {figure/Pictures/} }
\usepackage{subcaption,fullpage}
\usepackage{amsmath}
\usepackage{bm}
\usepackage[usenames]{color}

\newtheorem{exmp}{Example}[section]

\title{Topological Semantics for  Lumped Parameter Systems Modeling}

\author{Randi Wang \qquad Vadim Shapiro\\
Spatial Automation Laboratory\\
University of Wisconsin - Madison }

\date {\today}
\begin{document}
\maketitle


\begin{abstract}
{ 

Behaviors of many engineering systems are described by lumped parameter models that encapsulate the spatially distributed nature of the system into networks of lumped elements;  the dynamics of such a network is governed by a system of ordinary differential and algebraic equations.  Languages and  simulation tools  for modeling such systems differ in syntax,  informal semantics, and in the methods by which such systems of equations are generated and simulated, leading to numerous interoperability challenges.  Logical extensions of SysML aim specifically at unifying a subset of the underlying concepts in such languages.  

We propose to unify semantics of all such systems using standard notions from algebraic topology.   In particular, Tonti diagrams classify all physical theories in terms of physical laws (topological and constitutive) defined over a pair of  dual cochain complexes and may be used to describe different types of lumped parameter systems.   We  show that all possible methods for generating the corresponding state equations within each physical domain correspond to paths over Tonti diagrams.  We further propose a generalization of Tonti diagram that captures the behavior and supports canonical generation of state equations for multi-domain lumped parameter systems.

The unified semantics provides a basis for greater interoperability in systems modeling, supporting automated translation, integration, reuse, and numerical simulation of models created in different authoring systems and applications. Notably, the proposed algebraic topological semantics is also compatible with spatially and temporally distributed models that are at the core of modern CAD and CAE systems. 

}


\end{abstract}
\section{Introduction}

\subsection{Motivation}

Lumped parameter models are  commonly used to describe behaviors of many engineering systems \cite{kluever2015dynamic}. 
In such systems,  spatially and temporally distributed physical phenomena are approximated by a finite network of abstract components that store, dissipate, or transform energy;  the phenomena-specific constitutive properties of the components (e.g. generalized impedances) are estimated by domain integrals from  the actual system by a process of  ``reticulation''\cite{paynter1961analysis}.   Bond graphs \cite{paynter1961analysis},  linear graphs \cite{rowell1997system}, Modelica \cite{elmqvist1997introduction} and Simulink/Simscape \cite{chaturvedi2009modeling} are commonly used physical modeling languages for creating, editing, and simulating lumped parameter models.  
These languages  may differ widely  in their syntax,  but have similar (though not identical) semantics that specifies interconnectivity and constitutive relations of individual components in a system;   the resulting model of a system  created in such languages is then compiled into a system of state equations that may be  numerically solved to simulate the system's dynamic response.  Formally,  the dynamic behavior of a lumped parameter model is  described by the state equations,  a set of ordinary differential equations (ODEs) or differential algebraic equations (DAEs), whose solutions depend on the initial conditions.  

Generally speaking,  all modeling languages can handle the same broad class of problems but with non-trivial differences in  system types, representations, state equation derivation  and simulation mechanisms \cite{karnopp1990system, rowell1997system, fritzson1998modelica,xue2013system}.   For example,  in Simulink, components exchange numeric information uni-directionally and are not subject to conservation laws;   by contrast, the energy flow  between components is  bi-directional in  other languages, satisfying conservation laws.      While every linear graph model may be represented by a bond graph,  the converse statement is not true \cite{perelson1976bond}.  Furthermore,  parallel/serial junctions in a bond graph do not specify  ordering of branches in the junctions,  which implies that every bond graph in fact corresponds to a family of (dynamically) equivalent graph-based models in other languages.  Many languages generate state equations using efforts and flows as variables, but their integral forms may  also be used,  for example in bond graphs\cite{karnopp2012system}.   The above and other differences in syntax and  semantics of system modeling languages  lead to challenges in exchange, translation, and composition of models created in these languages.  Such interoperability difficulties  are only likely to increase due to ubiquitous and growing adoption  of physical modeling languages by industry and standards organizations.    

Conceptually,  there are two possible approaches to dealing with semantic interoperability issues:  ``point-to-point'' correspondence\footnote{Such correspondence may take a form of direct translation or using APIs.} between models created in different languages,  or standardization on a single neutral format that can be translated to/from models in any such languages.   The first approach is more practical but is problematic because it requires  $O(n^2)$ such translators, which is not only expensive, but  discourages development of new languages and simulation solutions.   The second approach is similar in spirit to STEP for product models,  which requires the neutral format to be formally defined and include the superset of models present in any such language.   
These difficulties can be observed  in a recent effort to extend System Modeling Language (SysML) with packages for direct communication with multiple simulation tools \cite{bock2017extension}.    Recently proposed Functional Mockup Interface (FMI) \cite{otter2010functional}  attempts to sidestep the semantic interoperability issues by supporting model exchange and integration via standardized XML and C-code interfaces;  this approach simply shifts the responsibility for semantic correctness of these tasks to the authoring systems and users.   

Irrespectively of the selected approach,  semantic interoperability requires establishing formal correspondence between concepts and constructs in distinct modeling languages.   This is the main goal of this paper.    Our approach relies on tools from algebraic topology and well known classification of physical theories developed over the years by Tonti \cite{tonti1975formal,tonti2013mathematical}, Roth \cite{roth1955application},  Branin \cite{branin1966algebraic}, Kron \cite{kron1963diakoptics}, and others.  
Importantly,  this classification generalizes to higher-dimensional physical models,  suggesting that the proposed framework may be extended to include spatially-distributed models represented by three-dimensional solid models, partial differential equations and finite element methods.

\subsection{Outline}

Section 2 briefly surveys and compares the main system modeling languages and surveys related work.  In Section 3, we summarize the well known algebraic topological model of physical systems;  this model  serves as the basis for Tonti's diagrams that classify physical variables, laws, and theories \cite{tonti2013mathematical}.   Tonti's classification is a starting point for the unified semantics proposed in this paper.   The main results of the paper are contained in Sections 4 and 5. 

Single domain lumped parameters systems are examined in detail in Section 4.  Because all such systems are isomorphic in terms of their physical variables, laws, and Tonti's diagrams, we only discuss the models of electrical systems -- with understanding that the discussion also applies to lumped parameters systems in other physical domains.
In particular, we show that paths over the corresponding Tonti diagram correspond to all possible ways to generate state equations for an electrical network, including well known node-potential and mesh-current methods that date to Maxwell \cite{Maxwell}.  While the discussion and the semantics proposed in Section 4 apply to {\em any one} single-domain lumped-parameter system model, in Section 5 we consider the more general multi-domain systems (that may combine electrical, mechanical, hydraulic, thermal, and other single-domain subsystems).  We show that all such systems may be represented by an algebraic topological model in terms of generalized state variables associated with extended and generalized Tonti diagrams.  
We conclude in Section 6 with discussion of practical applications and consequences of the proposed semantic unification, open issues, and possible extensions.


\section{Background}

\subsection{Brief history of lumped-parameter  modeling}

Informally,  the  significance of lumped-parameter models and languages can be attributed to two main factor factors.   Lumped parameter models are the simplest models of system behavior  (usually,  but not always dynamic)  that encapsulate and abstract away  all detailed geometric information.  This allows modeling behavior of systems without dealing with detailed embodiment of such systems -- either because these details are too complex or not known (for example in early design stages).    Secondly,  the lumped parameter models in different physical domains (electrical, mechanical motion, fluid flow, etc.) have identical mathematical structure  that unifies all such models.   Below we briefly summarizes the conceptual evolution of these developments.  

The efforts to classify and unify different physical models and theories date back at least  to  James Clerk Maxwell,  who expressed the desire to establish a formal analogy
between various physical quantities based on their mathematical form
\cite{Maxwell}, and continued through  20$^{\mbox{\small th}}$ century (for example see \cite{Dantzig}).
Maxwell also introduced the so called {\em mesh\/} and {\em node\/} methods for solving electrical networks \cite{Bamberg&90,Maxwell} that we will also use to derive state equations for lumped parameter systems in Section 4. 

Advances in analysis and simulation of (analog) electrical systems  motivated many to
use such analogies  in order to leverage the advances in electrical engineering in other 
physical domains  \cite{Nickle25,Olson}.  In particular,  Nickle proposed to predict the response of a mechanical dynamical system by translating it into a dynamically equivalent  analog electrical circuit \cite{Nickle25}.
This electrical-network approach to modeling and simulation of mechanical
systems became widely used and culminated in the work of Gabriel Kron, an influential and  controversial electrical engineer, who showed that most distributed mechanical systems may also be modeled  and efficiently simulated by analog electrical networks  \cite{Kron44,Kron57}.   Kron's work influenced  many others, including F. Branin, P. Roth,  E. Tonti,  and H. Paynter.    

More specifically,  Branin recognized that the versatility of electrical networks is an indication of the combinatorial and topological nature of the classical vector calculus \cite{Branin66}.   Roth identified specific algebraic topological model for stationary electrical networks in terms of  chain and cochain  sequences \cite{Roth55},  whereas Branin  generalized Roth diagram to two dual cochain sequences and revealed that this model underlies general three-dimensional network models.    Tonti \cite{tonti2013mathematical} built on these insights to develop a complete classification of space-time physical theories in terms of their underlying topological structures and codified this classification in terms of  concise (Tonti) diagrams.   We will review the Tonti diagrams in Section 3 and we will adopt them as a convenient way to describe and visualize  models of lumped parameter systems.  
More details of these developments can be found in Appendix D of \cite{tonti2013mathematical}.

Increasingly, these algebraic topological models  are finding applications in computational modeling of engineering systems.  For example, several researchers proposed that all geometric and physical computations should be based on a common (co)chain complex model \cite{palmer1993chain,dicarlo2009chain,dicarlo2014linear}, and that a vast majority of physical laws may be enforced as topological invariants of common  transformations used in automated design systems \cite{ramaswamy2003combinatorial}.   Algebraic topological structure has been used explicitly in variety of modeling systems spanning diverse areas such interactive physics \cite{chard2000multivector}, computer graphics \cite{egli2004chain}, and manufacturing systems \cite{bjorke1995manufacturing}.


\subsection {Languages for lumped parameter modeling}

Paynter's work \cite{Paynter} originally focused on methodology for use of lumped parameters models for design and analysis of engineering systems and culminated in the language of bond graphs.   He described a systematic method of converting a general system to a simplified lumped parameter model by a process of ``reticulation''  (creating a network model through discretization of a spatially-distributed system) and developed a symbolic language of {\em bond graphs\/} for describing such models.   A major conceptual and practical significance of bond graphs is that they model all lumped parameter systems in terms of small number of abstract generalized variables:  efforts, flows,  their derivatives, integrals, and constitutive relationships.   This allowed bond graphs to represent lumped parameter models of complex physically heterogeneous systems that are coupled using abstract transformer and gyrator elements and can be easily translated into a system of ordinary differential equations.  

    \begin{figure}[!htb]
        \centering
        \begin{subfigure}[b]{0.4\textwidth}
          \centering
          \includegraphics[width=.6\linewidth]{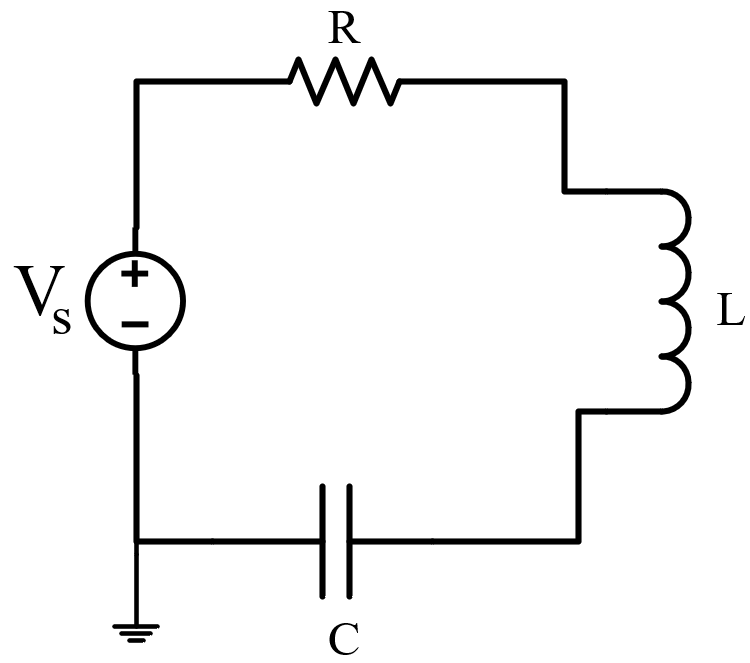}
          \caption{A simple RLC electrical circuit}
          \label{fig:Comparison_RLC problem}
        \end{subfigure} \par\medskip

        \centering
        \begin{subfigure}[b]{0.4\textwidth}
          \centering
          \includegraphics[width=.5\linewidth]{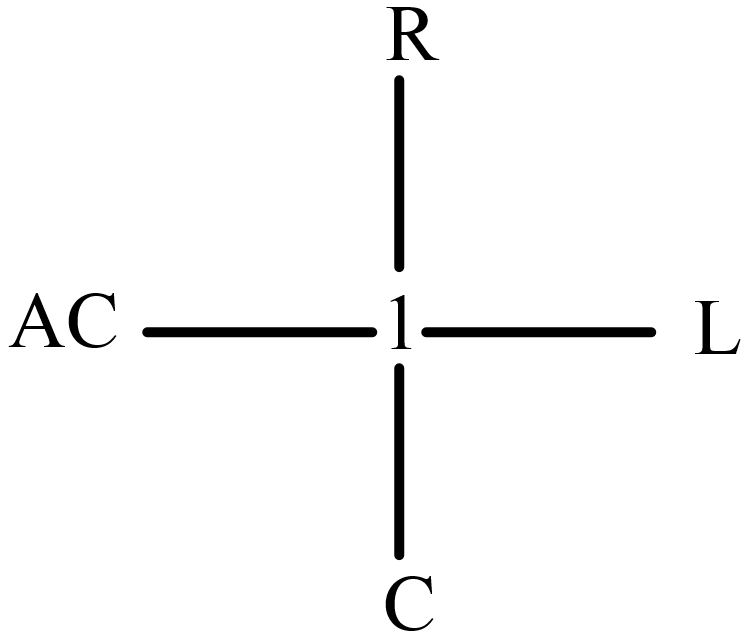}
          \caption{bond graph model}
          \label{fig:Comparison_bond graph}
        \end{subfigure}
        \centering
        \begin{subfigure}[b]{0.4\textwidth}
          \centering
          \includegraphics[width=.5\linewidth]{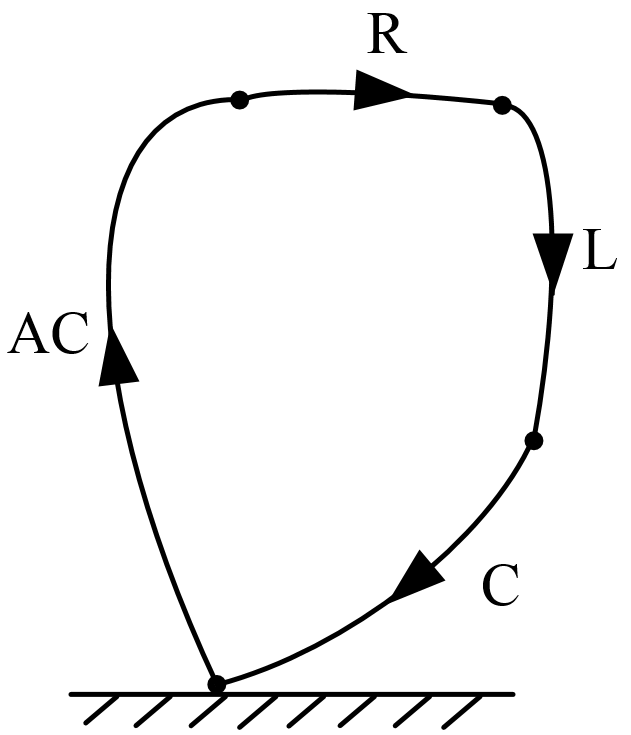}
          \caption{linear graph model}
          \label{fig:Comparison_linear graph}
        \end{subfigure} \par\medskip
         
        \begin{subfigure}[b]{0.4\textwidth}
          \centering
          \includegraphics[width=.7\linewidth]{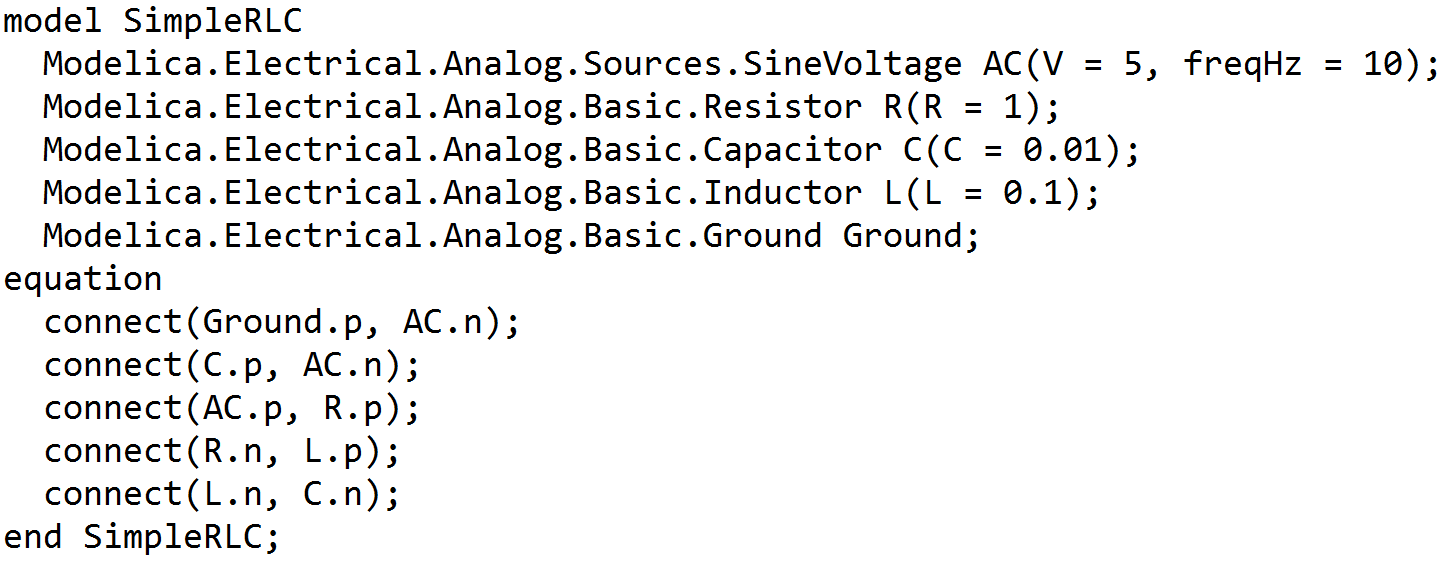}
          \caption{Modelica model}
          \label{fig:Comparison_Modelica}
        \end{subfigure}
        \begin{subfigure}[b]{0.4\textwidth}
          \centering
          \includegraphics[width=\linewidth]{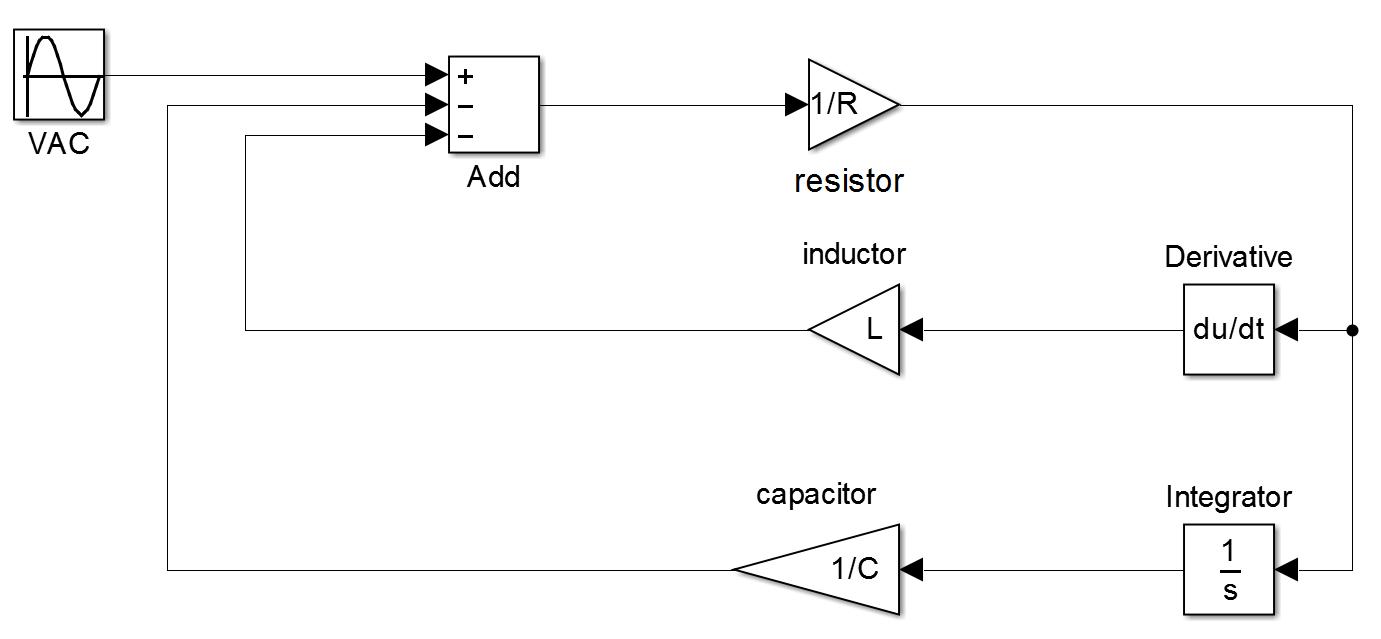}
          \caption{Simulink model}
          \label{fig:Comparison_Simulink}
        \end{subfigure}
        \caption{Comparison of different physical modeling languages for a simple electrical circuit problem}
        \label{fig:Comparison}
        \end{figure}      
Arguably, all lumped parameter models and languages are based on  concepts similar to those used in bond graphs and are now a standard textbook material \cite{karnopp2012system,rowell1997system}.  Lumped parameter models abstract a spatially and temporaly distributed system by  a network of components.  The purpose of the network is to model the  flow of energy through the system's components as a function of  time;  detailed geometry of the system is accounted for by integral properties of the components.    Formally,  the lumped parameter model is described in terms of ordinary differential and algebraic equations with variables that are specific to physical phenomena modeled by the network.   However, all such models are isomorphic based on the well known `analogies' between different physical theories, as summarized in Table \ref{table:analogy}.    In each physical domain,  the power is a product of two generalized variables that are classified as across or through (in case of Modelica and linear graph models), or effort and flow (in case of bond graphs).    The network includes abstract sources of these variables,  and  transforms the energy by storing or dissipating in three types elements that are commonly described by analogy with electrical networks as generalized resistors, inductances, and capacitors.  In addition, the energy may be transformed between different physical domains using transformer and gyrator components in the network.   More general components with complex behavior may be incorporated into the network, but all such components may be modeled as a composition of these basic components listed above \cite{paynter1961analysis}.  The formal reason for these analogies is the common underlying algebraic topological structure that we will discuss in Section 3.

\begin{table}[ht]
\caption{Analogy of physical variables}
\begin{center}
\label{table:analogy}
\resizebox{\linewidth}{!}{%
\begin{tabular}{c c c c c}
\hline
Variables type & Effort & Flow & Momentum & Displacement\\
\hline
Mechanical (translation) & Force & Velocity & Momentum & Displacement\\ 
Mechanical (rotational)  & Torque & Angular velocity & Angular momentum & Angle \\
Electrical & Voltage & Current & Flux linkage variable & Charge \\
Hydraulic & Pressure & Volume flow rate & Pressure momentum & Volume \\
\hline
\end{tabular}}
\end{center}
\end{table}
    
The differences between various  modeling languages stem from their approach to describing the basic components,  their connectivity in the network,  and the method by which the governing system of equations is generated.   Thus, bond graph uses 0 and 1 junctions to represent parallel or serial connections of components; linear graph uses directed edges to connect components (nodes), while Modelica/Simulink uses connectors to link library components.  Behavior of components is described in terms of state variables that are related by constitutive relations: bond graph associates effort, flow and constitutive relationships to 1-ports;  linear graph associates across and through variables to directed edges, while Modelica and Simulink associate these variables with blocks.  
 
The governing state equations is a system of ODEs (or more generally DAEs) that are generated from such descriptions  by methods that differ in the choice of state variables and generation algorithms.    Of course, all such equations describe identical dynamical behavior of the modeled system that is  determined by 
initial conditions.   To be concrete, let us  briefly compare how  such languages would be used to model a simple example of a serial RLC electrical circuit shown in Figure  \ref{fig:Comparison_RLC problem}.    Figure \ref{fig:Comparison} shows how the same circuit would be described in four different languages:   bond graphs, linear graphs, Modelica,  and Simulink.   As we already noted,  electrical circuits themselves have also been used as a general purpose modeling language for modeling lumped parameter system, for example in  \cite{kron1963diakoptics}, but they have been largely superceded by more modern languages in use today.  The following is a brief summary of steps to generate state equations for the circuit in each of the languages.  

\textbf{Bond graph} : (1) assign causality to the bond graph; (2) select input and energy state variables; (3) generate an initial set of system equations; (4) simplify initial equations to state-space form \cite{karnopp2012system}, giving 
    \begin{equation} \label{Comparison_RLC_bondgraph}
    \frac{d}{{dt}}\left\{ {\begin{array}{*{20}{c}}
    {{q_C}}\\
    {{p_L}}
    \end{array}} \right\} = \left[ {\begin{array}{*{20}{c}}
    0&{\frac{1}{L}}\\
    { - \frac{1}{C}}&{ - \frac{R}{L}}
    \end{array}} \right]\left\{ {\begin{array}{*{20}{c}}
    {{q_C}}\\
    {{p_L}}
    \end{array}} \right\} + \left\{ {\begin{array}{*{20}{c}}
    0\\
    {V_{AC}}
    \end{array}} \right\}
    \end{equation}
    where, ${{q_C}}$ is the time integral of the current of capacitor, ${{p_L}}$ is the time integral of the voltage of inductor. ${V_{AC}}$ represents the voltage of voltage source.

        \textbf{Linear graph}: (1) generate a normal tree from the linear graph; (2) select the primary and secondary variables, system order and state variables; (3) formulate constitutive, continuity and compatibility equations then simplify these equations to state equations \cite{koenig1967analysis}, giving 
            \begin{equation} \label{Comparison_RLC_lineargraph}
    \frac{d}{{dt}}\left\{ {\begin{array}{*{20}{c}}
    {{i_L}}\\
    {{v_C}}
    \end{array}} \right\} = \left[ {\begin{array}{*{20}{c}}
    { - \frac{1}{L}}&{ - \frac{R}{L}}\\
    0&{\frac{1}{C}}
    \end{array}} \right]\left\{ {\begin{array}{*{20}{c}}
    {{i_L}}\\
    {{v_C}}
    \end{array}} \right\} + \left\{ {\begin{array}{*{20}{c}}
    {\frac{1}{L}}\\
    0
    \end{array}} \right\}{V_{AC}}
    \end{equation}
    where, ${{i_L}}$ represents the current of inductor, ${{v_C}}$ represents the voltage of capacitor. ${{V_{AC}}}$ represents the voltage of voltage source.

        \textbf{Modelica/Simulink}: (1) based on the established model in Modelica/Simulink, extract constitutive equations of each component from library; (2) select dynamic variables; (3) express the problem as the smallest possible ODE system \cite{elmqvist1997introduction, chaturvedi2009modeling}.  In case of Modelica,  we get 
    \begin{equation} 
    \label{Comparison_RLC_Modelica}
    \left\{ {\begin{array}{*{20}{c}}
    {der(L.i)}\\
    {der(C.v)}
    \end{array}} \right\} = \left[ {\begin{array}{*{20}{c}}
    { - \frac{1}{L}}&{ - \frac{R}{L}}\\
    0&{\frac{1}{C}}
    \end{array}} \right]\left\{ {\begin{array}{*{20}{c}}
    {L.i}\\
    {C.v}
    \end{array}} \right\} + \left\{ {\begin{array}{*{20}{c}}
    {\frac{1}{L}}\\
    0
    \end{array}} \right\}{V_{AC}} 
    \end{equation}
    where, $der(*)$ represents the derivative of $*$ in terms of time. $L.i$ is the current of inductor, $C.v$ is the voltage of capacitor.  Similarly, for Simulink, we obtain 
    \begin{equation} 
    \label{Comparison_RLC_Simulink}
    L\frac{{di}}{{dt}} = {V_{AC}} - Ri - \frac{1}{C}\int\limits_{ - \infty }^t {i(t)dt}
    \end{equation}
    
It is easy to verify  that these equations are equivalent in terms of their solutions,  but are distinct in appearance due to the different choices of state variables and procedures by which they are generated in each modeling system.  For instance, linear graph uses across (effort) and through (flow) variables,  but bond graph uses their integral forms as state variables. 

\subsection {Interoperability of lumped parameter languages and systems}

Broadly speaking, interoperability subsumes the problems of exchange and integration of simulation models created in different systems.  The latter  often manifests itself as the need for co-simulation \cite{gomes2017co} and/or for executable semantics in system modeling tools.    For example, system engineering languages, such as SysML \cite{friedenthal2014practical} need to integrate with simulation models in order to predict the modelled system behavior;  however, each specific  tool needs a different interface,  as proposed in \cite{qamar2009designing, sindico2011integrating} for 
Matlab/Simulink \cite{qamar2009designing, sindico2011integrating}, in \cite{omg2009sysml, johnson2012integrating} for Modelica,  and for bond graph in \cite{turki2005sysml}. 
An effort to overcome these challenges  was recently described in  \cite {bock2017extension}, where a SysML extension is proposed specifically for the purpose of generating such interfaces automatically.  One of our goals  in this effort is to provide formal semantics for this and other interoperability efforts.  
Without such formal semantics, integration of distinct simulation tools requires a non-trivial software development effort that must resolve individual assumptions and differences of distinct models.  The purpose of the recently proposed FMI standard  \cite{blochwitz2011functional} is to streamline and simplify such efforts through generation of uniform C code.  Unifying semantics of distinct models would support automatic generation of such FMI interfaces.

Model to model conversion is an effective method for achieving interoperability between different modeling languages.  The conversion between the bond graph and linear graph models has been studied in the 70's. Ort and Martens proposed a topological procedure for converting the bond graph to the linear graph by identifying the correspondence of bonds in a bond graph and edges of a linear graph \cite{ort1974topological}.  While theoretically not every bond graph has a corresponding  linear graph \cite{perelson1976bond},  Birkett gave a deterministic cut-and-paste method for converting any physical bond graph to the corresponding linear graph model \cite{birkett1989mathematical}.   Bond graphs can also be converted to equivalent models in Simulink or Modelica. Specifically, \cite{calvo2011analysis} used bond graphs through Simulink to analyze dynamic systems by transforming bond graphs to equivalent block diagrams; researchers in \cite{broenink1997bond, broenink1999object} observed that translating non-causal bond graph models to Modelica is in principle a straightforward process, even if causal assignment cannot be specified in Modelica.   The reverse conversion from Modelica to bond graphs has been studied in \cite{d2006bond}. Strictly speaking, not every Modelica model can be translated into a bond graph, because power continuity (the energy balance) is strictly enforced by the junction structure of bond graph but does  not have to be enforced by interconnection of components in Modelica \cite{borutzky2009bond}.  Similarly, not all Simulink models may be translated into Modelica models,  but translation of selective Simulink models to equivalent Modelica models was considered in \cite{dempsey2003automatic}.  Our goal is to identify a subset of physical models that is supported by all of these languages and provide unified semantics that would remove any ambiguities in such conversions and other interoperability tasks.  
 
A fundamental limitation underlying all of the above approaches is that none of them offer a systematic path towards interoperability with full 3D (spatially distributed) models governed by partial differential equations and discretized by finite element, volume, or difference methods.  This area of research is in its infancy and exemplified by rather  limited efforts.  For example,  in \cite{natsupakpong2010determination}  finite element model of 3D deformable objects are converted to the corresponding lumped mass-spring models by minimizing the difference of stiffness matrices of these two models.  Others proposed various methods for associating geometric information with components in the lumped parameter systems.  Thus, integral properties from solid models in CAD systems may be mapped to lumped parameters, and mating constraints in assembly models are mapped to transformers in the corresponding Modelica models \cite {engelson1999design, bunus2000mechanical, engelson2003mechanical}.  Such methods also allow translating Adams rigid body simulation models to Modelica code \cite{bowles2001feasibility}.  Additional constraints and annotations may be introduced in order to explicitly maintain bilateral correspondence between parameters of lumped parameter  and three-dimensional models \cite{canedo2016maintaining,engelson2004easy,hoeft2008design}.  By contrast,  the semantics of lumped parameter models proposed in this papers naturally extends to full 3D models of spatially distributed phenomena.   

\section{Algebraic topological models of lumped parameter systems}

Algebraic topological interpretations of network models and various types of electrical circuits are well known in literature \cite{branin1966algebraic,tonti2013mathematical} and are now a standard textbook material \cite{bamberg1991course}.  We now apply such an interpretation to lumped parameter models and show that it provides  a neutral standard semantics for all such systems. Superficially,  all algebraic topological formulations are identical, but important semantic difference emerge in details.   As a starting point,  we adopt Tonti's classification \cite{tonti2013mathematical} of physical theories in terms of their algebraic topological models; however, specific requirements of the lumped parameter system require significant extensions and modifications that we discuss below. Strictly speaking,  a proper setting for physical modeling is 4D spacetime.  But,  most engineering models are set in space$\times$time,  where space and time models are treated separately. 

\subsection{Lumped parameter models as cochains}
    
In algebraic topological view of physics, physical properties are distributed in spacetime over finite chunks of space called  $p$-dimensional cells, or $p$-cells,  ($p=0,1,2...$) that fit together to form a cell complex that decomposes the undelying physical space.   Many choices of cells are possible;  they can be open or closed, $p$-simplicies,  $p$-balls, or $p$-manifolds;  specific choices are dictated by convenience and applications and define the type of cell complex\cite{hatcher2001algebraic}.  All cells are endowed with orientation,  or sense of direction, which becomes important in order to properly assign signs to physical properties associated with cells.  

As we already saw, all lumped parameter models are formulated using 2-dimensional cells complexes:  0-cells (nodes), 1-cells (edges), and 2-cells (cycles or ``meshes'').   These complexes are abstract in the sense that geometric coordinates or shapes of the cells are immaterial;  only their connectivity carries important physical information.\footnote{This is in stark contrast to spatially distributed physical phenomena modeled governed by partial differential equation where geometry of cells becomes critical.}   The distribution of physical properties is described by  assigning their types and quantities to the individual cells in this complex.   The formal mechanism for doing so requires discretizing the property ${g}$ over $p$-cells $e_\alpha ^p$ as a \textit{$p$-cochain\/} ${C^p}$, a formal sum 

\begin{equation} \label{pcochain}
{C^p} = \sum\limits_{\alpha  = 1}^{{n_p}} {{g_\alpha }} e_\alpha ^p
\end{equation}

The relation between physical properties is governed by two types of fundamental laws: metric laws and topological laws. Metric laws usually involve measurement while topological laws relate physical properties associated with space and its boundary. Topological laws can be formulated using formal linear coboundary $\delta$ operations on cochains. Specifically, coboundary $\delta_p$ operating on a \textit{$p$}-cochain produces a \textit{$(p+1)$}-cochain by transferring and adding the coefficient of the \textit{$p$}-cochain  to its cofaces (Eq.\ref{cob_cochain}).   Formally, 
\begin{equation} \label{cob_cochain}
\delta_p ({C^p}) = \delta_p \left( {\sum\limits_{\alpha  = 1}^{{n_p}} {{g_\alpha }} e_\alpha ^p} \right) = \sum\limits_{\beta  = 1}^{{n_{p + 1}}} {\left( {\sum\limits_{\alpha  = 1}^{{n_p}} {{h_{\alpha \beta }} \cdot {g_\alpha }} } \right)}  \cdot e_\beta ^{p + 1},
\end{equation}
where $n_p$ represents the number of cells in the $p$-cochain. The incidence coefficient $h_{\alpha \beta} = [e_\alpha ^p,e_\beta^{p+1}] \in \{ 0, \pm 1\} $ is determined by relative orientation of   \textit{$p$}-cell ${e_\alpha ^p}$ and it cofaces \textit{$(p+1)$}-cell ${e_\beta ^{p+1}}$ \cite{tonti2013mathematical}.  If $e_\beta ^{p+1}$ is not a coface of ${e_\alpha ^p}$, then ${h_{\alpha \beta }} = 0$;  otherwise, if the orientations of ${e_\alpha ^p}$ and ${e_\beta ^{p+1}}$ are consistent, then ${h_{\alpha \beta }} = + 1$, otherwise, ${h_{\alpha \beta }} = - 1$.  If we denote the usual $p$-incidence matrix as ${\bf A} = \left[ h_{\beta\alpha}\right]$,  then the coboundary operator $\delta_p$ is commonly represented by its transpose ${\bf A}^T$. 

Informally, the coboundary operations capture the essence of balance, equilibrium, conservation, compatibility, and other topological laws.  For cochains on finite cell complexes, coboundary operators $\delta_p, p=1,2,3$ correspond to the usual vector calculus operators of gradient, curl, and divergence respectively.  The vector calculus identities $\nabla  \times (\nabla \phi ) = 0$, and $\nabla  \cdot (\nabla  \times {\bf{F}}) = 0$   are simply instances of the Poincare lemma stating that $\delta_{p} (\delta_{p-1} ()) = {\bf{0}}$, where $\bf{0}$ denotes a null cochain.  A collection of cochains and coboundary operators on a cell complex form a cochain complex\cite{hatcher2001algebraic}.

\subsection{Physical theories as Tonti diagrams} 

Every physical theory is conceptualized in terms of relationships between two types of dual physical quantities that are referred to by various authors as configuration/source\cite{tonti2013mathematical}, through/across\cite{branin1966algebraic}, or effort/flow\cite{karnopp1990system}. In what follows we will adopt Tonti's convention and distinguish between configuration type variables, that are modeled as cochains on primary cell complex decomposition of space, and source variables that are modeled as cochains on the dual cell complex decomposition of the same space.  This notion is illustrated in Figure \ref{fig:simplex_model},  where the primal cell complex is shown in blue the dual cell complex is shown in black.  (The actual geometry of cells is irrelevant for this discussion.)  
Each $p$-dimensional cell in the primal cell complex is dual to a unique $(n-p)$-dimensional cell in the dual cell complex.  
By duality, it also follows that the coboundary operator ${\widetilde \delta_{p}}$ on the dual cell complex can be represented by the transpose of  $\delta_p = {\bf A}^T$, or simply ${\bf A}$.

This conceptualization of physical quantities in terms cochains on dual cell complexes is not arbitrary:  it arises  from first principles based on how the postulated quantities are measured.  In each case,  the measurement process implies the intrinsic dimension of the associated quantity (e.g., displacements are measured at a point, currents are measured across the surface, voltage drop is measured along a path, and so on). The decision whether a particular quantity belongs to the primal or dual complex is determined by the oddness principle that requires change of sign under change of orientation of the relevant cell. The primal cells are are endowed with inner orientation, while the dual cells are oriented relative to the containing (outer) space.   The reader is referred to \cite{tonti2013mathematical} for detailed discussion of these concepts.

\begin{figure}[!htb]
        \centering
        \includegraphics[width=\linewidth]{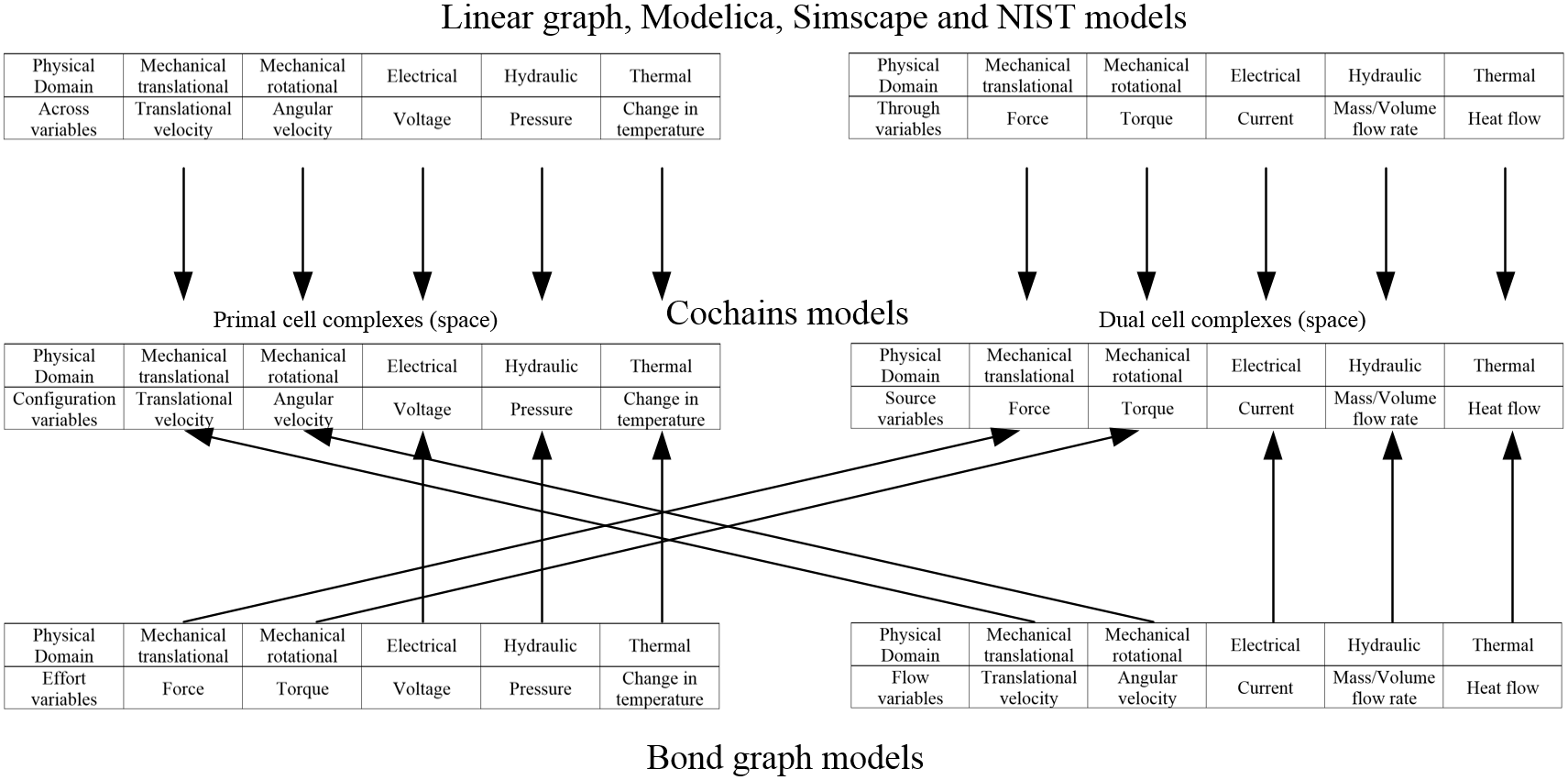}
        \caption{Categorization of physical variables category in different approaches to physical modeling}
        \label{fig:category_comparison}
    \end{figure}

Figure \ref{fig:category_comparison} shows the correspondence between the primal and dual cochains of variables and naming conventions used by different authors and lumped-parameter modeling systems.  We note that the adopted classification in terms of primal and dual cochains is consistent with the conventions in linear graphs \cite{rowell1997system}, Modelica \cite{elmqvist1997introduction}, Simscape \cite{chaturvedi2009modeling} and NIST models \cite{bock2017extension}, but differs slightly from that in
bond graphs \cite{kluever2015dynamic}.

\begin{figure}[!htb]
        \centering
        \includegraphics[width=.5\linewidth]{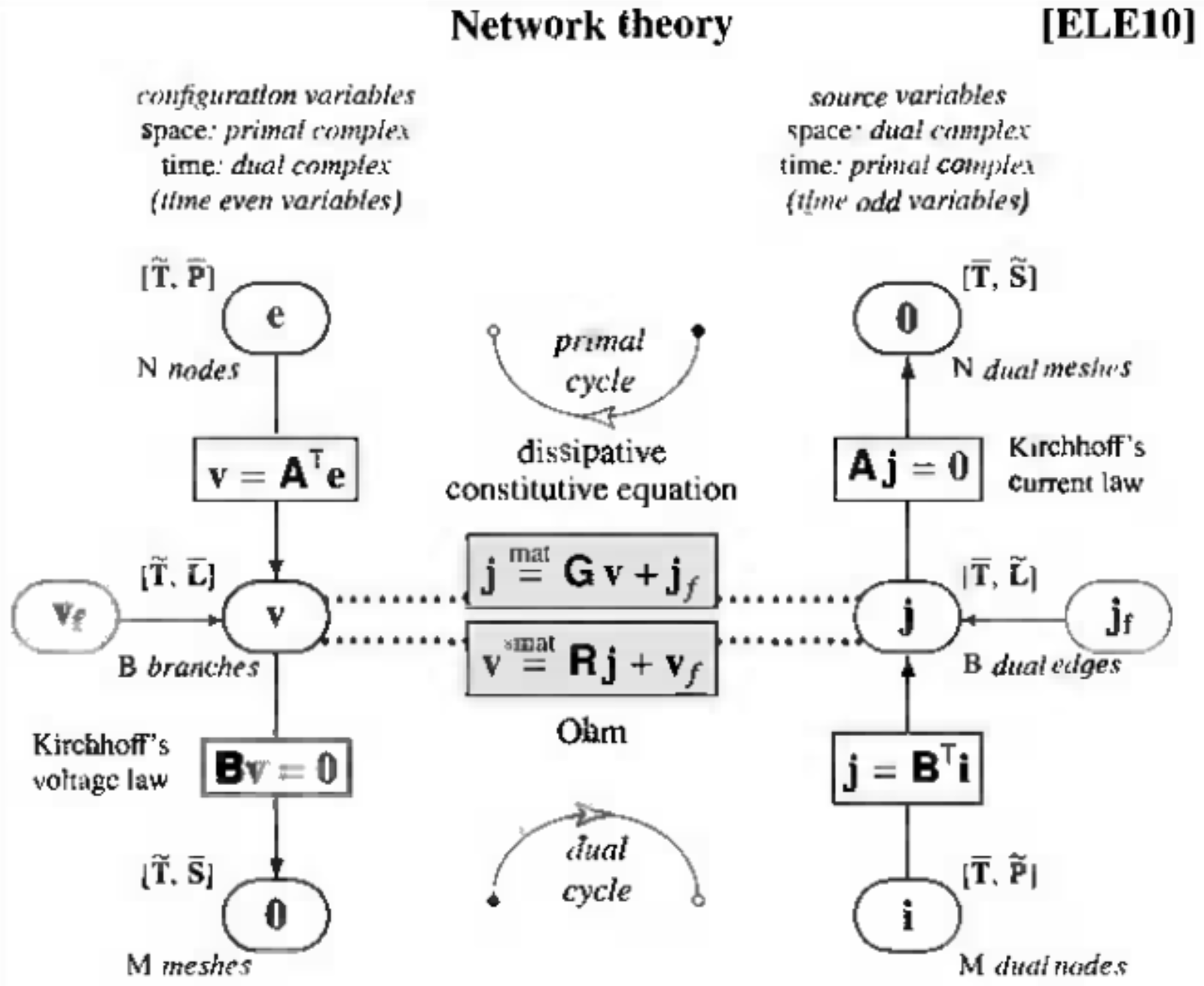}
        \caption{Tonti Diagram of network theory - constitutive equations are modified to account for voltage and current sources}
        \label{fig:ELE10_Sec3}
    \end{figure}

Physical laws (topological and metric) relate different types of variables within each physical theory.  Tonti proposed a systematic method for representing these laws using a diagram that can be considered an evolved combination of the so-called Roth diagrams \cite{Roth70} in terms of cochain sequences and ``Maxwell house'' diagram to represent all topological and metric relationships in electromagnetism \cite{bossavit1998computational}.  A simple example of such a diagram is the Tonti diagram of electrical (static) network theory  is shown in (Figure \ref{fig:ELE10_Sec3}).  It describes the network systems that satisfy Kirchhoff Current Law (KCL) and Kirchhoff Voltage Law (KVL)  using a pair of cochains complexes dual to each other \cite{tonti2013mathematical}. 

The diagram consists of two vertical sequences corresponding to the primal (left) and dual (right) cochain complexes, ordered by dimension.  The vertical arrows correspond to the coboundary operations, going down for primal cochains and going up for dual cochains.  Formally, the two sequences are exact\footnote{The cochain sequence is exact if it satisfies $\delta_{p} \circ \delta_{p-1} = 0$ ($p\geq 1$)} and form two dual cochain complexes:
\begin{equation}\label{primal-sequence}
	{\rm primal:} \ \ {\bf e}^0 
	\stackrel{\delta_0} {\longrightarrow}
	{\bf v}^1
	\stackrel{\delta_1} {\longrightarrow}
	{\bf 0}^2 
\end{equation}
\begin{equation}\label{dual-sequence}
	{\rm dual:} \ \ {\bf 0}^2 
	\stackrel{\widetilde \delta_1} {\longleftarrow}
	{\bf j}^1
	\stackrel{\widetilde \delta_0} {\longleftarrow}
	{\bf i}^0
\end{equation}

The measured relationships between dual quantities are represented by the horizontal links in the diagram.  In the case of network theory,  as shown in Figure  \ref{fig:ELE10_Sec3}, the primal variable are node potentials $\bf{e}$ associated with 0-cells, voltage drops $\bf{v}$ and sources associated with 1-cells, and voltage drops associated with 2-cells (meshes or cycles) that are identically $\bf{0}$ as the consequence of KVL.   The  cochains of adjacent dimensions satisfy topological laws expressed by the corresponding coboundary operations depicted as down-facing vertical arrows.  Thus, 1-cochain of voltage drops ${{\bf{v}}^{{1}}} = {{\bf{A}}^T}{{\bf{e}}^{{0}}}$  is implied by the coboundary operation $\delta_{0}$ on 0-cochain of node potentials ${{\bf{e}}^0}$;  and KVL is just a restatement of the Poicare lemma.  Similarly,  the dual source (current) variables:  0-cochain $\bf{i}$, 1-cochains $\bf{j}$, and 2-cochain $\bf{0}$  are indicated in the right branch of the diagram,   related by the sequence of two coboundary operations, indicated as arrows going up and expressing KCL.  

The constitutive relation between 1-cochain of voltage drops ${{\bf{v}}}$ and 1-cochain of currents $\bf{j}$ satisfies Ohm's Law (or its inverse).  The diagram also reveals two (dual) methods of generating the governing state equations for network models, depending on the choice of state variables. The two methods are indicated by primal and dual `cycles' which refer to two different ways of composing topological and metric laws.   For instance, in the primal cycle,  1-cochain voltage drop $\bf{v}$ is obtained by coboundary operation on 0-cochain node potential $\bf{e}$. Using the Ohm's law, $\bf{v}$ is converted to 1-cochain branch currents $\bf{j}$ in the dual cell complex, where coboundary operation $\bf{A}$ on $\bf{j}$ equals zero. The physical meaning of the latter coboundary operation $\bf{A}$ is that the algebraic sum of branch currents of a dual loop equals zero, as required by KCL.   Similarly, the dual cycle relies on KVL to generate the dual state equations for the same system.

\begin{figure}[!htb]
        \centering
        \includegraphics[width=.55\linewidth]{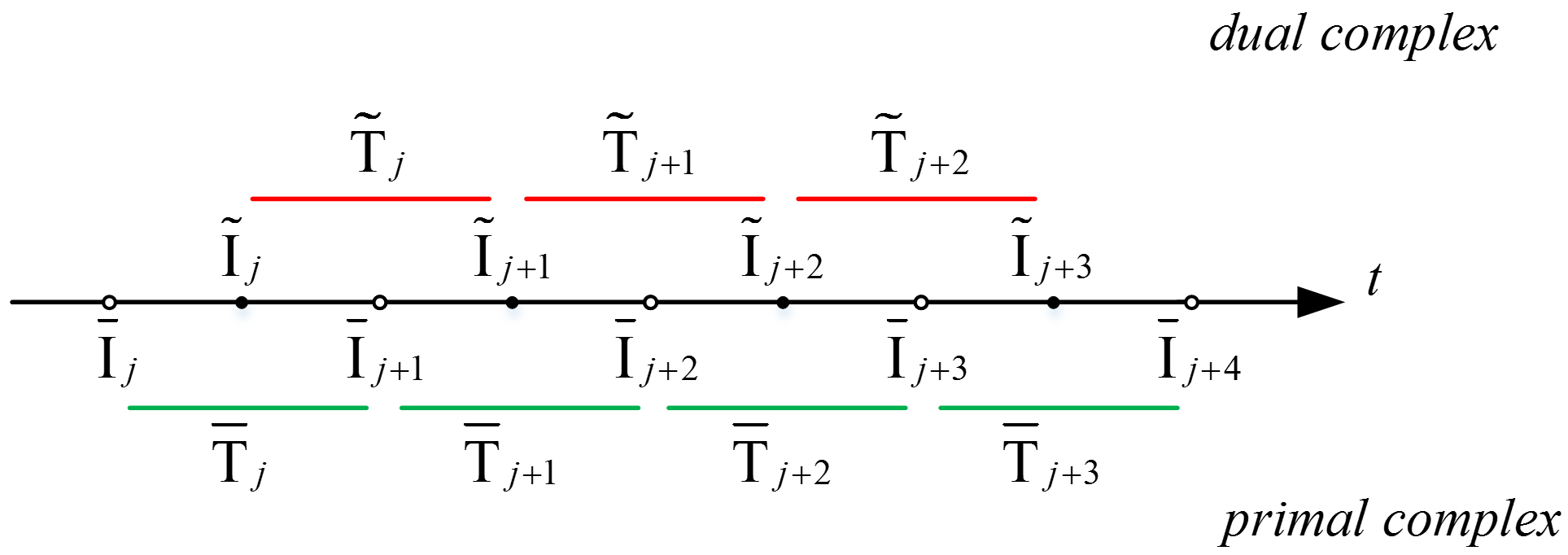}
        \caption{Cell complexes model of time}
        \label{fig:TimeElement}
        \end{figure}

Even though the diagram in  Figure \ref{fig:ELE10_Sec3} describes a static phenomena, the reader will notice that the  configuration and source variables are associated with time instances:  primal instances for source variables and dual instances for configuration variables.  This distinction becomes critical in dynamic physical models where,  once again, the primary time elements (0-dimensional instances and 1-dimensional intervals) are distinguished from the dual time elements based on the oddness principle that requires sign change under reversal of motion \cite{tonti2013mathematical}.
Strictly speaking, a proper setting for all physical models is a 4D spacetime, which we chose to represent as direct product of space and time.  In other words, for each type of spatial variable, we can also consider its behavior in time which is represented by a pair of dual 1-dimensional time complexes, as shown in  Figure \ref{fig:TimeElement}.   Here 0-cells ${\overline I _*}$ and 1-cells ${\overline T _*}$  represent primal time instances and intervals, while 0-cells ${\widetilde I_*}$ and 1-cells ${\widetilde T_*}$  represent the dual  time instances and intervals, respectively. 

    \begin{figure}[!htb]
        \centering
        \includegraphics[width=.5\linewidth]{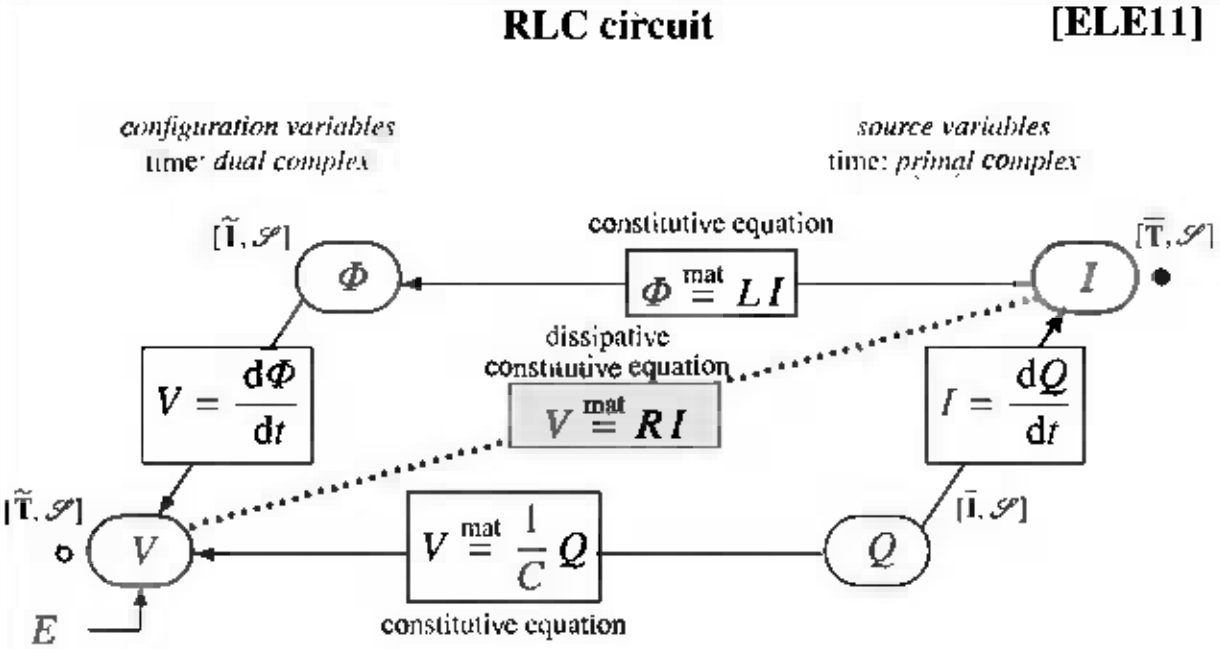}
        \caption{Tonti diagram of RLC circuit - only voltage sources are included}
        \label{fig:RLC_Tonti_ELE11}
    \end{figure}

\begin{figure}[!htb]
        \begin{subfigure}{.5\textwidth}
              \centering
              \includegraphics[width=.8\linewidth]{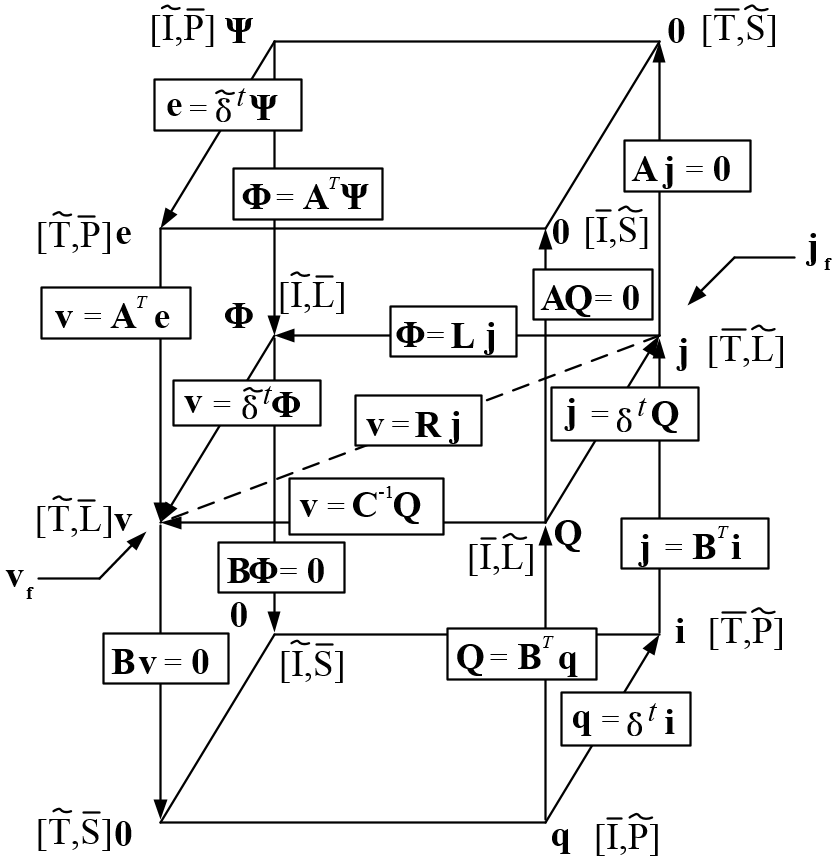}
              \caption{Matrix operators on dual cell complexes}  
              \label{fig:Overall_RLC_M}
        \end{subfigure}%
        \begin{subfigure}{.5\textwidth}
              \centering
              \includegraphics[width=.8\linewidth]{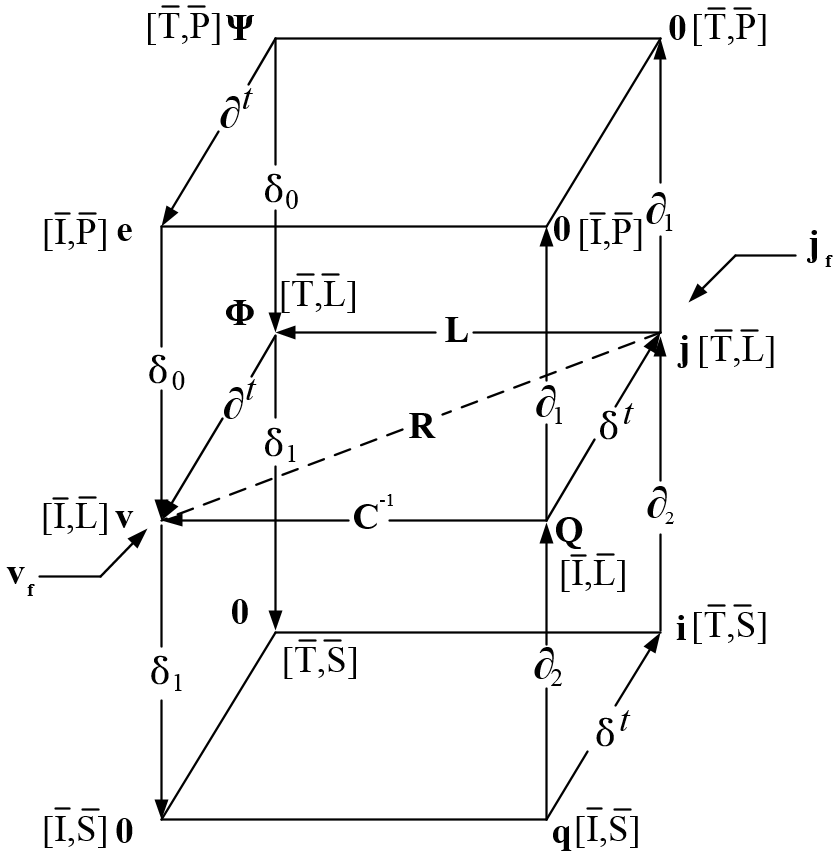}
              \caption{Topological and constitutive relationships on a single cell complex}
              \label{fig:Overall_RLC_T}
        \end{subfigure}
            \caption{Extended Tonti diagram for RLC network system - with voltage and current sources}
            \label{fig:Overall_RLC}
    \end{figure}

The introduction of time cell complexes in time has two consequences.   First,  it identifies the usual time derivative with the corresponding coboundary operator $\delta^t$. Thus, if $\bf{q}$ is a 0-cochain of primal (or dual) time instances, then   ${\overline I}$ (respectively, ${\widetilde I}$) is the 1-cochain of time intervals of $\bf{q}$ defined by ${\delta _0^t}\bf{q}$ (respectively ${\widetilde \delta _0^t}\bf{q}$).  On a finite cell complex, ${\delta _0^t}$ become the time difference operation.  
Secondly, all space Tonti diagrams now acquire an additional time dimension, giving rise to horizontal sections of the diagram.  One such section is shown in   Figure \ref{fig:RLC_Tonti_ELE11},  which corresponds to the Tonti diagram of RLC circuit systems \cite{tonti2013mathematical}.
Here, the usual differential relations $V = {{d\Phi } \mathord{\left/
 {\vphantom {{d\Phi } {dt = }}} \right.
 \kern-\nulldelimiterspace} {dt}}$ and $I = {{dQ} \mathord{\left/
 {\vphantom {{dQ} {dt}}} \right.
 \kern-\nulldelimiterspace} {dt}}$  are consequence of  topological relations ${\bf{V}} = \tilde \delta _0^t{\bf{\Phi }}$ and ${\bf{I}} = \delta _0^t{\bf{Q}}$, respectively, where $\Phi$ is magnetic flux and $Q$ is electric charge.  Two new constitutive equations describe the capacitance relation between electric charge ${{{Q}}}$ and voltage drop ${{{v}}}$ and the inductance relation between magnetic flow ${{{\Phi}}}$ and currents ${{{j}}}$.
The second time derivative is the result of composition of two first time derivatives.   For instance, the second order equation differential equation $V = {{d\Phi } \mathord{\left/
 {\vphantom {{d\Phi } {dt = }}} \right.
 \kern-\nulldelimiterspace} {dt = }}{{Ld\left( I \right)} \mathord{\left/
 {\vphantom {{Ld\left( I \right)} {dt = }}} \right.
 \kern-\nulldelimiterspace} {dt = }}{{L{d^2}Q} \mathord{\left/
 {\vphantom {{L{d^2}Q} {d{t^2}}}} \right.
 \kern-\nulldelimiterspace} {d{t^2}}}$ 
 in Figure \ref{fig:RLC_Tonti_ELE11} can be expressed as  
 ${\bf{V}} = \tilde \delta _0^t{\bf{\Phi }} = {\bf{L}}\tilde \delta _0^t{\bf{I}} = \tilde \delta _0^t{\bf{L}} \delta _0^t{\bf{Q}}$ \cite{tonti2013mathematical, ferretti2015cell}.
Finally,  we note that the vertical space  diagram in Figure \ref{fig:ELE10_Sec3} and the horizontal time diagram in Figure \ref{fig:RLC_Tonti_ELE11} can be combined into a single three-dimensional diagram, as described in Section 4 and shown in Figure \ref{fig:Overall_RLC_M}.

In addition to physical quantities in Figure \ref{fig:ELE10_Sec3} and \ref{fig:RLC_Tonti_ELE11}, the diagram in  Figure \ref{fig:Overall_RLC_M} includes  magnetic flux potential ${{{\Psi}}}$ and  mesh electric charge ${{{q}}}$ that are related by coboundary operators to magnetic flux $\Phi$ and electric charge $Q$ respectively.  In other words, the extended Tonti diagram includes two additional cochain complexes defined by the two sequences:   
\begin{equation}\label{primal-sequence-t}
	{\rm primal:} \ \ {\bf \Psi}^0 
	\stackrel{\delta_0} {\longrightarrow}
	{\bf \Phi}^1
	\stackrel{\delta_1} {\longrightarrow}
	{\bf 0}^2 
\end{equation}
\begin{equation}\label{dual-sequence-t}
	{\rm dual:} \ \ {\bf 0}^2 
\stackrel{\widetilde \delta_1} {\longleftarrow}
	{\bf Q}^1
\stackrel{\widetilde \delta_0} {\longleftarrow}
	{\bf q}^0
\end{equation}

 Additional horizontal arrows indicate the corresponding time coboundary (derivative) operations.  
With the added constitutive, topological equations and physical variables, the extended Tonti diagram includes all possible quantities, as well as constitutive and topological laws of RLC electrical circuit systems.   In the following section, we explicitly distinguish between the constitutive and topological laws as they are commonly used in practice in a single cell complex network model that is shown in Figure \ref{fig:Overall_RLC_T}.

 
\subsection{Dual cochain complexes on a single cell complex} 

\begin{figure}[!htb]
        \centering
        \includegraphics[width=.8\linewidth]{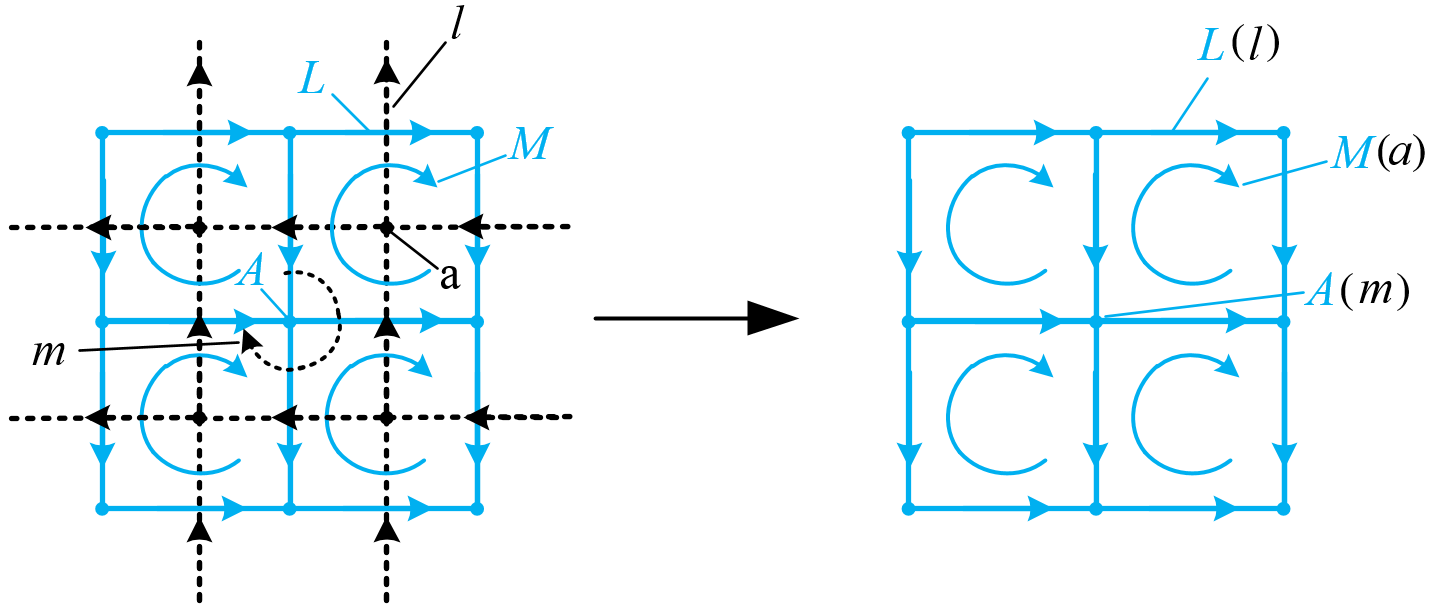}
        \caption{Dual cochain complexes on a single cell complex}
        \label{fig:simplex_model}
    \end{figure}

The ultimate goal of physical modeling is numerical simulation which is usually performed on a discretization of spacetime.  If one were to accept that every physical theory is formulated in terms of dual cochain complexes,  it would be reasonable to expect that most modeling and simulation tools are also formulated in terms of dual discretizations (one for configuration variables and another one for the source variable) as illustrated  in Figure \ref {fig:simplex_model} (on the left). In fact, such dual discretizations are often advocated in literature as more natural and numerically stable alternatives, for example, in mimetic discretization schemes \cite{castillo2013mimetic},  cell methods \cite{tonti2014starting,ferretti2015cell},  discrete exterior calculus \cite{hirani2003discrete} and other modeling approaches. However, the vast majority of numerical schemes appear to be based on a single discretization of space, which supports evaluation of both primal and dual cochains.  For example, in most finite element, finite difference, and finite volume methods, all configuration and source variables are associated with cells (often nodes) in the underlying mesh, and their duality is hidden within the numerical scheme itself \cite{Mattiussi}.   
 
In lumped parameter models, dual discretizations are particularly counter-intuitive,  since all spatially distributed properties have already been integrated (lumped) and only connectivity of the underlying cell complex remains visible.  That connectivity often directly corresponds to the physical embedding; for example, a single electrical network carries both voltage and current information.   Similarly,  general network model is a single cell complex where primal and dual cochains are represented.  The mapping of dual cochains on the primal cell complex is straightforward and is accomplished by mapping the dual $(n-p)$-cells to their corresponding primal $p$-cells, as shown in Figure \ref{fig:simplex_model}.
Consider how this mapping would work for the cochains in Figure \ref{fig:ELE10_Sec3}.  
With this mapping, node potentials $\bf{e}$, voltage drops $\bf{v}$ and sources, and voltage drops of cycles  $\bf{0}$ are still  associated with primal 0-cells (e.g. \textit{A}), 1-cells (e.g. \textit{L}), and 2-cells (e.g. \textit{M}) respectively.  However, the dual source (current) variables:  mesh currents $\bf{i}$, branch currents $\bf{j}$, and  currents merging at nodes $\bf{0}$ which were originally associated with dual 0-cells (e.g. \textit{a}), 1-cells (e.g. \textit{l}), and 2-cells (e.g. \textit{m}) are now  associated with primal 2-cells (e.g. \textit{M}), 1-cells (e.g. \textit{L}), and 0-cells (e.g. \textit{A}) respectively.

With this mapping, all $(n-p)$-coboundary operations in dual cell complexes would become the $p$-boundary operations $\partial$ in the primal cochain complex, which operate on a \textit{$p$}-cochains and produce a \textit{$(p-1)$}-cochains:
 \begin{equation} \label{b_cochain}
\partial_p ({C^p}) = \partial_p (\sum\limits_{\alpha  = 1}^{{n_p}} {{g_\alpha }} e_\alpha ^p) = \sum\limits_{\beta  = 1}^{{n_{p - 1}}} {\left( {\sum\limits_{\alpha  = 1}^{{n_p}} {{h_{\alpha \beta }} \cdot {g_\alpha }} } \right)}  \cdot e_\beta ^{p - 1}, 
\end{equation} 
which is similar to Eq.\ref{cob_cochain}, except that the coefficients are transferred from $p$-cells to their $(p-1)$-faces.   This implies that the coboundary operation ${\widetilde \delta _{n - p}}$ on the dual cells complex is mapped into the boundary operation $\partial_p$ on the primal cell complex,\footnote{This observation also justifies use of $\partial_p$ operator, usually reserved for  $p$-chains, on $p$-cochains} i.e., 
        \begin{equation} \label{primalB_dualCoB}
        {\partial _p} = {\widetilde \delta _{n - p}}
        \end{equation}
This explains, for example, why the  KCL on the primal complex is described by condition ${\partial _1}{{\bf{j}}^1} = {\bf{0}}$, stating that the branch currents must add up to zero at every node.   

The single complex network model is summarized by a new type of Tonti diagram shown in Figure \ref{fig:Overall_RLC_T}.   Here the dual (source) cochains have been mapped to  the corresponding cochains on the primal cell complex to form the dual cochain complex with boundary $\partial_p$ operators replacing the original dual ${\widetilde \delta _{n - p}}$ coboundary operators.  In other words,  the dual cochain sequences  (\ref{dual-sequence}) and (\ref{dual-sequence-t}) become respectively 

\begin{equation}\label{dual-sequence-primal}
	{\bf i}^2 
	\stackrel{\partial_2} {\longrightarrow}
	{\bf j}^1
	\stackrel{\partial_1} {\longrightarrow}
	{\bf 0}^0
\end{equation}
\begin{equation}\label{dual-sequence-t-primal}
	{\bf q}^2 
	\stackrel{\partial_2} {\longrightarrow}
	{\bf Q}^1
	\stackrel{\partial_1} {\longrightarrow}
	{\bf 0}^0
\end{equation}
  
In the rest of the report, we will assume that an algebraic topological model of a lumped parameter system is described by such single (primal) cell complex and a corresponding Tonti diagram with four  cochain complexes (correspong to the four vertical `legs' of the diagram) and all relationships between them.\footnote{Strictly speaking, use $\partial_p$ operators give rise to chain complexes, but it should be clear that these chain complexes on the primal cell complex are isomorphic to the cochain complexes over the dual cell complex.}    The two primal cochain complex include:   (1) 0-cochain node potentials ${{\bf{e}}^0}$, 1-cochain voltage drops ${{\bf{v}}^1}$, 2-cochain of mesh voltages (which is ${\bf{0}}$ by KVL); and (2) 0-cochain magnetic flux potentials ${{\bf{\Psi}}^0}$, 1-cochain magnetic fluxes  ${{\bf{\Phi}}^1}$, and 2-cochain of mesh magnetic fluxes and 2-cochain of mesh voltages (which is ${\bf{0}}$ by KVL).  The cochains in each complex are related by the spatial coboundary operators,  while the corresponding $p$-cochains in the two complexes are related by the time boundary operator.   Similarly,   the two dual cochain complexes are:  (1) 0-cochain of node currents (which is ${\bf{0}}$ by KCL), 1-cochain currents ${{\bf{j}}^1}$, 2-cochain mesh currents ${{\bf{i}}^2}$;  and (2) 0-cochain of node electric charges,  1-cochain electric charges ${{\bf{Q}}^1}$, 2-cochain mesh electric charges ${{\bf{q}}^2}$.   The cochains in each of the dual complexes related by the spatial boundary operators, while the corresponding $p$-chains in the two dual complexes are related by the time coboundary operators.  

\section{Single-domain lumped parameter systems} 
\label{single}


In this section, we will show how to use Tonti diagrams to describe lumped parameter systems and introduce an automated method of generating system state equations. We will focus on lumped parameter models of a single physical domain, exemplified by classical RLC electrical circuit systems.  Application to other physical domains is immediate, since all such models are isomorphic. 

\subsection{Static systems}

\begin{figure}[!htb]
		\centering
		\includegraphics[width=.35\linewidth]{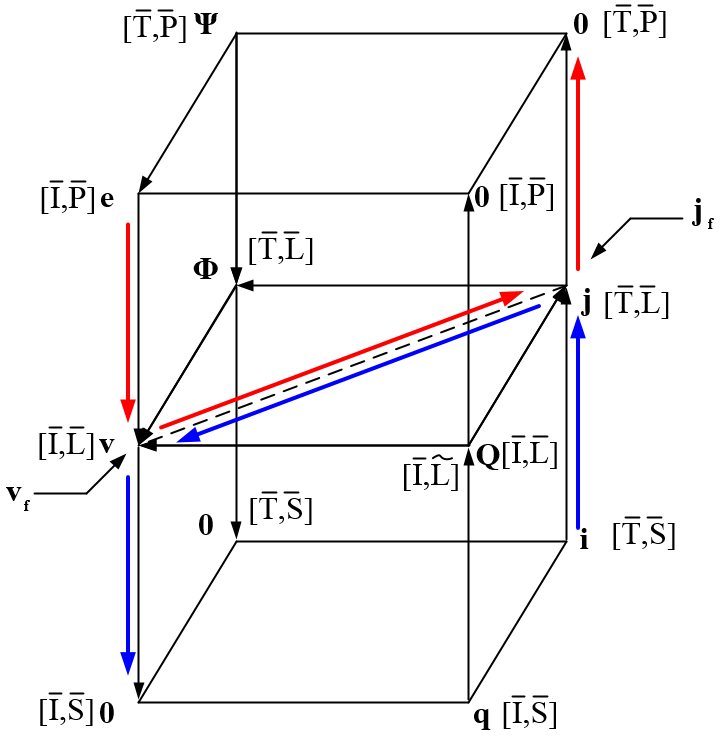}
		\caption{Paths corresponding to primal and dual cycle in Figure \ref{fig:ELE10_Sec3}}
		\label{fig:routeEx_static}
	\end{figure}

     Classical single-domain RLC electrical circuits consist of five types of physical elements: resistors, capacitors, inductors, voltage sources and current sources. We will first consider a special case of static resistive circuits that contain only constant resistors,  as well as voltage and current sources; later we will extend the approach to general dynamic electrical circuits. The algebraic topological structure of static electrical circuits relies only on the two dual cochain complexes modeled over a single cell complex.  The primal cochain complex includes 0-cochain of node potentials ${{\bf{e}}^0}$, 1-cochain of voltage drops ${{\bf{v}}^1}$, and 2-cochain of mesh voltages (which is ${\bf{0}}$ by KVL), related by the coboboundry operators;  the dual cochain complex consists of 0-cochain of node currents  (which is ${\bf{0}}$ by KCL),   1-cochain of branch currents ${{\bf{j}}^1}$, and 2-cochain mesh currents ${{\bf{i}}^2}$ related by the boundary operators. The topological and constitutive relations between these cochains are given by the diagram in Figure \ref{fig:ELE10_Sec3},  and two methods of generating the equations are indicated by primal and dual `cycles' respectively in the diagram.  In the context of the more general model, these cycles correspond to red and blue paths in the extended Tonti diagram as shown in (Figure \ref{fig:routeEx_static}).  Each path is defined by a sequence of the arrows in the diagram indicating composition of the corresponding physical laws.
     
For example, if we use the red path in Figure \ref{fig:routeEx_static}, then the 0-cochain ${{\bf{e}}^0}$ is selected as the state variable. The system state equation can be generated by composition of  three  steps starting with a 0-cochain ${{\bf{e}}^0}$ of node potentials.  First,  the downward red arrow indicates that the node potentials give rise to voltage drops associated with incident branches using the coboundary operator 
${\bf{v}}^1 = {\delta _0}{\bf{e}}^0$.   The second lateral red arrow correspond to the constitutive relation ${{\bf{j}}^1}  = {\bf{G}}\left( {{{\bf{v}}^1} +{\bf{v}}_{\rm{f}}^1} \right) +  {\bf{j}}_{\rm{f}}^1$, which accounts for the contribution of  1-cochains of voltage sources  ${\bf{v}}_{\rm{f}}^1$ and source currents ${\bf{j}}_{\rm{f}}^1$ in each branch of the electrical network.   Finally,  the upward vertical red arrow indicates the application of KCL ${\partial _1}{{\bf{j}}^1} = {\bf{0}}$ requiring that branch currents add up to zero.   The composition of the three laws yields:

\begin{equation} \label{node potential_0_positive}
\partial_1 \left( 
{\bf{G}} \left( 
\delta _0 {\bf{e}}^0 +{\bf{v}}_{\rm{f}}^1 
\right) 
+  {\bf{j}}_{\rm{f}}^1 
\right)
={\bf 0}
\end{equation}
Eq.\ref {node potential_0_positive} is usually written in a more traditional form as
\begin{equation} \label{node potential_0}
\partial_1 \left( 
{\bf{G}} \left( 
-\delta _0 {\bf{e}}^0 -{\bf{v}}_{\rm{f}}^1 
\right) 
+  {\bf{j}}_{\rm{f}}^1 
\right)
={\bf 0}
\end{equation}
As explained in \cite{tonti2013mathematical}, the minus sign in front of $\delta _0 {\bf{e}}^0$ is due to the largely historical assumption that the vertices (0-cells) of the primal cell complex are oriented as sinks.  The second minus sign in front of ${\bf{v}}_{\rm{f}}^1$ signifies the fact that a (positive) voltage source should be subtracted from the (positive) voltage drop in every branch.   


An alternative method for generating the system state equations follows the dual blue path in Figure \ref{fig:routeEx_static}.  The process starts with the dual 2-cochain of mesh currents ${{\bf{i}}^2}$ selected as the state variable and amounts to  composition of the analogous three steps:  boundary operator $\partial_2$ applied to mesh currents in order to generate branch currents, constitutive Ohm's law ${\bf R}$ that relates the branch currents to voltage drops, and coboundary operator $\delta_1$ applied to the voltage drops in accordance with KVL.  Taking into account the voltage and current sources, the process results in: 
\begin{equation} \label{meshCurrent_0}
{\delta _1}\left( {{\bf{R}} ( {\partial _2}{{\bf{i}}^2} - {\bf{j}}_{\rm{f}}^1} \right) + {\bf{v}}_{\rm{f}}^1)  = {\bf{0}}
\end{equation}


Collecting the terms with known voltage and current sources and moving them to the right hand side, the system state equations Eq.\ref{node potential_0} and  Eq.\ref{meshCurrent_0} transform to Eq.\ref{node potential} and Eq.\ref{meshCurrent}, respectively.  
\begin{equation} \label{node potential}
{\partial _1}{\bf{G}}{\delta _0}{{\bf{e}}^0} = {\partial _1}\left( {{\bf{j}}_{\rm{f}}^1 - {\bf{Gv}}_{\rm{f}}^1} \right)
\end{equation}
\begin{equation} \label{meshCurrent}
{\delta _1}{\bf{R}}{\partial _2}{{\bf{i}}^2} = {\delta _1}\left( { - {\bf{v}}_{\rm{f}}^1 + {\bf{Rj}}_{\rm{f}}^1} \right)
\end{equation}
When boundary and cobounadry operators are replaced by the corresponding incidence matrices describing a cell complex underlying a specific lumped parameter system,  the state equations become systems of linear equations that can be solved for the unknown state variables (node potentials ${\bf{e}^0}$ and mesh currents ${\bf{i}^2}$ respectively).

        
    \begin{figure}[!htb]
        \begin{subfigure}{.5\textwidth}
              \centering
              \includegraphics[width=.95\linewidth]{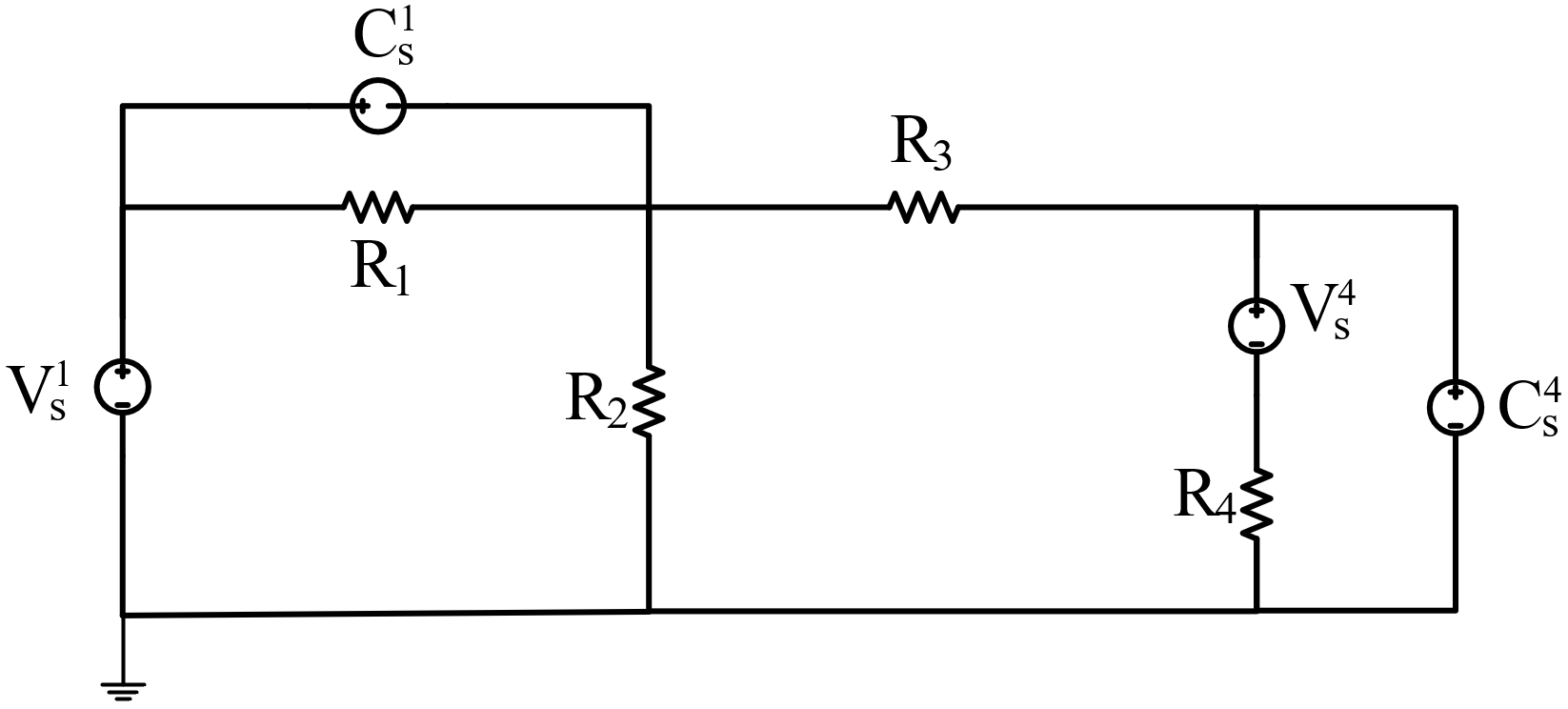}
              \caption{An electrical circuit of pure resistance}
              \label{fig:R problem}
        \end{subfigure}%
        \begin{subfigure}{.5\textwidth}
              \centering
              \includegraphics[width=.5\linewidth]{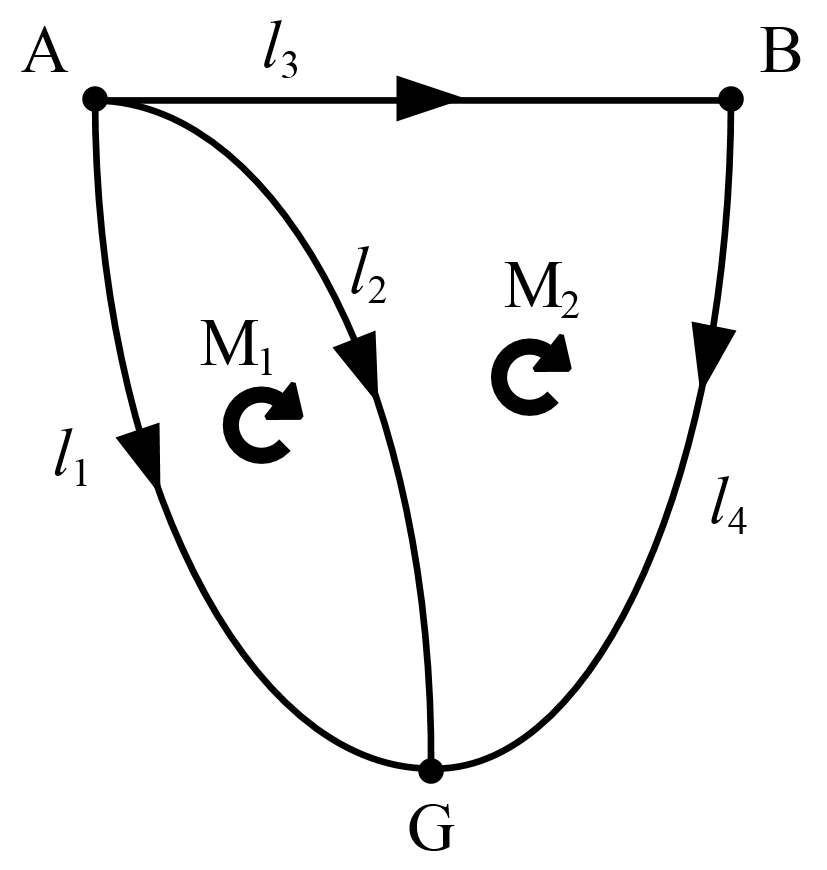}
              \caption{Topological structure}
              \label{fig:RTopological structure}
        \end{subfigure}
            \caption{An electrical circuit of pure resistance and its topological structure}
            \label{fig:R}
    \end{figure}   

\begin{exmp}
We will use an example of static electrical circuit in Figure \ref{fig:R problem} to illustrate the derivation of Eq.\ref{node potential} and Eq.\ref{meshCurrent} in concrete setting.   The shown electrical network contains four constant resistors  ${R_1} = 6\Omega $, ${R_2} = 3\Omega $, ${R_3} = 2\Omega $, ${R_4} = 4\Omega $,  two constant current sources $C_s^1 = 3A$, $C_s^4 = 1A$ and two constant voltage sources $V_s^1 = 8V$, $V_s^4 = 4V$.  Topologically,  the network is a 2-dimensional complex shown in Figure \ref{fig:RTopological structure} and consisting of three 0-cells (A,B,G), four 1-cells(${l_1}$,${l_2}$,${l_3}$,${l_4}$) and two 2-cells(${M_1}$,${M_2}$).  

The algebraic topological model of the electrical circuit contains: primal 0-cochain node potentials (${{\bf{e}}^0} = {{e}_A} \cdot A + {{e}_B} \cdot B + {{{e}}_G} \cdot G$), primal 1-cochain voltage drops (${{\bf{v}}^1} = {V_1} \cdot {l_1} + {V_2} \cdot {l_2} + {V_3} \cdot {l_3} + {V_4} \cdot {l_4}$), dual 1-cochain of branch currents (${{\bf{j}}^1} = {j_1} \cdot {l_1} + {j_2} \cdot {l_2} + {j_3} \cdot {l_3} + {j_4} \cdot {l_4}$), dual 2-cochain of mesh currents (${{\bf{i}}^2} = {i_1} \cdot {M_1} + {i_2} \cdot {M_2}$) and two cochains that are always ${\bf{0}}$: the dual 0-cochain of node currents and the primal 2-cochain of loop voltage drops.  In order to get unique solution of the state equations, we consider 0-cell G as the reference node,  implying the boundary condition of $e_G=0$.  Following the red path on the Tonti diagram generates equation  Eq.\ref{node potential} with individual terms as follows:  

\begin{equation} \label{boundary10}
{\partial _1} = \left[ {\begin{array}{*{20}{c}}
{ - 1}&{ - 1}&{ - 1}&0\\
0&0&{ + 1}&{ - 1}\\
{ + 1}&{ + 1}&0&{ + 1}
\end{array}} \right] 
\end{equation}

\begin{equation} \label{boundaryTranspose}
{\delta _0} = \partial _1^T
 \end{equation}   
 
		\begin{equation} \label{InverseResistanceMatrix} 
		{\bf{G}} = \left[ {\begin{array}{*{20}{c}}
		{R_1^{ - 1}}&0&0&0\\
		0&{R_2^{ - 1}}&0&0\\
		0&0&{R_3^{ - 1}}&0\\
		0&0&0&{R_4^{ - 1}}
		\end{array}} \right] = \left[ {\begin{array}{*{20}{c}}
		{{1 \mathord{\left/
 		{\vphantom {1 6}} \right.
		 \kern-\nulldelimiterspace} 6}}&0&0&0\\
		0&{{1 \mathord{\left/
		 {\vphantom {1 3}} \right.
		 \kern-\nulldelimiterspace} 3}}&0&0\\
		0&0&{{1 \mathord{\left/
		 {\vphantom {1 2}} \right.
 		\kern-\nulldelimiterspace} 2}}&0\\
		0&0&0&{{1 \mathord{\left/
		 {\vphantom {1 4}} \right.
		 \kern-\nulldelimiterspace} 4}}
		\end{array}} \right]
		\end{equation} 

        \begin{equation} \label{currentsource}
{\bf{j}}_{\rm{f}}^1 = {\left[ {\begin{array}{*{20}{c}}
{C_s^1}&0&0&{ - C_s^4}
\end{array}} \right]^T} = {\left[ {\begin{array}{*{20}{c}}
3&0&0&{ - 1}
\end{array}} \right]^T}
\end{equation}
        
\begin{equation} \label{voltagesource}
{\bf{v}}_{\rm{f}}^1 = {\left[ {\begin{array}{*{20}{c}}
{V_s^1}&0&0&{V_s^4}
\end{array}} \right]^T} = {\left[ {\begin{array}{*{20}{c}}
8&0&0&4
\end{array}} \right]^T}   
\end{equation}

Substituting  Eq.\ref{boundary10} $\sim$ Eq.\ref{voltagesource} into Eq.\ref{node potential} we obtain a linear system of equations that has the solution of  ${{\bf{e}}^0} = {\left[ { - {1 \mathord{\left/
 {\vphantom {1 {2,\;{7 \mathord{\left/
 {\vphantom {7 {3,\;0}}} \right.
 \kern-\nulldelimiterspace} {3,\;0}}}}} \right.
 \kern-\nulldelimiterspace} {2,\;{7 \mathord{\left/
 {\vphantom {7 {3,\;0}}} \right.
 \kern-\nulldelimiterspace} {3,\;0}}}}} \right]^T}$. 
Once the 0-cochain node potentials ${{\bf{e}}^0}$ is known, it is easy to obtain 1-cochain voltage drops ${{\bf{v}}^1}$ and 1-cochain branch currents ${{\bf{j}}^1}$ by using the topological equation ${{\bf{v}}^1} =  - \left( {{\delta _0}{{\bf{e}}^0}} \right)$ and the constitutive equation ${{\bf{j}}^1} - {\bf{j}}_{\rm{f}}^1 = {\bf{G}}\left( {{{\bf{v}}^1} - {\bf{v}}_{\rm{f}}^1} \right)$.

Similarly following the  blue path in the Tonti diagram, the generated system state equation Eq.\ref{meshCurrent} would involve:  

        \begin{equation} \label{coboundary12}
{\delta _1} = \left[ {\begin{array}{*{20}{c}}
{ - 1}&{ + 1}&0&0\\
0&{ - 1}&{ + 1}&{ + 1}
\end{array}} \right]
        \end{equation}
        
        \begin{equation} \label{coboundaryTranspose}
\partial _2 = {\delta _1}^T
 		\end{equation}  

        \begin{equation} \label{ResistanceMatrix} 
        {\bf{R}} = \left[ {\begin{array}{*{20}{c}}
		{{R_1}}&0&0&0\\
		0&{{R_2}}&0&0\\
		0&0&{{R_3}}&0\\
		0&0&0&{{R_4}}
		\end{array}} \right] = \left[ {\begin{array}{*{20}{c}}
		6&0&0&0\\	
		0&3&0&0\\
		0&0&2&0\\
		0&0&0&4
		\end{array}} \right]
        \end{equation} 

Substituting Eq.\ref{currentsource} $\sim$ Eq.\ref{ResistanceMatrix} into Eq.\ref{meshCurrent}, we obtain another system of linear equations that yields the solutions for the mesh currents ${{\bf{i}}^2} = {\left[ {\begin{array}{*{20}{c}}
{ - {{19} \mathord{\left/
 {\vphantom {{19} {12}}} \right.
 \kern-\nulldelimiterspace} {12}}}&{ - {{17} \mathord{\left/
 {\vphantom {{17} {12}}} \right.
 \kern-\nulldelimiterspace} {12}}}
\end{array}} \right]^T}$. The 1-cochain branch currents ${{\bf{j}}^1}$ and 1-cochain voltage drops ${{\bf{v}}^1}$ are immediately obtained by applying the topological relation
${{\bf{j}}^1} = {\partial _2}{{\bf{i}}^2}$  and the constitutive equation ${{\bf{v}}^1} - {\bf{v}}_{\rm{f}}^1 = {\bf{R}}\left( {{{\bf{j}}^1} - {\bf{j}}_{\rm{f}}^1} \right)$.
\end{exmp}

\subsection{Dynamic systems}

\begin{figure}[!htb]
        \centering
        \begin{subfigure}[b]{0.24\textwidth}
              \centering
              \includegraphics[width=\linewidth]{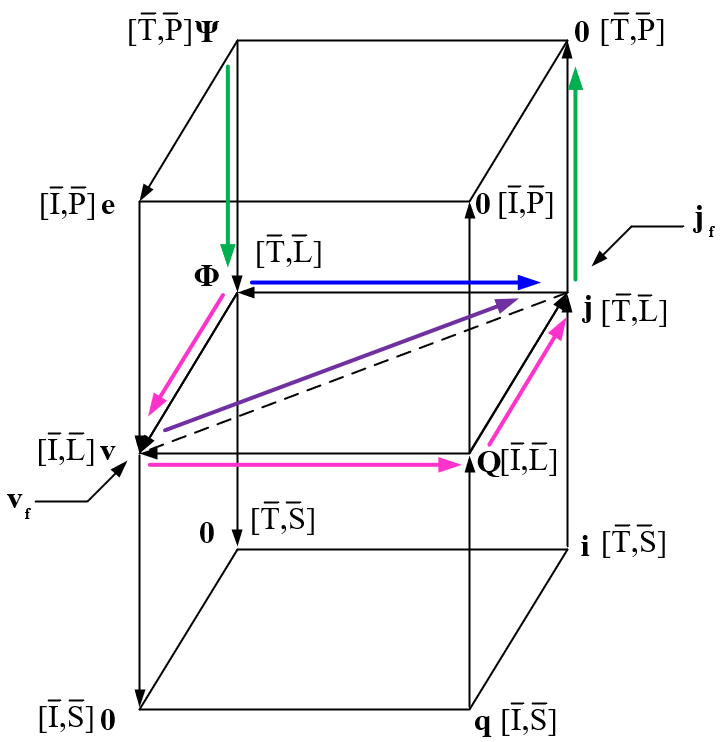}
              \caption{State variable: ${{\bf{\Psi }}^0}$ }
              \label{fig:state_variable_psi0}
        \end{subfigure}
        \begin{subfigure}[b]{0.24\textwidth}
              \centering
              \includegraphics[width=\linewidth]{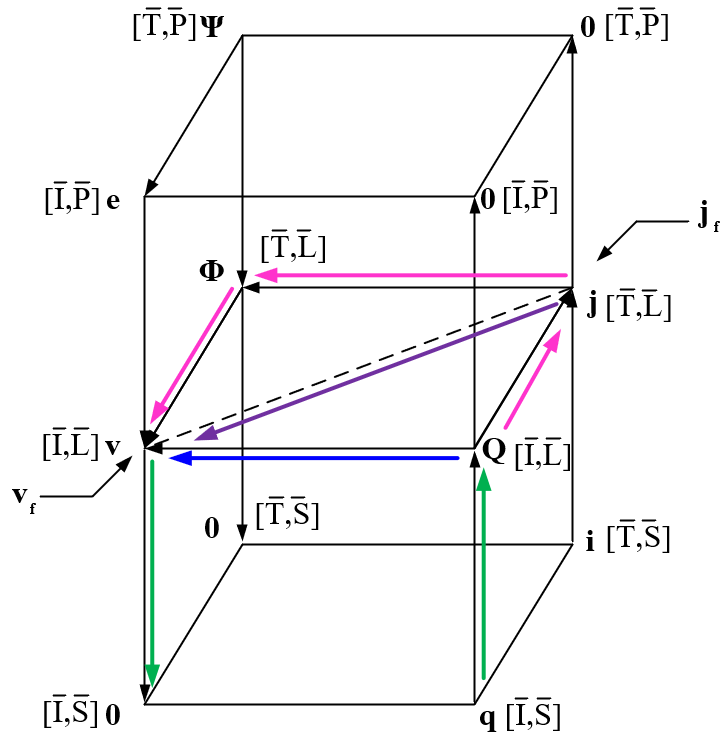}
              \caption{State variable:  ${{\bf{q}}^2}$ }
              \label{fig:state_variable_q0}
        \end{subfigure} 
        \begin{subfigure}[b]{0.24\textwidth}
          \centering
          \includegraphics[width=\linewidth]{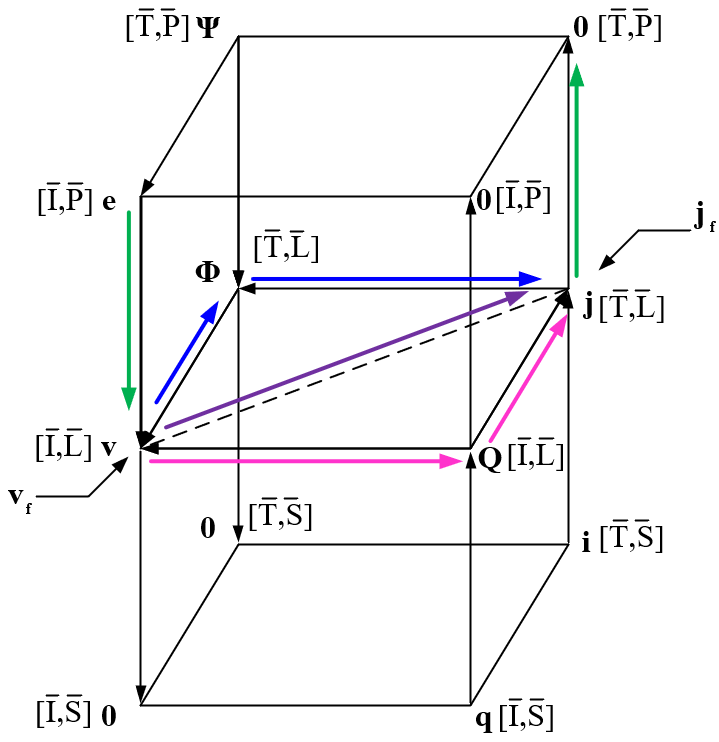}
          \caption{State variable: ${{\bf{e}}^0}$}
          \label{fig:state_variable_e0}
        \end{subfigure}
        \begin{subfigure}[b]{0.24\textwidth}
          \centering
          \includegraphics[width=\linewidth]{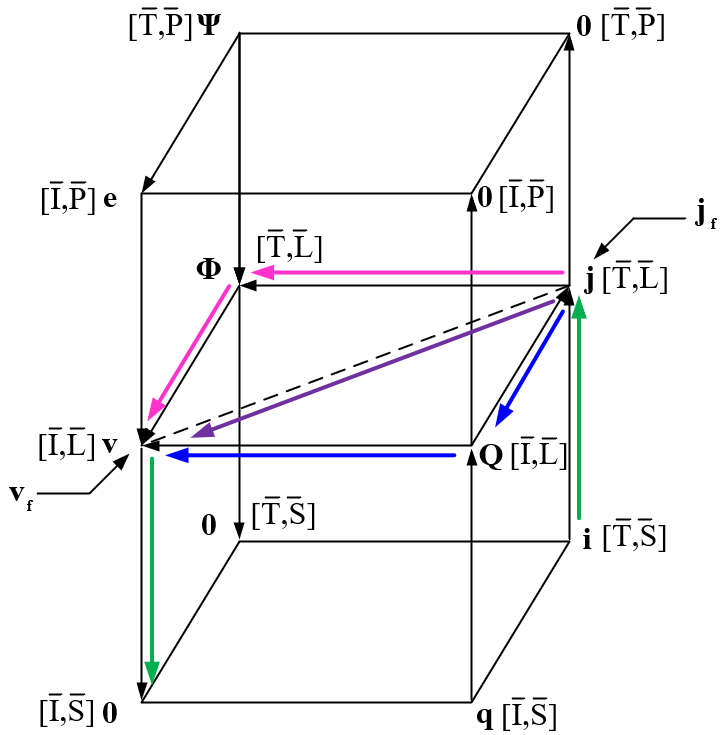}
          \caption{State variable: ${{\bf{i}}^2}$}
          \label{fig:state_variable_i0}
			\end{subfigure} \par\medskip
        \begin{subfigure}[b]{0.24\textwidth}
              \centering
              \includegraphics[width=\linewidth]{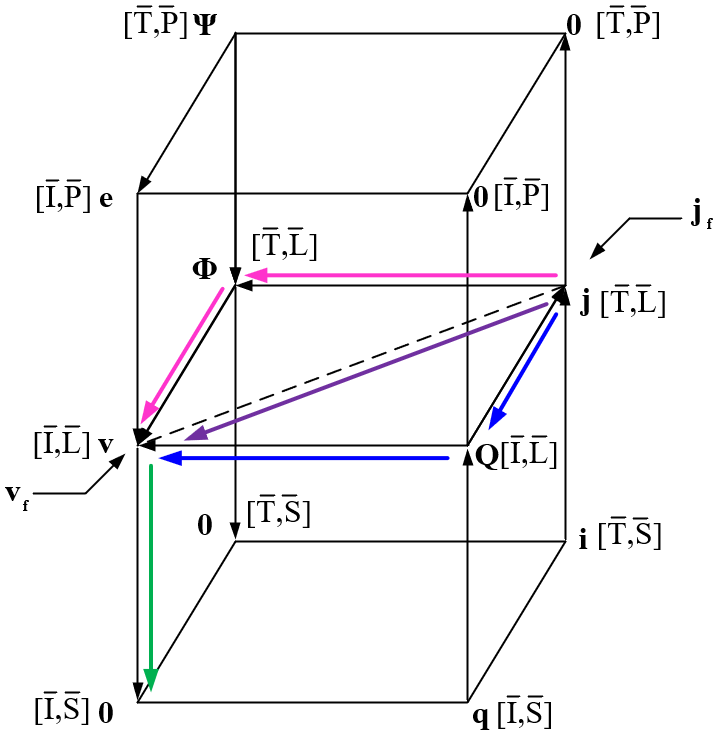}
              \caption{State variable: ${{\bf{j}}^1}$}
              \label{fig:state_variable_j1}
        \end{subfigure}
        \begin{subfigure}[b]{0.24\textwidth}
              \centering
              \includegraphics[width=\linewidth]{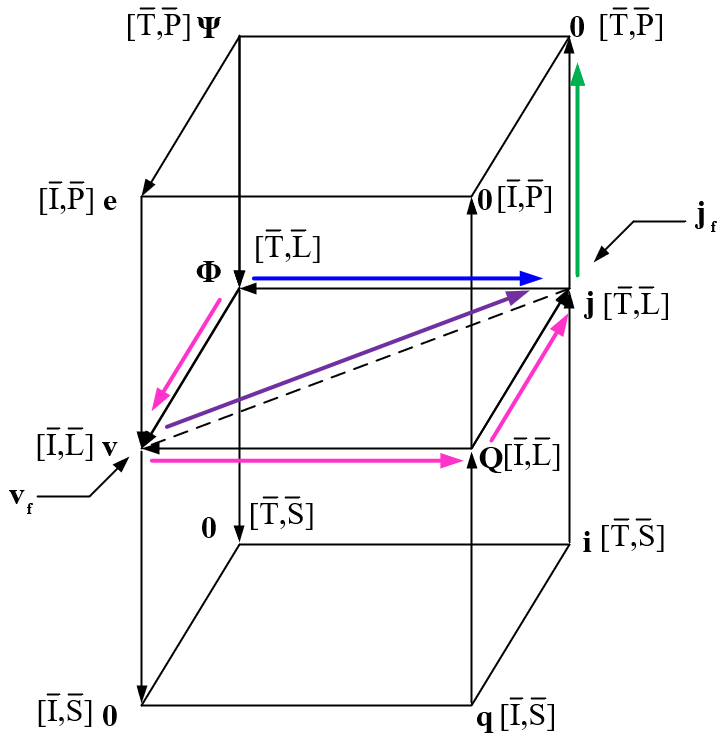}
              \caption{State variable: ${{\bf{\Phi}}^1}$}
              \label{fig:state_variable_phi1}
        \end{subfigure} 
        \begin{subfigure}[b]{0.24\textwidth}
              \centering
              \includegraphics[width=\linewidth]{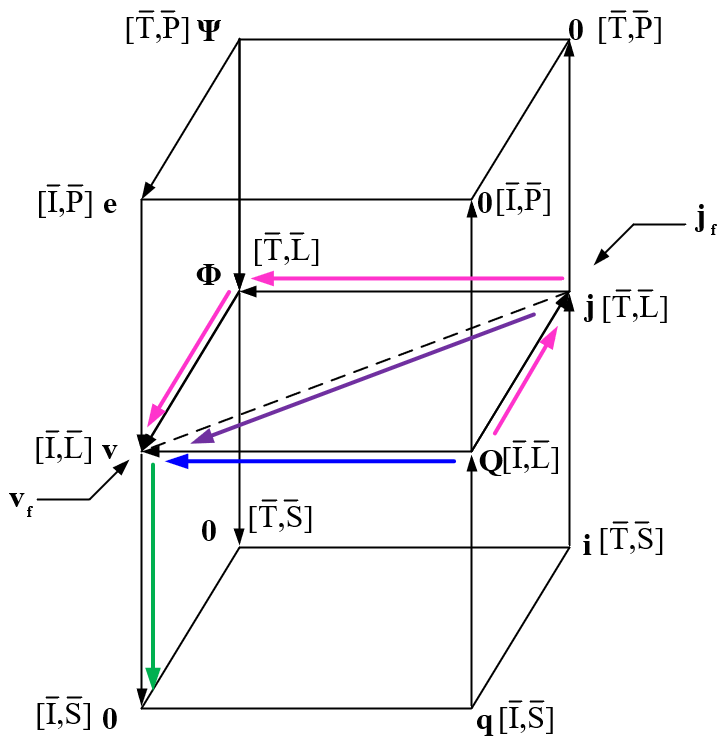}
              \caption{State variable: ${{\bf{Q}}^1}$}
              \label{fig:state_variable_Q1}
        \end{subfigure}
        \begin{subfigure}[b]{0.24\textwidth}
              \centering
              \includegraphics[width=\linewidth]{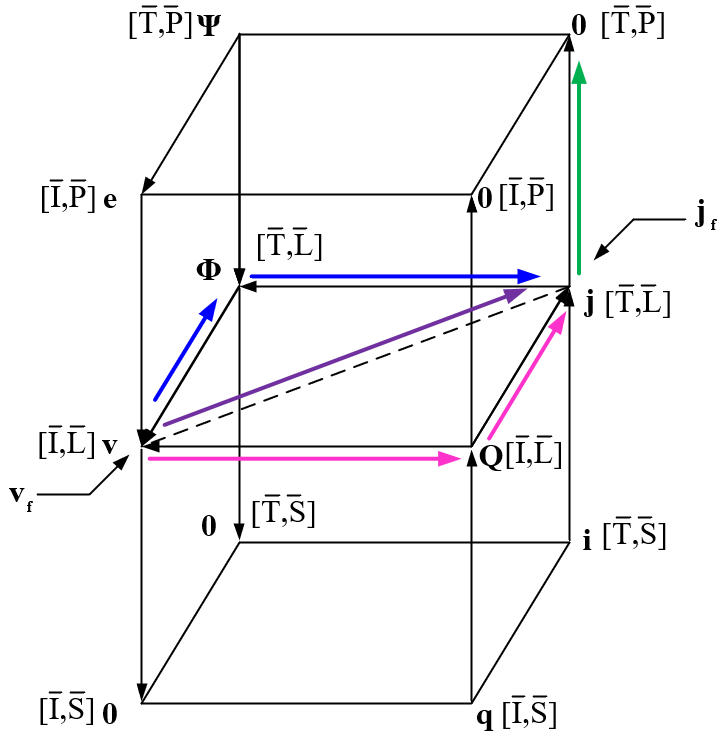}
              \caption{State variable: ${{\bf{v}}^1}$}
              \label{fig:state_variable_v1}
        \end{subfigure}
            \caption{State equation generation paths on the extended Tonti diagram}
            \label{fig:paths and state equations}
        \end{figure}

In this section, we will extend the above approach to general dynamic electrical circuits. The algebraic topological structure of dynamic electrical circuits relies on cochains from all four cochain complexes modeled over a single cell complex.   The topological and constitutive relations between these cochains are given by the diagram in Figure \ref{fig:Overall_RLC_T} and eight different methods of generating the state equations are indicated by paths in the diagrams shown in Figure \ref{fig:paths and state equations}.  Just as with  static systems, each path is a sequence of the arrows indicating composition of the corresponding physical laws.  In contrast to static systems, the middle horizontal section of the diagram allows  three alternative (pink, purple and blue) paths relating the primal $1$-cochain of voltage drops ${{\bf{v}}^1}$ and the dual 1-cochain of branch currents ${{\bf{j}}^1}$ corresponding to capacitance, resistance, and inductance constitutive relationship respectively.  The presence of alternative paths indicate superposition of the corresponding equations generated by each path.

For example, if we use the paths in Figure \ref{fig:state_variable_psi0} to generate the state equations, then the 0-cochain ${{\bf{\Psi}}^0}$ is selected as the state variable. The system state equation can be generated by composition of five physical laws (two topological and three constitutive)  starting with a 0-cochain ${{\bf{\Psi}}^0}$.  First, coboundary operator in space $\delta_0$ applied to potential magnetic fluxes ${{\bf{\Psi }}^0}$ in order to generate magnetic fluxes ${{\bf{\Phi}}^1}$.  Now the path splits in two:   the blue arrow corresponds to constitutive law ${\bf{L^{-1}}}$ (of the inductor) that relates the magnetic fluxes to branch currents of inductors;   the pink arrow takes the magnetic fluxes to generate voltage drops ${\bf{v}}$ via the boundary operator in time $\partial _1^t$.  From here the path splits in two again:  the purple arrow corresponds to the constitutive Ohm's law ${\bf{G}}$ that relates the voltage drops to branch currents of resistors; continuing along the pink path, the constitutive capacitance law ${\bf{C}}$ relates the voltage drops to electric charges ${{\bf{Q}}^1}$ of capacitors is followed by the time coboundary operator $\delta _0^t$ applied to electric charges of capacitors to generate branch currents of capacitors.   Note that the three paths corresponding to the three constitutive laws merge into a single 1-cochain of branch currents ${\bf{j}}^1$,  which is then transformed one more time by the upward green arrow corresponding to KCL ${\partial _1}{{\bf{j}}^1}= {\bf{0}}$.  Taking into account the voltage and current sources, above processes results in following state equations:

\begin{equation} \label{mesh_current_state_eq_0}
{\partial _1}\left( {\delta _0^t{\bf{C}}\left( { - \partial _1^t\left( {{\delta _0}{{\bf{\Psi }}^0}} \right) - {\bf{v}}_{\rm{f}}^1} \right) + {{\bf{R}}^{ - 1}}\left( { - \partial _1^t\left( {{\delta _0}{{\bf{\Psi }}^0}} \right) - {\bf{v}}_{\rm{f}}^1} \right) + {{\bf{L}}^{ - 1}}\left( { - \left( {{\delta _0}{{\bf{\Psi }}^0}} \right) - \delta _0^t{\bf{v}}_{\rm{f}}^1} \right) + {\bf{j}}_{\rm{f}}^1} \right) = {\bf{0}}
\end{equation}

Collecting the terms with known voltage and current sources and moving them to the right hand side, the system state equations Eq.\ref{mesh_current_state_eq_0}  can be written in a more intuitive form as  
\begin{equation} \label{mesh_current_state_eq}
\begin{split}
 {{\partial _{1}}} \left( {\underbrace { \delta _0^t {\bf{C}} \partial _1^t{{{{\delta_{0}}} }}{{\bf{\Psi }}^0}}_{\substack{\text{currents of C}}} + \underbrace { {{{\bf{R}}^{ - 1}}} \partial _1^t{{ {{\delta_{0}}} }}{{\bf{\Psi }}^0}}_{\substack{\text{currents of R}}} + \underbrace {{{{\bf{L}}^{ - 1}}} {{ {{\delta_{0}}} }}{{\bf{\Psi }}^0}}_{\substack{\text{currents of L} }}} \right)  =  {{\partial _{1}}} \left[ {\underbrace {{\bf{j}}_{\bf{f}}^1}_{\substack{\text{current sources} }} - \underbrace {\left( {\delta _0^t {\bf{C}}  +  {{{\bf{R}}^{ - 1}}} +  {{{\bf{L}}^{ - 1}}} {{{\delta _0^t} }}} \right){\bf{v}}_{\bf{f}}^1}_{\substack{\text{equivalent current sources} \\ \text{generated from}\\ \text{voltage sources}}}} \right]
\end{split}
\end{equation}

Other methods for generating the system state equation follow the different paths in Figure \ref{fig:state_variable_q0} - Figure \ref{fig:state_variable_v1}. For example, in Figure \ref{fig:state_variable_q0}, the process starts with the dual 2-cochain of mesh charge ${{\bf{q}}^2}$ selected as the state variable and amounts to another composition of the five physical laws indicated by the corresponding paths.  The blue, purple, and pink path corresponds to the three constitutive laws (capacitance, resistance, and inductance respectively), relating the branch electric charges to the branch voltage drops.  The two green arrows correspond to the $\partial_2$ operator transforming mesh charges ${{\bf{q}}^2}$ to branch charges ${{\bf{Q}}^1}$ and application of KVL (${\delta _1}{{\bf{v}}^1} = {\bf{0}}$).  Putting it all together and taking into account the voltage and current sources, the composition procedures results in: 

\begin{equation} \label{node_potential_state_eq_0}
{\delta _1}\left( {\partial _1^t{\bf{L}}\left( {\delta _0^t{\partial _2}{{\bf{q}}^2} - {\bf{j}}_{\rm{f}}^1} \right) + {\bf{R}}\left( {\delta _0^t{\partial _2}{{\bf{q}}^2} - {\bf{j}}_{\rm{f}}^1} \right) + {{\bf{C}}^{ - 1}}\left( {{\partial _2}{{\bf{q}}^2} - \partial _1^t{\bf{j}}_{\rm{f}}^1} \right) + {\bf{v}}_{\rm{f}}^1} \right) = {\bf{0}}
\end{equation}
or in a more intuitive form similar to Eq.\ref{mesh_current_state_eq}:
\begin{equation} \label{node_potential_state_eq}
\begin{split}	
{\delta _{1}}\left( {\underbrace {\partial _1^t {\bf{L}} \delta _0^t\partial _2{{\bf{q}}^2}}_{\substack{\text{voltage drops of L} }} + \underbrace { {\bf{R}} \delta _0^t\partial _2{{\bf{q}}^2}}_{\substack{\text{voltage drops of R} }} + \underbrace { {{{\bf{C}}^{ - 1}}} \partial _2{{\bf{q}}^2}}_{\substack{\text{voltage drops of C} }}} \right) ={\delta _{1}}\left[ {\underbrace {-{\bf{v}}_{\bf{f}}^1}_{\substack{\text{voltage sources}}} + \underbrace {\left( {\partial _1^t {\bf{L}}+  {\bf{R}}} +{\bf{C}}^{ - 1} \partial _1^t \right){\bf{j}}_{\bf{f}}^1}_{\substack{\text{equivalent voltage drops} \\ \text{generated from}\\ \text{current sources}}}} \right]
\end{split}
\end{equation}


It can be observed that Eq.\ref{mesh_current_state_eq} and  Eq.\ref{node_potential_state_eq} respectively represent the current and voltage equilibrium of the system.  The generated system state equations may be viewed as algebraic with coboundary operators interpreted  as finite difference operators on a finite cell complex.  However,  as we already observed, in lumped parameter systems   space and time are treated separately, and discretization of time is often delayed until a particular numerical integration scheme is chosen.  In this case,  viewing boundary $\partial_1^t$ and coboundary $\delta_0^t$ operations as differentiation in time syntactically transforms  Eq.\ref{mesh_current_state_eq} and  Eq.\ref{node_potential_state_eq} to a more familiar form:  

\begin{equation} \label{mesh_current_state_AE_simpler_ode} 
\begin{split}		
{{\partial _{1}}} \left( \underbrace  {{\bf{C}} {{ {{\delta _{0}}} }}{{{\bf{\ddot \Psi }}}^0}}_{\substack{\text{currents of C}}} +  \underbrace { {{{\bf{R}}^{ - 1}}} {{{{\delta _{0}}} }}{{{\bf{\dot \Psi }}}^0}}_{\substack{\text{currents of R}}} +  \underbrace { {{{\bf{L}}^{ - 1}}}{{{{\delta _{0}}} }}{{\bf{\Psi }}^0}}_{\substack{\text{currents of L}}} \right) = {{\partial _{1}}} \left[ {\underbrace {{\bf{j}}_{\bf{f}}^1}_{\substack{\text{current sources}}} - \underbrace {\left( { {\bf{C}} {\bf{\dot v}}_{\bf{f}}^1 + {{{\bf{R}}^{ - 1}}} {\bf{v}}_{\bf{f}}^1 +  {{{\bf{L}}^{ - 1}}} \int {{\bf{v}}_{\bf{f}}^1dt} } \right)}_{\substack{\text{equivalent current sources} \\ \text{generated from}\\ \text{voltage sources}}}} \right]
\end{split}
\end{equation}		

\begin{equation} \label{node_potential_method_AE_simpler_ode} 
\begin{split}
{\delta _{1}} \left( \underbrace { {\bf{L}}\partial _{2}{{{\bf{\ddot q}}}^2}}_{\substack{\text{voltage drops of C}}} + \underbrace { {\bf{R}} \partial _{2}{{{\bf{\dot q}}}^2}}_{\substack{\text{voltage drops of R}}} + \underbrace { {{{\bf{C}}^{ - 1}}} \partial _{2}{{\bf{q}}^2}}_{\substack{\text{voltage drops of L}}} \right) = {\delta _{1}}\left( {\underbrace { - {\bf{v}}_{\bf{f}}^1}_{\substack{\text{voltage sources}}} + \underbrace { {\bf{L}} {\bf{\dot j}}_{\bf{f}}^1 +  {\bf{R}} {\bf{j}}_{\bf{f}}^1 +  {{{\bf{C}}^{ - 1}}} \int {{\bf{j}}_{\bf{f}}^1dt} }_{\substack{\text{equivalent voltage drops} \\ \text{generated from}\\ \text{current sources}}}} \right)
\end{split}
\end{equation}

\begin{exmp} 

 \begin{figure}[!htb]
        \begin{subfigure}{.5\textwidth}
              \centering
              \includegraphics[width=0.6\linewidth]{Simple_RLC.png}
              \caption{A simple RLC electrical circuit}
              \label{fig:Comparison_RLC problem_copy}
        \end{subfigure}%
        \begin{subfigure}{.5\textwidth}
              \centering
              \includegraphics[width=.5\linewidth]{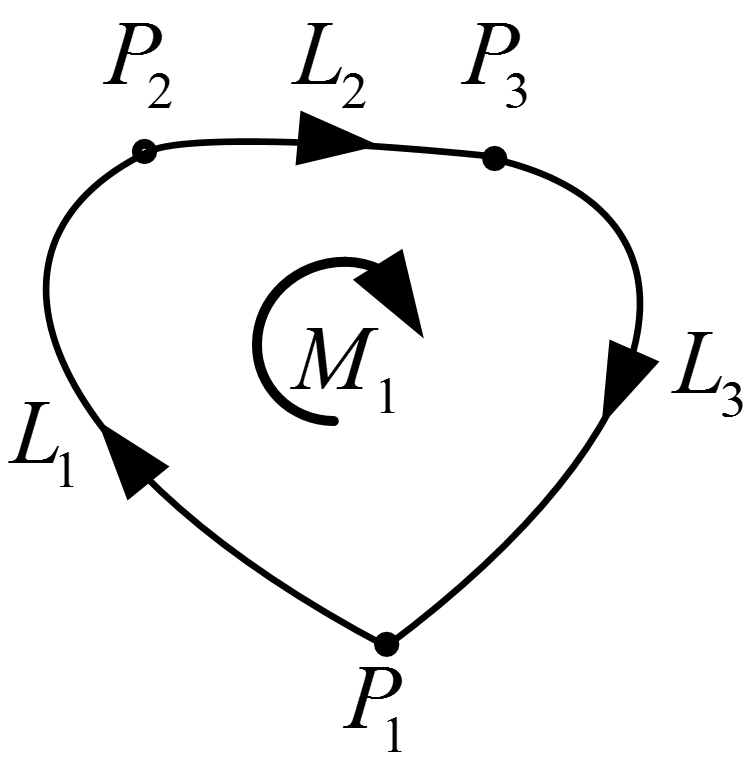}
              \caption{Topological structure}
              \label{fig:Simple_RLC_Topoglical_Structure}
        \end{subfigure}
            \caption{A simple RLC electrical circuit and its topological structure}
            \label{fig:RLC_Simulation}
    \end{figure}

We will use a very simple example of RLC electrical circuit (Figure \ref{fig:Comparison_RLC problem}) to illustrate the derivation of Eq.\ref{node_potential_method_AE_simpler_ode}. A copy of Figure \ref{fig:Comparison_RLC problem} is shown in Figure \ref{fig:Comparison_RLC problem_copy}. The shown electrical network contains one resistor R, one capacitor C, one inductor L and one voltage source $V_s$. Topologically,  the network is a 2-dimensional cell complex shown in Figure \ref{fig:Simple_RLC_Topoglical_Structure} and consists of three 0-cells (${P_1}$,${P_2}$,${P_3}$), three 1-cells (${L_1}$,${L_2}$,${L_3}$) and one 2-cell (${M_1}$).

The algebraic topological model of the electrical circuit contains: primal 0-cochain node potentials (${{\bf{e}}^0} = {e_{{1}}} \cdot {P_1} + {e_{{2}}} \cdot {P_2} + {e_{{3}}} \cdot {P_3}$), primal 0-cochain magnetic flux potentials (${{\bf{\Psi }}^0} = {\psi _{{1}}} \cdot {P_1} + {\psi _{{2}}} \cdot {P_2} + {\psi _{{3}}} \cdot {P_3}$), primal 1-cochain voltage drops (${{\bf{v}}^1} = {v_{{1}}} \cdot {L_1} + {v_{{2}}} \cdot {L_2} + {v_{{3}}} \cdot {L_3}$), primal 1-cochain magnetic fluxes  (${{\bf{\Phi }}^1} = {\phi _{{1}}} \cdot {L_1} + {\phi _{{2}}} \cdot {L_2} + {\phi _{{3}}} \cdot {L_3}$),  dual 1-cochain currents (${{\bf{j}}^1} = {j_{{1}}} \cdot {L_1} + {j_{{2}}} \cdot {L_2} + {j_{{3}}} \cdot {L_3}$), dual 1-cochain electric charges (${{\bf{Q}}^1} = {Q_{{1}}} \cdot {L_1} + {Q_{{2}}} \cdot {L_2} + {Q_{{3}}} \cdot {L_3}$), dual 2-cochain mesh currents (${{\bf{i}}^2} = i \cdot {M_1}$), dual 2-cochain mesh electric charges (${{\bf{q}}^2} = q \cdot {M_1}$), and four cochains that are always ${\bf{0}}$: 2-cochain loop voltage drops, 2-cochain loop magnetic fluxes, 0-cochain node currents and 0-cochain of node electric charges. Following the paths in Figure \ref{fig:state_variable_q0} generates Eq.\ref{node_potential_method_AE_simpler_ode}, with individual terms as follows:

\begin{equation} \label{SimpleRLC_Cob}
{\delta _1} = \left[ {\begin{array}{*{20}{c}}
1&1&1
\end{array}} \right]
\end{equation} 

\begin{equation} \label{cobT}
\partial _2 = {\delta _1}^T
\end{equation}  

\begin{equation} \label{SimpleRLC_global_L}
{\bf{L}} = \left[ {\begin{array}{*{20}{c}}
0&0&0\\
0&L&0\\
0&0&0
\end{array}} \right]
\end{equation}

\begin{equation} \label{SimpleRLC_global_R}
{\bf{R}} = \left[ {\begin{array}{*{20}{c}}
R&0&0\\
0&0&0\\
0&0&0
\end{array}} \right]
\end{equation}

\begin{equation} \label{SimpleRLC_global_Cinverse}
{{\bf{C}}^{ - 1}} = \left[ {\begin{array}{*{20}{c}}
0&0&0\\
0&0&0\\
0&0&{{C^{ - 1}}}
\end{array}} \right]
\end{equation}

\begin{equation} \label{SimpleRLC_global_Vs}
{\bf{v}}_{\bf{f}}^1 = {\left[ {\begin{array}{*{20}{c}}
{-V_{s}}&0&0
\end{array}} \right]^T}
\end{equation} 

\begin{equation} \label{SimpleRLC_global_Cs}
{\bf{j}}_{\bf{f}}^1 = {\left[ {\begin{array}{*{20}{c}}
0&0&0
\end{array}} \right]^T}
\end{equation}

Substituting Eq.\ref{SimpleRLC_Cob} $\sim$ Eq.\ref{SimpleRLC_global_Cs} into Eq.\ref{node_potential_method_AE_simpler_ode},  we obtain the system state equations as shown in Eq.\ref{SimpleRLC_state_equation_node_potential} . Since there is only one 2-cell, the mesh current equals the branch current, so Eq.\ref{SimpleRLC_state_equation_node_potential} is equivalent to the system state equation generated by the Simulink model shown in Eq.\ref{Comparison_RLC_Simulink}.

	\begin{equation} \label{SimpleRLC_state_equation_node_potential}
	L{{\ddot q}} + R{{\dot q}} + {C^{ - 1}}{q} = {V_{s}}
	\end{equation}

\end{exmp}

\begin{exmp}

        \begin{figure}[!htb]
        \centering
        \begin{subfigure}{0.5\textwidth}
          \centering
          \includegraphics[width=.8\linewidth]{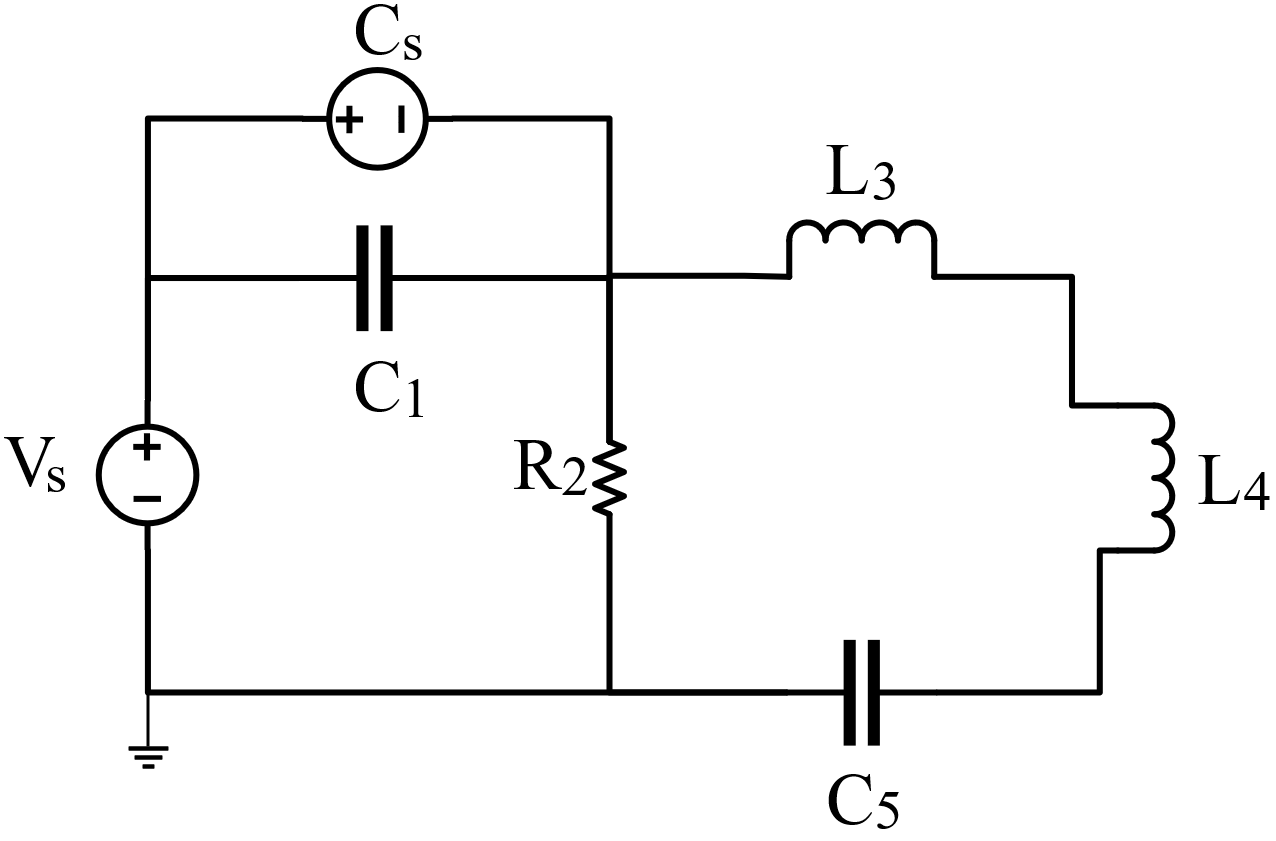}
          \caption{An RLC electrical circuit}
          \label{fig:RLC problem}
        \end{subfigure}%
        \begin{subfigure}{0.3\textwidth}
          \centering
          \includegraphics[width=.8\linewidth]{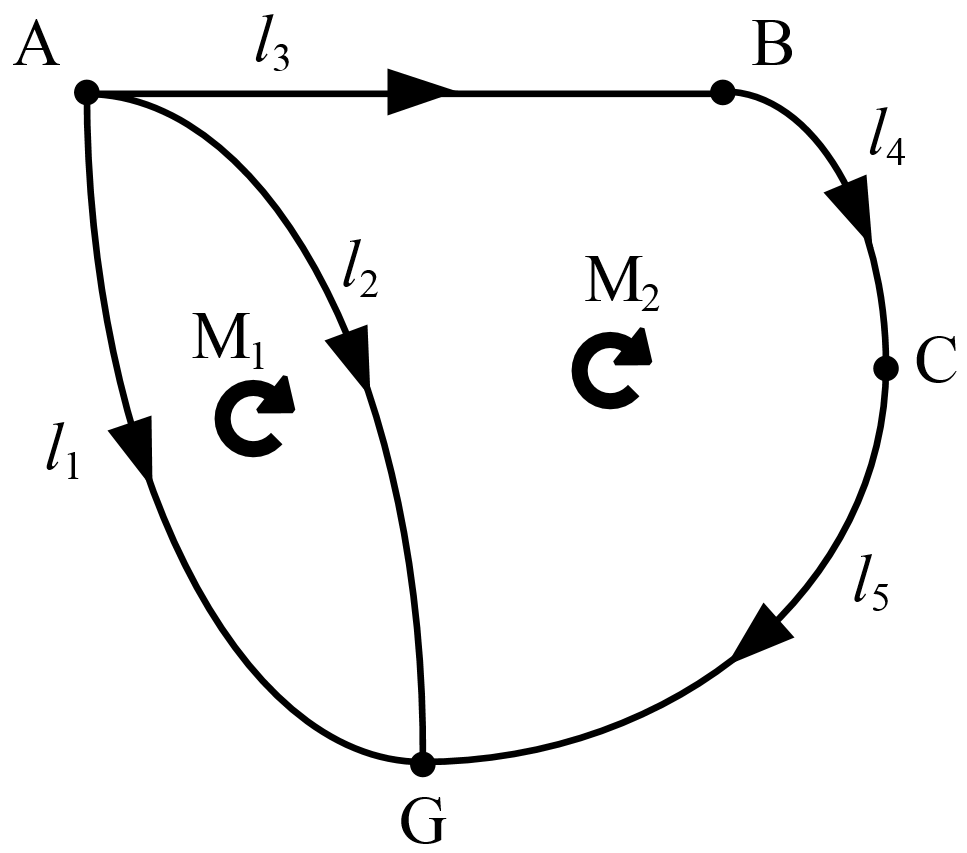}
          \caption{Topological structure}
          \label{fig:RLCproblemtopologicalstructure}
        \end{subfigure}\par\medskip
        \caption{An RLC electrical circuit and its topological structure}
        \label{fig:RLC}
        \end{figure} 
        
We will use another example of electrical circuit in Figure \ref{fig:RLC problem} to illustrate the derivation of Eq.\ref{mesh_current_state_AE_simpler_ode}  . The shown electrical circuit contains one resistor $R_2$, two capacitors $C_1$, $C_5$, two inductors $L_3$, $L_4$, one current source $C_s$ and one voltage source $V_s$. Topologically, the network is a 2-dimensional cell complex shown in Figure \ref{fig:RLCproblemtopologicalstructure}, and consisting of  four 0-cells (A,B,C,G), five 1-cells(${l_1}$,${l_2}$,${l_3}$,${l_4}$,${l_5}$) and two 2-cells(${M_1}$,${M_2}$). 

The algebraic topological model of the electrical circuit contains: primal 0-cochain node potentials (${{\bf{e}}^0} = {e_A} \cdot A + {e_B} \cdot B + {e_C} \cdot C + {e_G} \cdot G$), primal 0-cochain magnetic flux potentials (${{\bf{\Psi }}^0} = {\psi _A} \cdot A + {\psi _B} \cdot B + {\psi _C} \cdot C + {\psi _G} \cdot G$), primal  1-cochain voltage drops (${{\bf{v}}^1} = {v_1} \cdot {l_1} + {v_2} \cdot {l_2} + {v_3} \cdot {l_3} + {v_4} \cdot {l_4} + {v_5} \cdot {l_5}$), primal  1-cochain magnetic fluxes  (${{\bf{\Phi }}^1} = {\phi _{{1}}} \cdot {l_1} + {\phi _{{2}}} \cdot {l_2} + {\phi _{{3}}} \cdot {l_3}$), dual  1-cochain currents (${{\bf{j }}^1} = {j _1} \cdot {l_1} + {j _2} \cdot {l_2} + {j _3} \cdot {l_3} + {j _4} \cdot {l_4} + {j _5} \cdot {l_5}$), dual  1-cochain electric charges (${{\bf{Q}}^1} = {Q_1} \cdot {l_1} + {Q_2} \cdot {l_2} + {Q_3} \cdot {l_3} + {Q_4} \cdot {l_4} + {Q_5} \cdot {l_5}$), dual 2-cochain mesh currents (${{\bf{i}}^2} = {i_1} \cdot {M_1} + {i_2} \cdot {M_2}$), dual 2-cochain mesh electric charges (${{\bf{q}}^2} = {q_1} \cdot {M_1} + {q_2} \cdot {M_2}$), and four cochains that are always ${\bf{0}}$: 2-cochain loop voltage drops, 2-cochain loop magnetic fluxes, 0-cochain node currents and 0-cochain of node electric charges. In order to get unique solution of the state equations, we consider 0-cell G as the reference node, implying the boundary condition of ${e_G} = 0$. Following the paths in Figure \ref{fig:state_variable_psi0}, generates Eq.\ref{mesh_current_state_AE_simpler_ode}, with individual terms as follows:

\begin{equation} \label{restrictedboundary10RLC}
{\partial _1} = \left[ {\begin{array}{*{20}{c}}
{ + 1}&{ + 1}&0&0&{ + 1}\\
{ - 1}&{ - 1}&{ - 1}&0&0\\
0&0&{ + 1}&{ - 1}&0\\
0&0&0&{ + 1}&{ - 1}
\end{array}} \right] 
\end{equation}

\begin{equation} \label{trans}
{\delta _0} = \partial _1^T
\end{equation}

\begin{equation} \label{RLC_global_C}
{\bf{C}} = \left[ {\begin{array}{*{20}{c}}
{{C_1}}&0&0&0&0\\
0&0&0&0&0\\
0&0&0&0&0\\
0&0&0&0&0\\
0&0&0&0&{{C_5}}
\end{array}} \right]
\end{equation}

\begin{equation} \label{RLC_global_inverseR}
{{\bf{R}}^{ - 1}} = \left[ {\begin{array}{*{20}{c}}
0&0&0&0&0\\
0&{{1 \mathord{\left/
 {\vphantom {1 {{R_2}}}} \right.
 \kern-\nulldelimiterspace} {{R_2}}}}&0&0&0\\
0&0&0&0&0\\
0&0&0&0&0\\
0&0&0&0&0
\end{array}} \right]
\end{equation}

\begin{equation} \label{RLC_global_inverseL}
{{\bf{L}}^{ - 1}} = \left[ {\begin{array}{*{20}{c}}
0&0&0&0&0\\
0&0&0&0&0\\
0&0&{{1 \mathord{\left/
 {\vphantom {1 {{L_3}}}} \right.
 \kern-\nulldelimiterspace} {{L_3}}}}&0&0\\
0&0&0&{{1 \mathord{\left/
 {\vphantom {1 {{L_4}}}} \right.
 \kern-\nulldelimiterspace} {{L_4}}}}&0\\
0&0&0&0&0
\end{array}} \right]
\end{equation}

\begin{equation} \label{RLC_global_Vs}
{\bf{v}}_{\bf{f}}^1 = {\left[ {\begin{array}{*{20}{c}}
{{V_s}}&0&0&0&0
\end{array}} \right]^T}
\end{equation} 
			
\begin{equation} \label{RLC_global_Cs}
{\bf{j}}_{\bf{f}}^1 = {\left[ {\begin{array}{*{20}{c}}
{{C_s}}&0&0&0&0
\end{array}} \right]^T}
\end{equation} 
			
Substituting Eq.\ref{restrictedboundary10RLC} $\sim$ Eq.\ref{RLC_global_Cs} into Eq.\ref{mesh_current_state_AE_simpler_ode}, we obtain the system state equations as follows: 
\begin{equation} \label{RLC_state_equation_node_potential}
\begin{array}{l}
\left[ {\begin{array}{*{20}{c}}
{{C_1} + {C_5}}&{ - {C_1}}&0&{ - {C_5}}\\
{ - {C_1}}&{{C_1}}&0&0\\
0&0&0&0\\
{ - {C_5}}&0&0&{{C_5}}
\end{array}} \right]{{{\bf{\ddot \Psi }}}^0} + \left[ {\begin{array}{*{20}{c}}
{{1 \mathord{\left/
 {\vphantom {1 {{R_2}}}} \right.
 \kern-\nulldelimiterspace} {{R_2}}}}&{{{ - 1} \mathord{\left/
 {\vphantom {{ - 1} {{R_2}}}} \right.
 \kern-\nulldelimiterspace} {{R_2}}}}&0&0\\
{{{ - 1} \mathord{\left/
 {\vphantom {{ - 1} {{R_2}}}} \right.
 \kern-\nulldelimiterspace} {{R_2}}}}&{{1 \mathord{\left/
 {\vphantom {1 {{R_2}}}} \right.
 \kern-\nulldelimiterspace} {{R_2}}}}&0&0\\
0&0&0&0\\
0&0&0&0
\end{array}} \right]{{{\bf{\dot \Psi }}}^0}\\
 + \left[ {\begin{array}{*{20}{c}}
0&0&0&0\\
0&{{1 \mathord{\left/
 {\vphantom {1 {{L_3}}}} \right.
 \kern-\nulldelimiterspace} {{L_3}}}}&{ - {1 \mathord{\left/
 {\vphantom {1 {{L_3}}}} \right.
 \kern-\nulldelimiterspace} {{L_3}}}}&0\\
0&{ - {1 \mathord{\left/
 {\vphantom {1 {{L_3}}}} \right.
 \kern-\nulldelimiterspace} {{L_3}}}}&{{1 \mathord{\left/
 {\vphantom {1 {{L_3}}}} \right.
 \kern-\nulldelimiterspace} {{L_3}}} + {1 \mathord{\left/
 {\vphantom {1 {{L_4}}}} \right.
 \kern-\nulldelimiterspace} {{L_4}}}}&{ - {1 \mathord{\left/
 {\vphantom {1 {{L_4}}}} \right.
 \kern-\nulldelimiterspace} {{L_4}}}}\\
0&0&{ - {1 \mathord{\left/
 {\vphantom {1 {{L_4}}}} \right.
 \kern-\nulldelimiterspace} {{L_4}}}}&{{1 \mathord{\left/
 {\vphantom {1 {{L_4}}}} \right.
 \kern-\nulldelimiterspace} {{L_4}}}}
\end{array}} \right]{{\bf{\Psi }}^0} = \left[ {\begin{array}{*{20}{c}}
{{C_s} - {{\dot V}_s}{C_1}}\\
{ - {C_s} + {{\dot V}_s}{C_1}}\\
0\\
0
\end{array}} \right]
\end{array}
\end{equation}
\end{exmp}

\section{Multi-domain lumped parameter systems}

\subsection{Interactions of single-domain models}

Engineering systems are usually constructed as compositions of single-domain subsystems in order to perform complex engineering tasks.  Representative examples include electric motors (electro-mechanical systems), ovens (electro-thermal systems), hydraulic pumps (hydraulic-mechanical systems), etc.  We will refer to such systems as multi-domain systems, where lumped-parameter behavior of each single-domain is governed by an extended Tonti diagram as described in the previous section. It should be understood that multi-domain systems subsume the special case of homogeneous systems where all single-domain systems are of the same type, e.g. all electrical or all mechanical. 
Composition of two single-domain systems is associated with the process of energy conversion between the two systems, also called \textit{transduction} \cite{karnopp2012system}.  Specific means of transduction vary from system to system, but common examples of transducers (devices the perform transduction) include gears and levers, electrical transformers, motors, piezoelectric devices, hydraulic pumps, and so on.  

\begin{figure}[!htb]
        \centering
        \begin{subfigure}[b]{0.5\textwidth}
          \centering
          \includegraphics[width=0.5\linewidth]{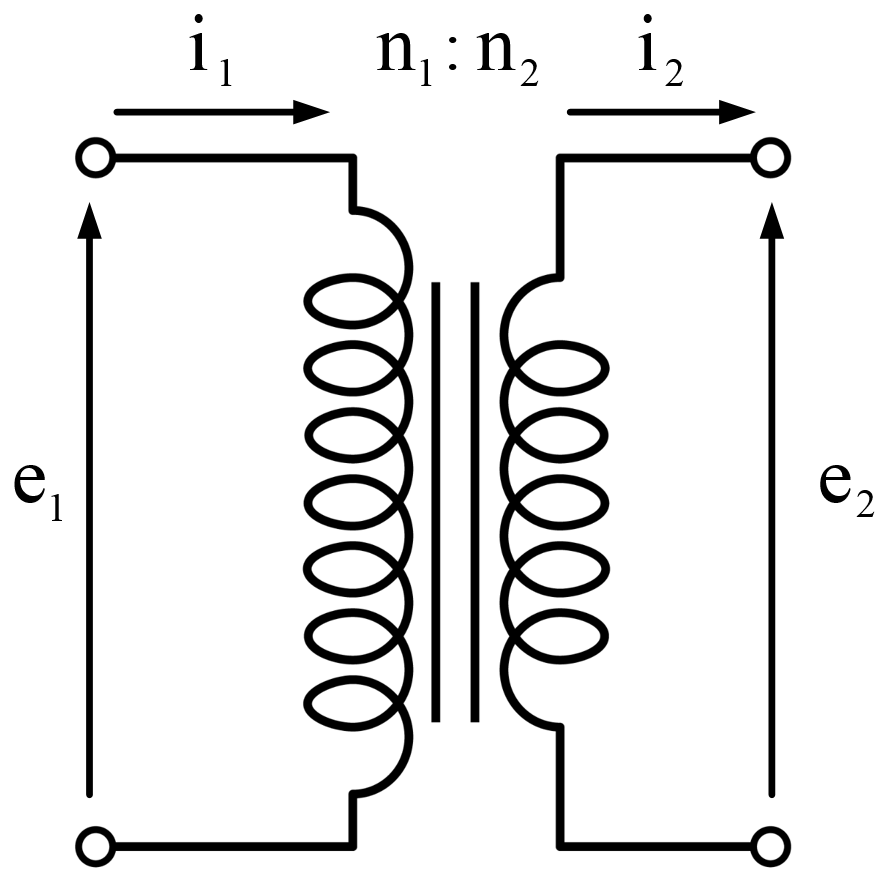}
          \caption{A schematic ideal electrical transformer}
          \label{fig:electrical_transformer}
        \end{subfigure}%
        \begin{subfigure}[b]{0.5\textwidth}
          \centering
          \includegraphics[width=.7\linewidth]{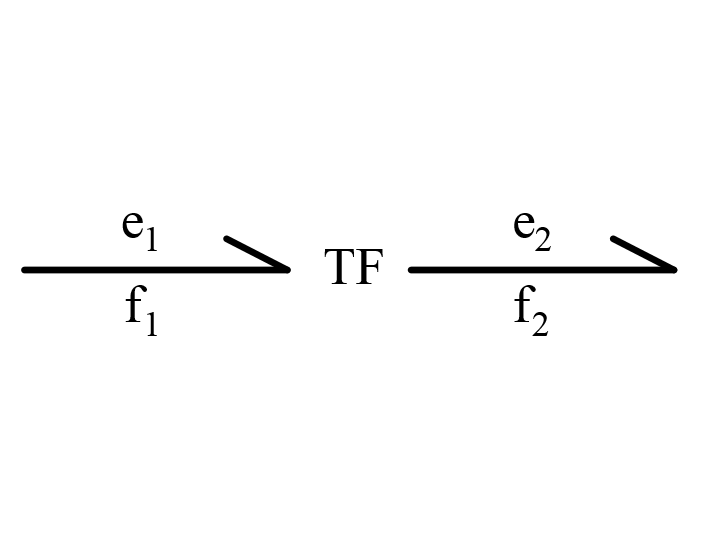}
          \caption{Bond graph model of transformers}
          \label{fig:bond_graph_transformer}
        \end{subfigure}\par\medskip
        
        \begin{subfigure}[b]{0.49\textwidth}
          \centering
          \includegraphics[width=.49\linewidth]{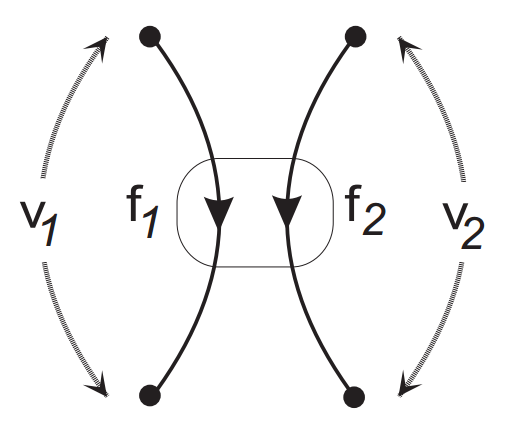}
          \caption{Linear graph model of transformers}
          \label{fig:linear_graph_transformer}
        \end{subfigure}
        \begin{subfigure}[b]{0.49\textwidth}
          \centering
          \includegraphics[width=\linewidth]{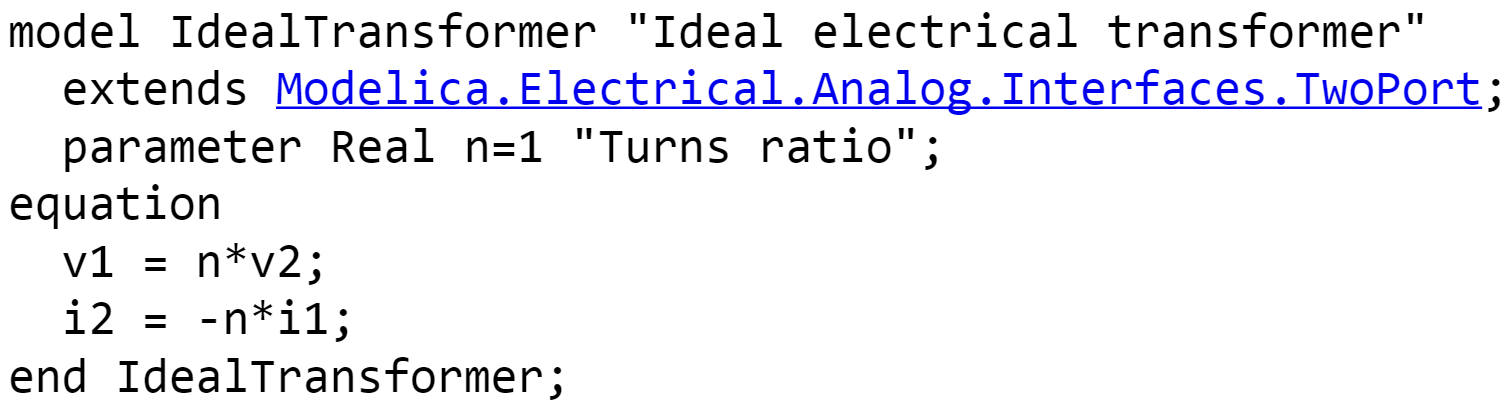}
          \caption{Modelica model of the ideal electrical transformer}
          \label{fig:Modelica_transformer}
        \end{subfigure}
        \caption{Comparison of transformers of different modeling languages}
        \label{fig:Comparison_transformer}
        \end{figure}

\begin{figure}[!htb]
        \centering
        \begin{subfigure}[b]{0.5\textwidth}
          \centering
          \includegraphics[width=0.7\linewidth]{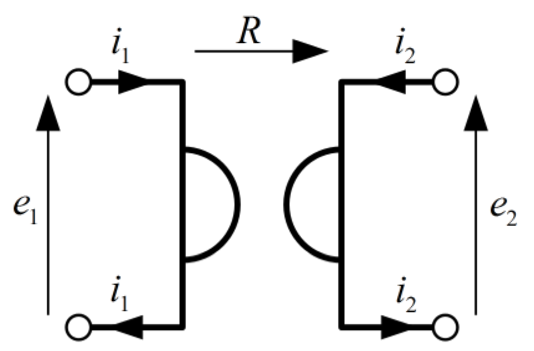}
          \caption{a schematic ideal electrical gyrator}
          \label{fig:electrical_gyrator}
        \end{subfigure}%
        \begin{subfigure}[b]{0.5\textwidth}
          \centering
          \includegraphics[width=.7\linewidth]{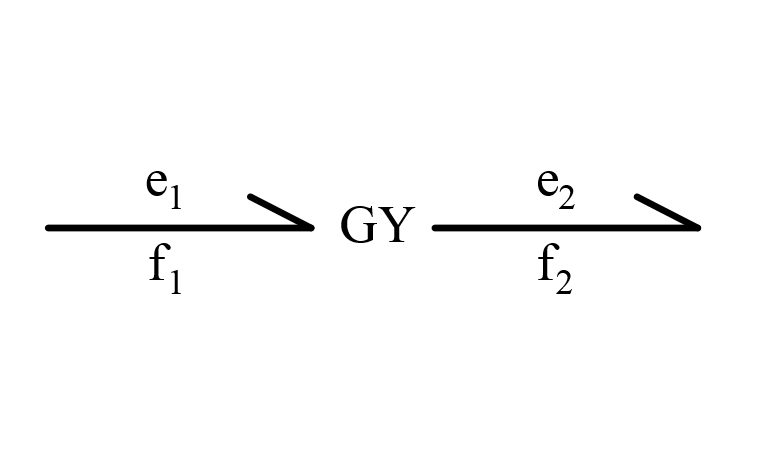}
          \caption{Bond graph model of gyrators}
          \label{fig:bond_graph_gyrator}
        \end{subfigure}\par\medskip
        
        \begin{subfigure}[b]{0.49\textwidth}
          \centering
          \includegraphics[width=.49\linewidth]{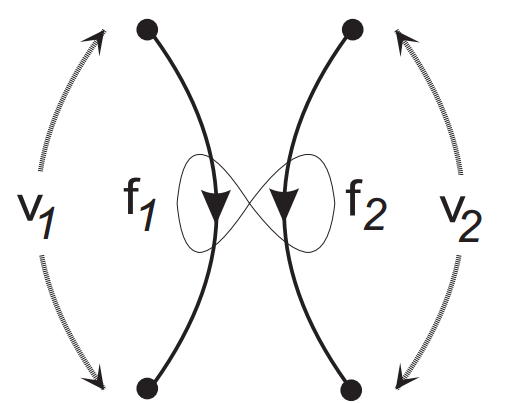}
          \caption{Linear graph model of gyrators}
          \label{fig:linear_graph_gyrator}
        \end{subfigure}
        \begin{subfigure}[b]{0.49\textwidth}
          \centering
          \includegraphics[width=\linewidth]{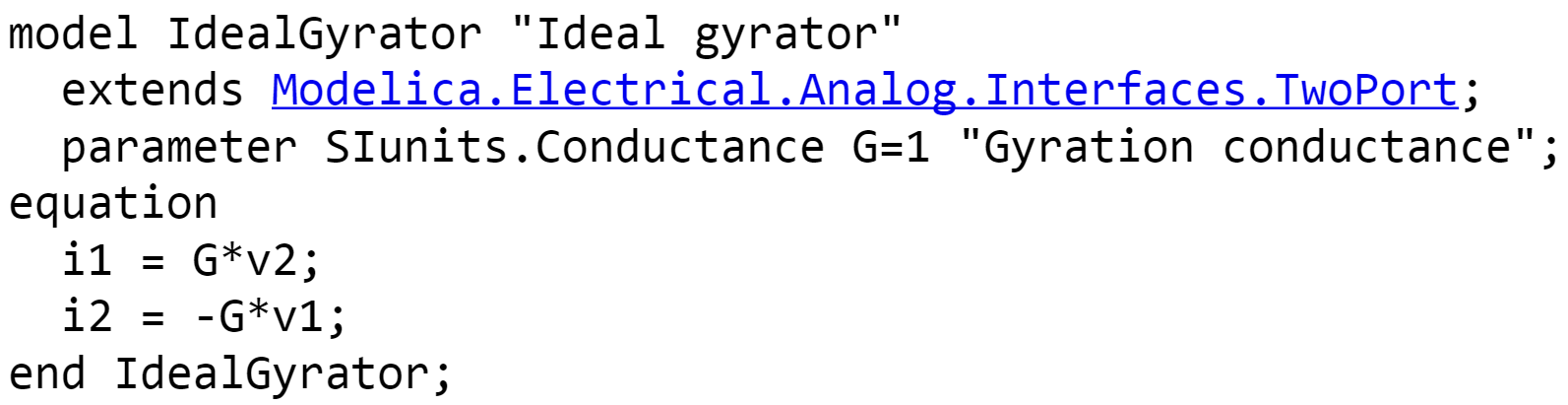}
          \caption{Modelica model of the ideal gyrator}
          \label{fig:Modelica_gyrator}
        \end{subfigure}
        \caption{Comparison of gyrators of different modeling languages}
        \label{fig:Comparison_gyrator}
        \end{figure}

Devices used to couple the same type variables of different physical domains are usually called transformers (e.g. electrical transformers), while devices coupling the dual type variables are called gyrators (e.g. electric motors) \cite{karnopp2012system}.  
Physical transformers and gyrators always have energy loss during the energy transduction, but if the leakage is small enough then the energy loss is usually neglected during modeling. Transformers and gyrators with no energy loss are called ideal transformers and gyrators \cite{karnopp2012system}.  System modeling languages generally use ideal  transformers and gyrators as abstract constructs 
in order to avoid modeling complex structures of physical transformers and gyrators devices. 
For instance, bond graphs use abstract 2-port transformers and gyrators which can be used to connect any 1-ports \cite{karnopp2012system}. Figure \ref{fig:Comparison_transformer} and Figure \ref{fig:Comparison_gyrator} show several transformer and gyrator models from different system modeling languages. 

\begin{figure}[!htb]
        \centering
        \begin{subfigure}[b]{0.5\textwidth}
          \centering
          \includegraphics[width=0.8\linewidth]{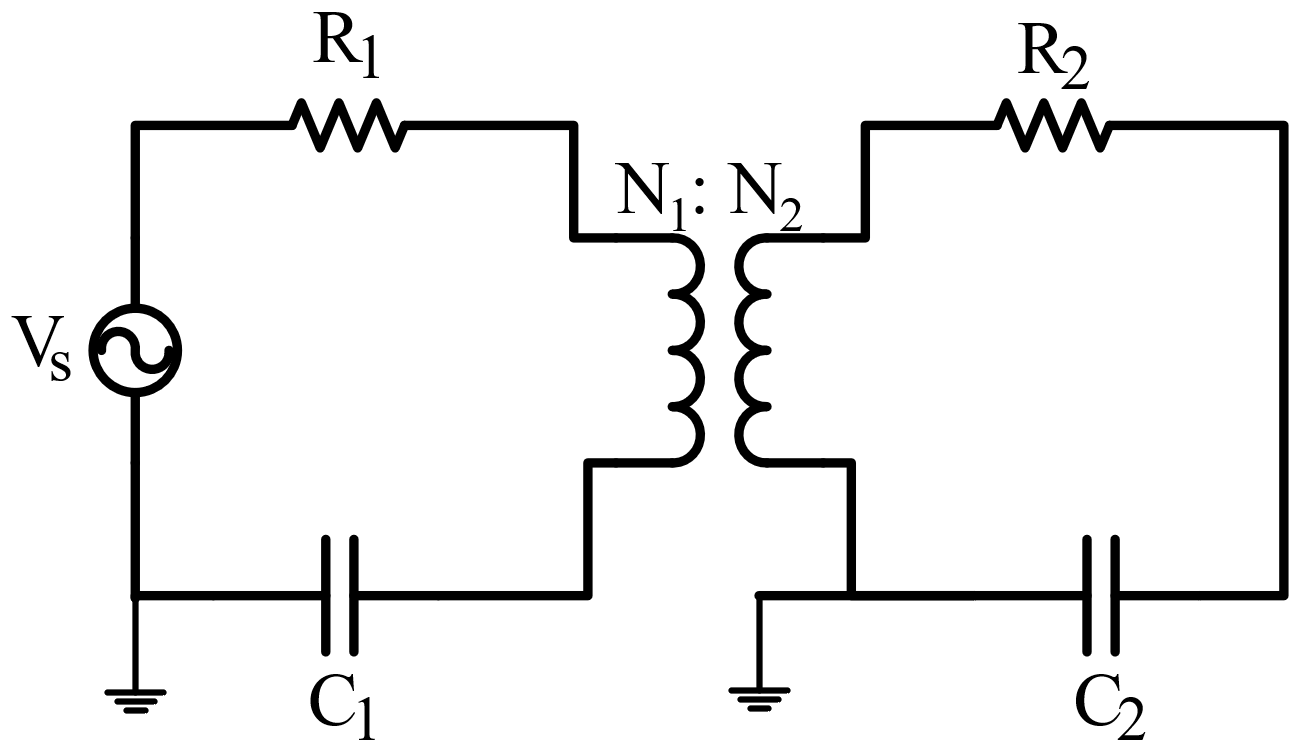}
          \caption{A multi-domain RLC electrical circuit}
          \label{fig:homo_RLC}
        \end{subfigure} \par\medskip

        \centering
        \begin{subfigure}[b]{0.49\textwidth}
          \centering
          \includegraphics[width=0.8\linewidth]{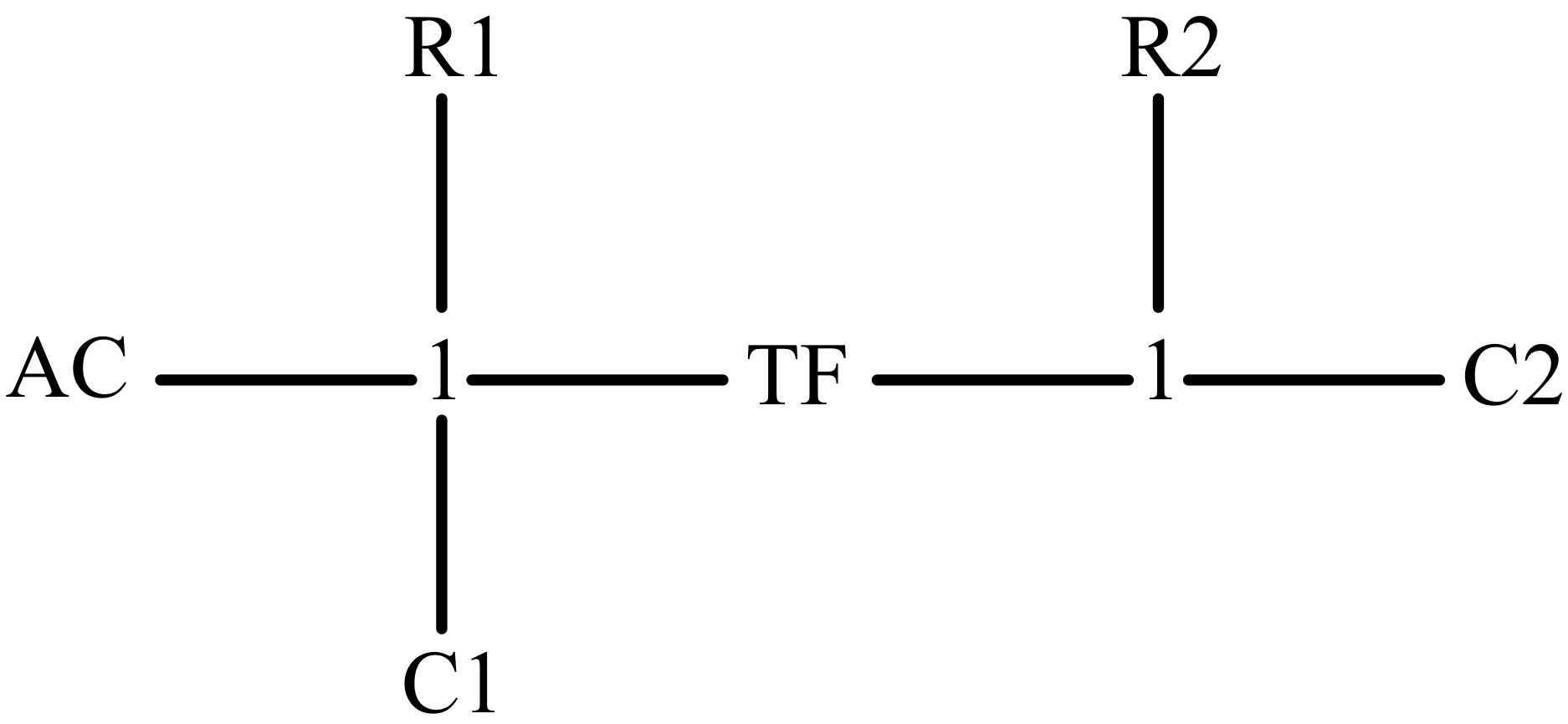}
          \caption{Bond graph model}
          \label{fig:homo_RLC_bond_graph}
        \end{subfigure}
        \centering
        \begin{subfigure}[b]{0.49\textwidth}
          \centering
          \includegraphics[width=0.6\linewidth]{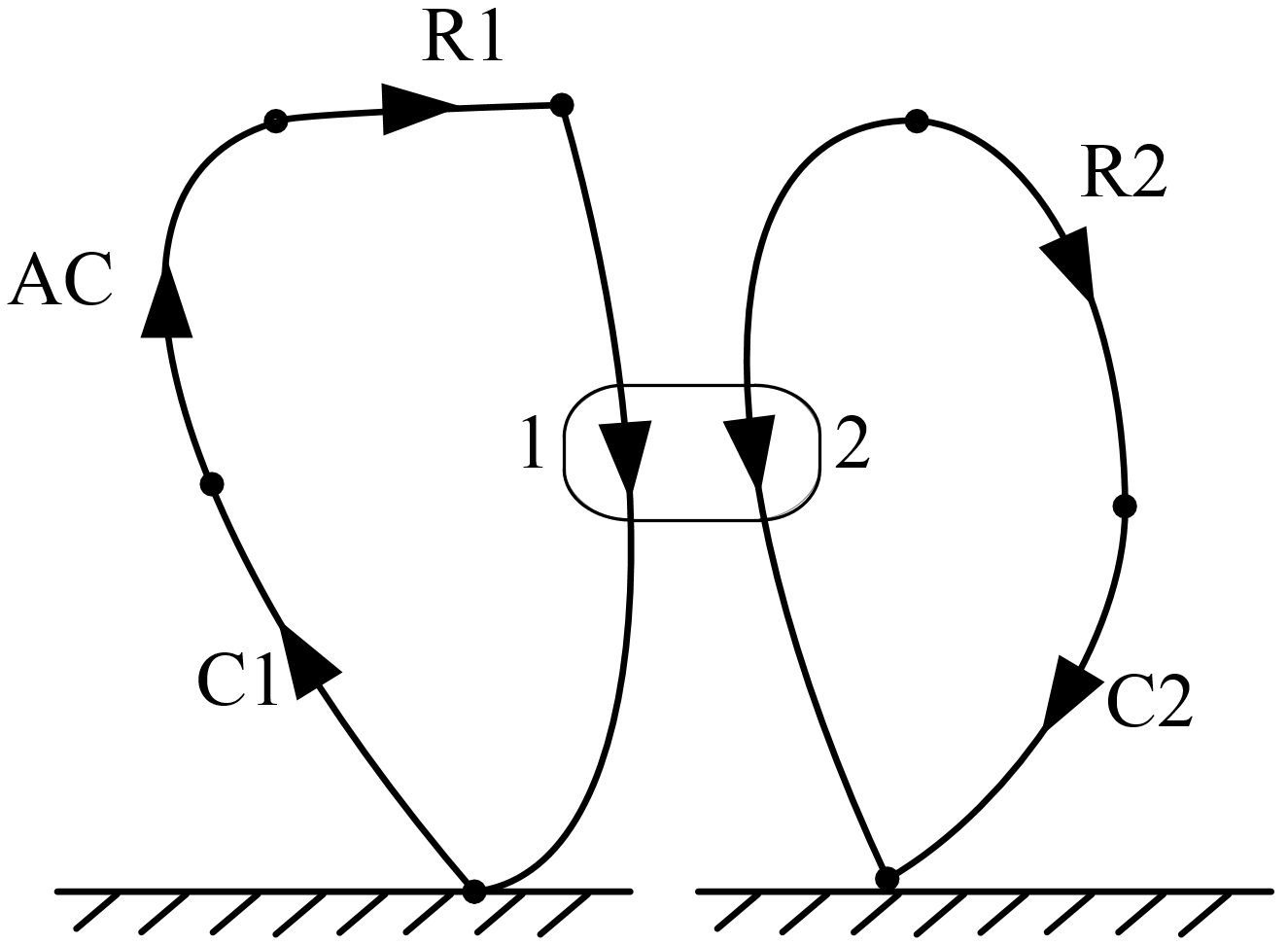}
          \caption{Linear graph model}
          \label{fig:homo_RLC_linear_graph}
        \end{subfigure} \par\medskip
         
        \begin{subfigure}[b]{0.49\textwidth}
          \centering
          \includegraphics[width=\linewidth]{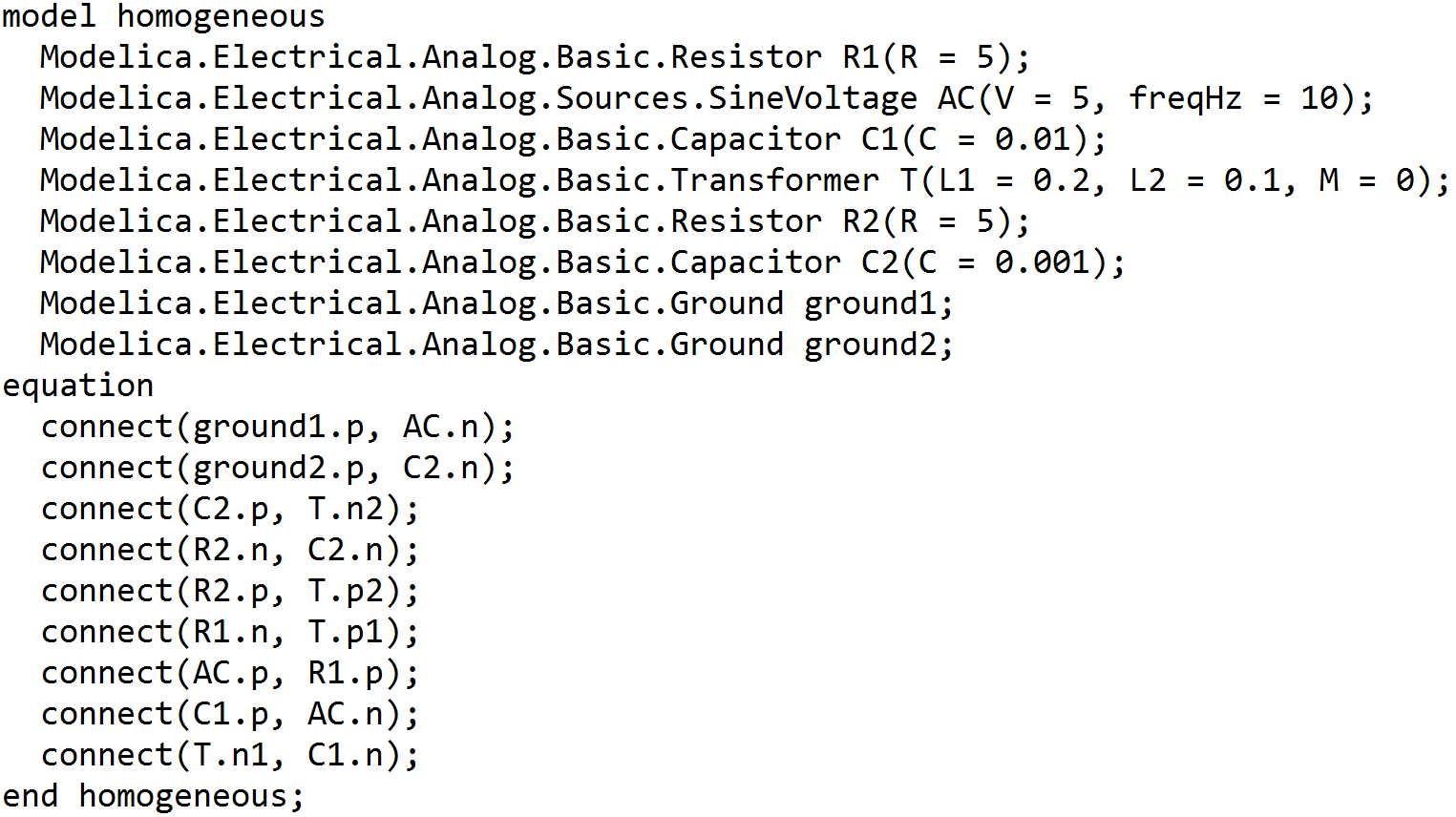}
          \caption{Modelica model}
          \label{fig:homo_RLC_Modelica}
        \end{subfigure}
        \begin{subfigure}[b]{0.49\textwidth}
          \centering
          \includegraphics[width=0.7\linewidth]{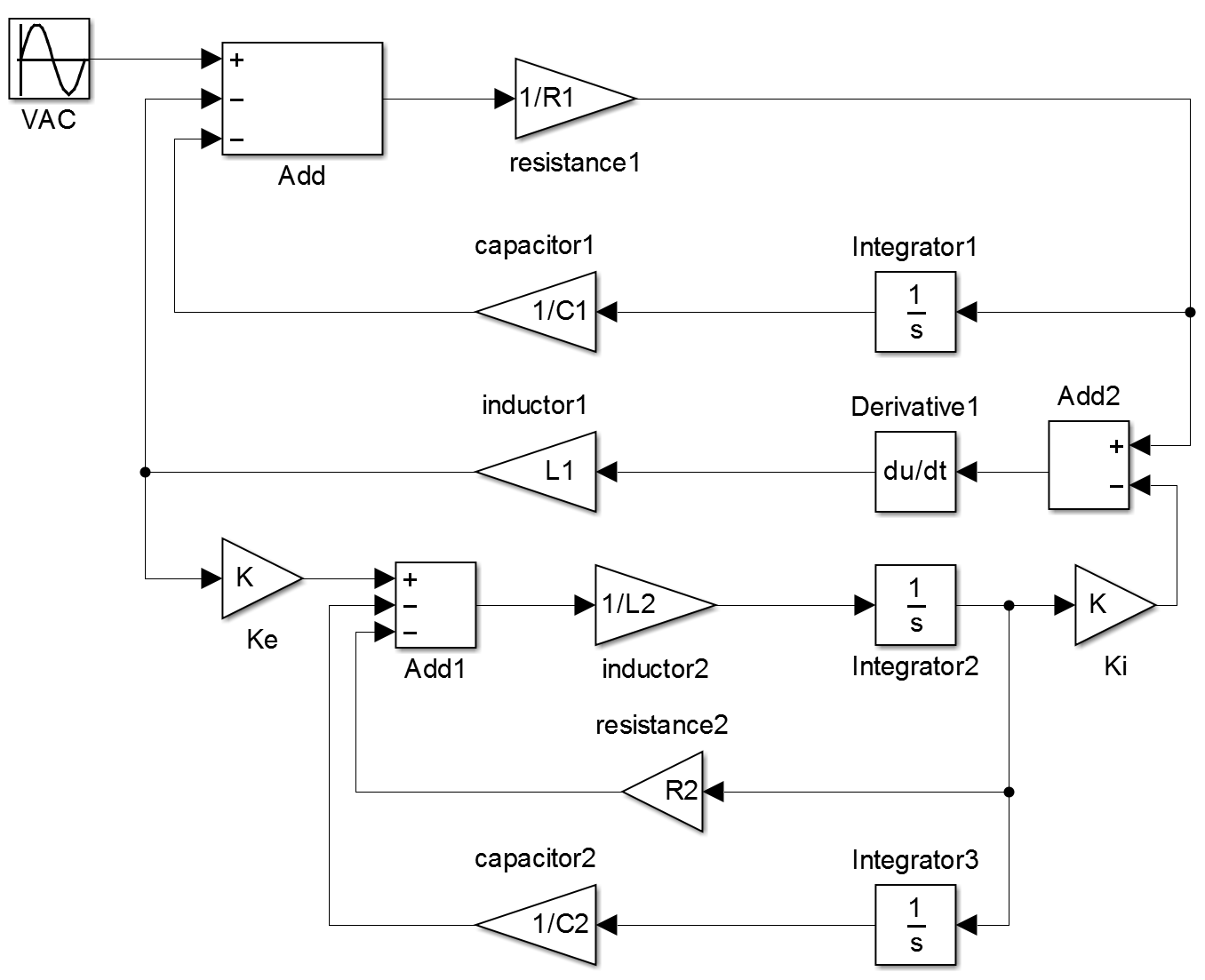}
          \caption{Simulink model}
          \label{fig:homo_RLC_Simulink}
        \end{subfigure}
        \caption{Different models of a multi-domain RLC electrical circuit}
        \label{fig:Comparison_homo_RLC}
        \end{figure}

\begin{figure}[!htb]
        \centering
        \begin{subfigure}[b]{0.5\textwidth}
          \centering
          \includegraphics[width=0.8\linewidth]{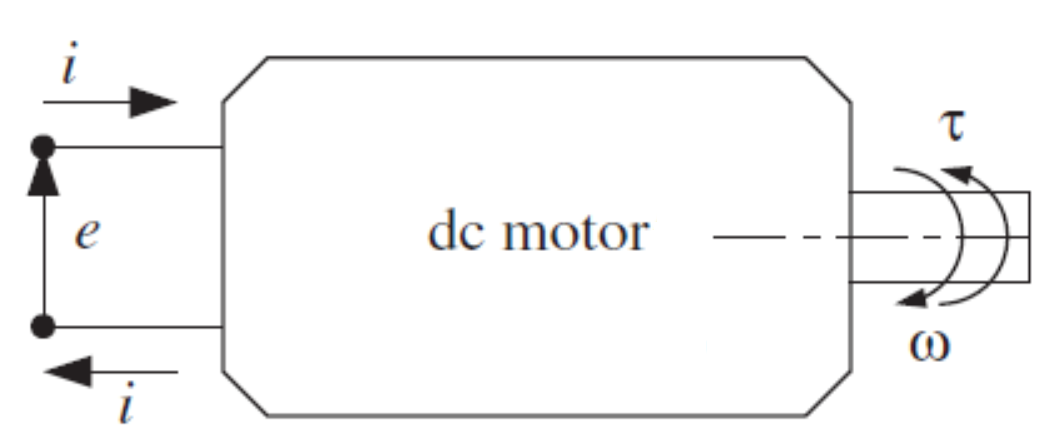}
          \caption{A schematic DC motor}
          \label{fig:Symbol_Motor}
        \end{subfigure} \par\medskip

        \centering
        \begin{subfigure}[b]{0.49\textwidth}
          \centering
          \includegraphics[width=0.8\linewidth]{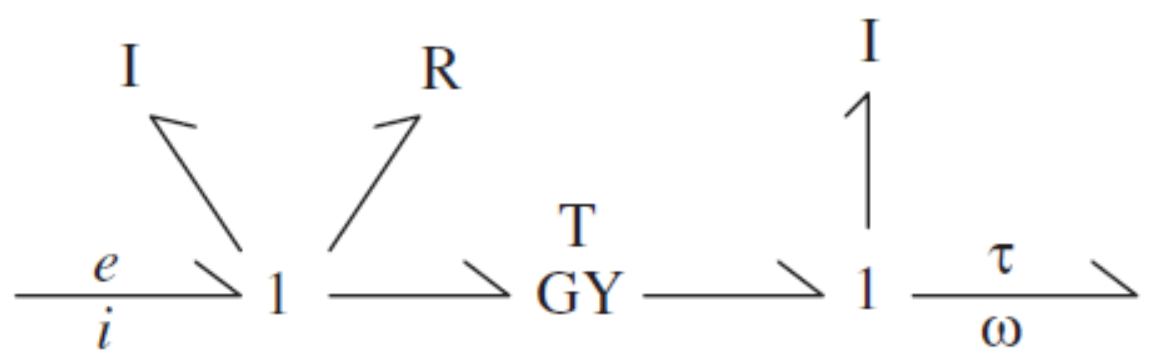}
          \caption{Bond graph model \protect\cite{karnopp2012system}}
          \label{fig:Motor_bond_graph}
        \end{subfigure}
        \centering
        \begin{subfigure}[b]{0.49\textwidth}
          \centering
          \includegraphics[width=0.8\linewidth]{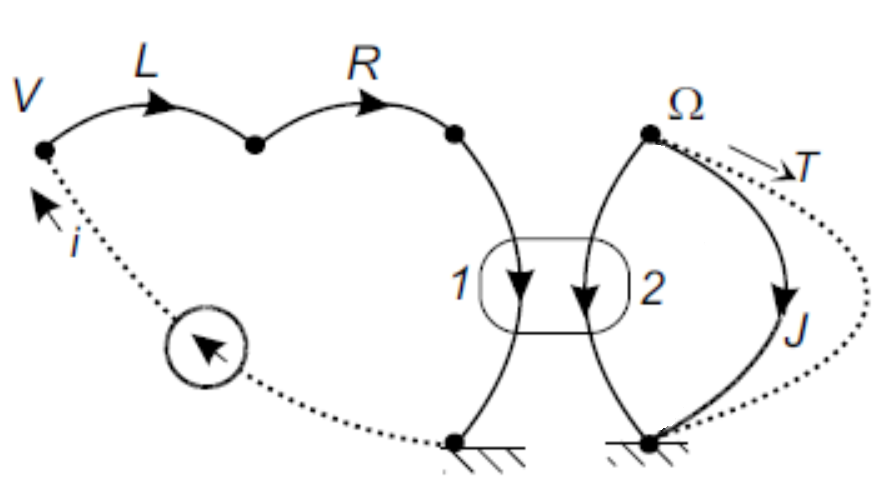}
          \caption{Linear graph model \protect\cite{rowell1997system}}
          \label{fig:Motor_linear_graph}
        \end{subfigure} \par\medskip
         
        \begin{subfigure}[b]{0.49\textwidth}
          \centering
          \includegraphics[width=\linewidth]{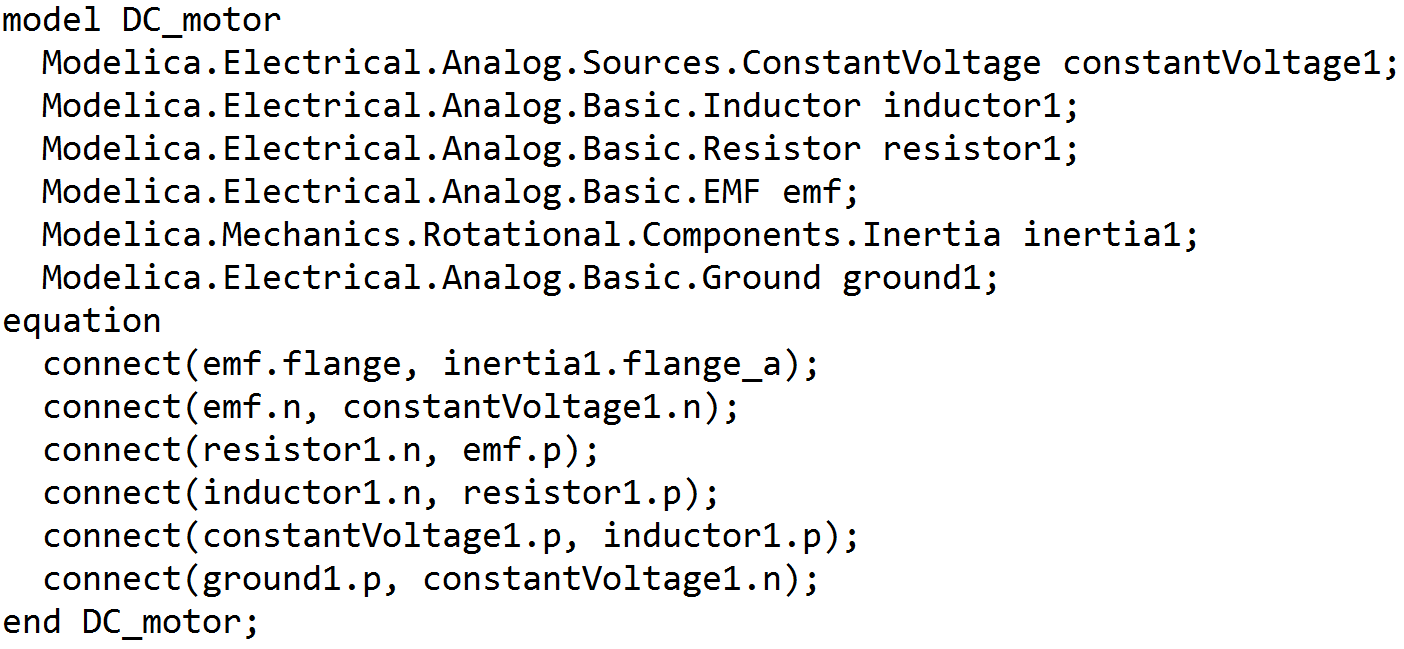}
          \caption{Modelica model}
          \label{fig:Motor_Modelica}
        \end{subfigure}
        \begin{subfigure}[b]{0.49\textwidth}
          \centering
          \includegraphics[width=0.8\linewidth]{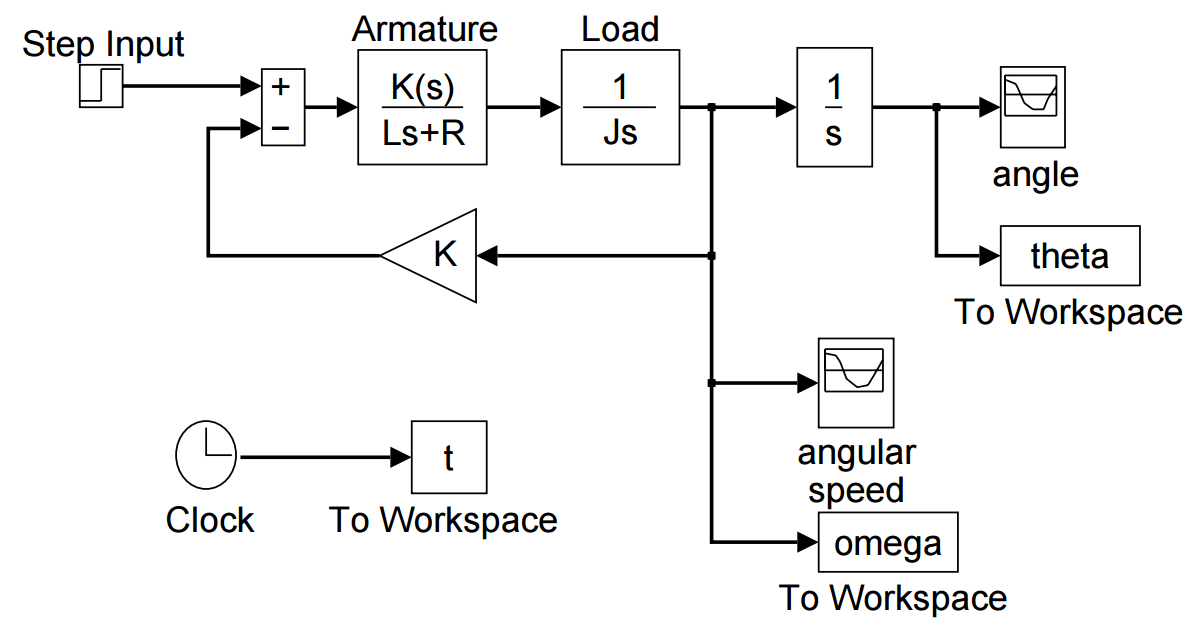}
          \caption{Simulink model \protect\cite{babuska1999matlab}}
          \label{fig:Motor_Simulink}
        \end{subfigure}
        \caption{Different models of a DC motor}
        \label{fig:Comparison_Motor}
        \end{figure}

Such abstract transformers and gyrators in each language can represent hundreds of complex transformer and gyrator devices. 
For example, Figures \ref{fig:Comparison_homo_RLC} and \ref{fig:Comparison_Motor} show two examples of multi-domain models represented by different modeling languages. Figure \ref{fig:Comparison_homo_RLC} is an example of a homogeneous multi-domain system, where all sub-domains are electrical domains. Figure \ref{fig:Comparison_Motor} is an example of a heterogeneous multi-domain system, where one electrical domain and one mechanical domain are connected by a DC motor. 

\begin{figure}[!htb]
		\centering
		\includegraphics[width=\linewidth]{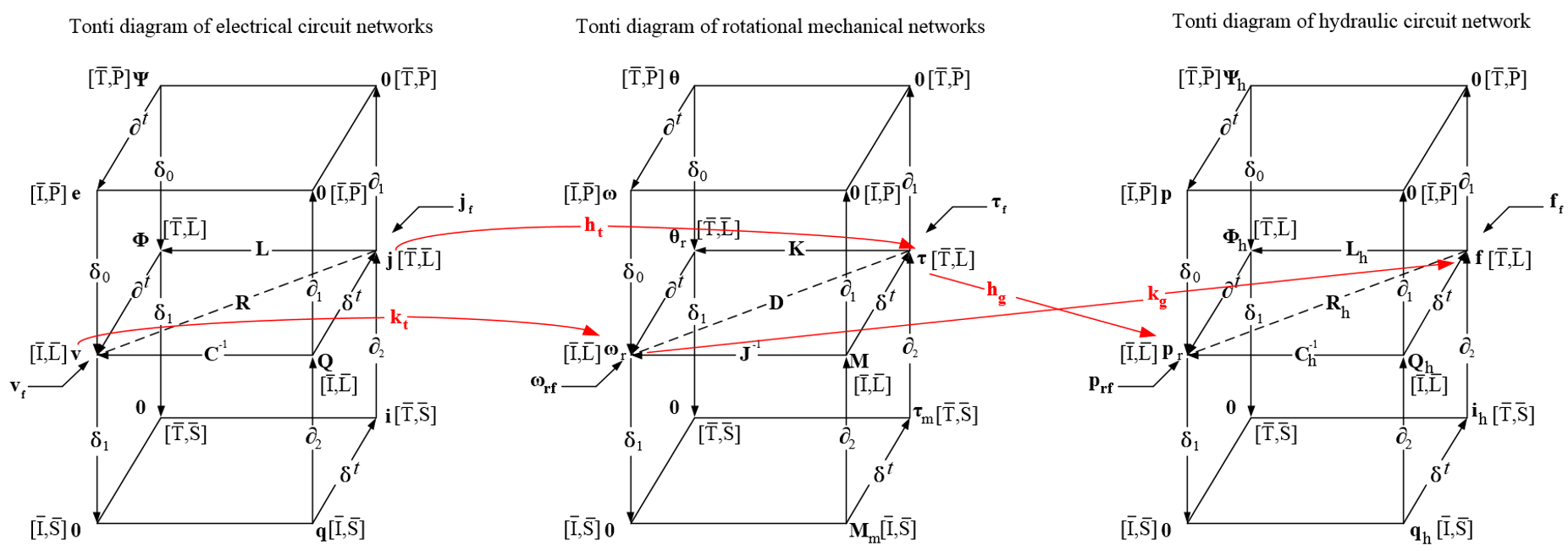}
		\caption{Behavior of three-domain system, a hydraulic pump driven by electrical motor, is abstracted by relations between three interacting extended Tonti diagrams.}
		\label{fig:interaction}		
		\medskip
		\begin{minipage}{\textwidth} 
		{\footnotesize{Commonly-used physical variables in the Tonti diagram of rotational mechanical networks: $\bm{\theta}$ - angle, ${\bm{\theta}}_r$ -  relative angle, $\bm{\omega}$ - rotational velocity, ${\bm{\omega}}_r$ - relative rotational velocity, $\bm{\tau}$ - torque, $\bf{M}$ - angular momentum, $\bf{K}$ - rotational stiffness, $\bf{D}$ - rotational damping coefficient and $\bf{J}$ - moment of inertia. Commonly-used physical variables in the Tonti diagram of hydraulic networks: ${{\bf{\Psi }}_h}$ - potential of hydraulic flux, ${{\bf{\Phi }}_h}$ - hydraulic flux, $\bf{p}$ - pressure, ${\bf{p}}_r$ - pressure drop, ${\bf{Q}}_h$ - flow of volume, $\bf{f}$ - flow rate, ${\bf{L}}_h$ -  hydraulic inductance, ${\bf{R}}_h$ - hydraulic resistance, ${\bf{C}}_h$ - hydraulic capacitance.
${{\bf{h}}_{{t}}}$ and ${{\bf{k}}_{{t}}}$ are transformer modulus. ${{\bf{k}}_{{t}}}$ is the ratio of relative rotational velocity of motor to voltage drop and ${{\bf{h}}_{{t}}}$ is the ratio of torque to current}. ${{\bf{h}}_{{g}}}$ and ${{\bf{k}}_{{g}}}$ are gyrator modulus. ${{\bf{k}}_{{g}}}$ is the ratio of flow rate to relative rotational velocity of motor and ${{\bf{h}}_{{g}}}$ is the ratio of pressure drop to torque.} 
\end{minipage}
	\end{figure}

In our proposed combinatorial model of lumped parameter systems,  physical transducer devices can be abstracted by additional relations between primal and dual variables in each of the subsystems;  such relations may be governed by additional constitutive, interaction, or conservation constrains imposed on the multi-domain system \cite{tonti2003classification}.  For example, consider the three-domain system consisting of a hydraulic pump that is controlled by an electrical motor.  Behaviors of the subsystems (electrical, rotational motion, and hydraulic) are described by the corresponding extended Tonti diagrams) that are further constrained as shown in Figure \ref{fig:interaction}.	

In principle, such a representation is sufficient for capturing the behavior of a multi-domain system.  Each Tonti diagram corresponds to a system of ordinary differential equations that are coupled by the transducer constraints.   When the constraints are algebraic, this representation corresponds to the usual system of differential-algebraic equations.   More complex transducer relationships may involve multiple physical variables as well as non-linear and differential constraints \cite{singleton1950theory}, resulting in more complex models of behaviors. 

However, this representation neither recognizes nor takes advantage of the fact that all single-domain behaviors are isomorphic, which allows to treat the whole multi-domain system as a collection of four constrained cochain complexes on  a \textit{single} cell complex model. 
Below we will define such a model, which takes a form of a generalized Tonti diagram.  We then show that, in the presence of two most common transducers:  ideal transformers and gyrators,  the governing equations for such a model may be generated by following the paths on the generalized Tonti diagram.  This results extends the result of Section 3 to multi-domain lumped parameter systems.


\subsection{Generalized Tonti diagram for multi-domain systems}  
 
Since lumped parameter models in different physical domain are isomorphic, so are their corresponding Tonti diagrams.   In this sense, a single Tonti diagram describes behavior of all lumped parameter systems, provided that the variable of the same space-time type are identified and generalized.  

Two most common generalizations are mechanical (generalized displacement-force model) and electrical (generalized voltage-current model). For the sake of consistency with the discussion in section \ref{single}, we will adopt the generalized electrical model.  For example, the electrical voltage, mechanical translational velocity and hydraulic pressure difference are all considered to be of the same type called the generalized voltage; the electrical resistors, mechanical dampers and hydraulic resistors are all identified as   generalized resistors, and so on. In order to emphasize the generalized nature of all physical quantities and to distinguish them from the actual physical electrical network model, we will choose a different set of symbols.   Specifically, the generalized Tonti diagram is defined by four exact cochain sequences on a single cell complex:  

\begin{eqnarray}\label{generalized=sequence}
	{\rm primal:} \ \ & {\bf p}^0 
	\stackrel{\delta_0} {\longrightarrow}
	{\bf a}^1
	\stackrel{\delta_1} {\longrightarrow}
	{\bf 0}^2 
\\
	 &{\bf d}^0 
	 \stackrel{\delta_0} {\longrightarrow}
	{\bf u}^1
	\stackrel{\delta_1} {\longrightarrow}
	{\bf 0}^2 
\\
	{\rm dual:}  \ \ &{\bf s}^2 
	 \stackrel{\partial_2} {\longrightarrow}
	{\bf t}^1
	\stackrel{\partial_1} {\longrightarrow}
	{\bf 0}^0
\\
	& {\bf n}^2 
	\stackrel{\partial_2} {\longrightarrow}
	{\bf m}^1
	\stackrel{\partial_1} {\longrightarrow}
	{\bf 0}^0,
\end{eqnarray}
where ${{\bf{p}}^0}$ is a 0-cochain generalized potentials, ${{\bf{a}}^1}$ is a 1-cochain generalized voltages, ${{\bf{d}}^0}$ is a 0-cochain generalized potential magnetic fluxes, ${{\bf{u}}^1}$ is a 1-cochain generalized magnetic fluxes, ${{\bf{s}}^2}$ is a 2-cochain generalized mesh currents, 
${{\bf{t}}^1}$ is a 1-cochain generalized currents, ${{\bf{n}}^2}$ is 2-cochain generalized mesh charges, and ${{\bf{m}}^1}$ is a 1-cochain generalized electric charges.   There are also four cochains that are always ${\bf{0}}$:  2-cochain of generalized mesh magnetic fluxes, 2-cochain of generalized mesh voltages, 0-cochain of generalized node currents and 0-cochain of generalized node electric charges.

\begin{figure}[!htb]
        \begin{subfigure}{.5\textwidth}
              \centering
              \includegraphics[width=.8\linewidth]{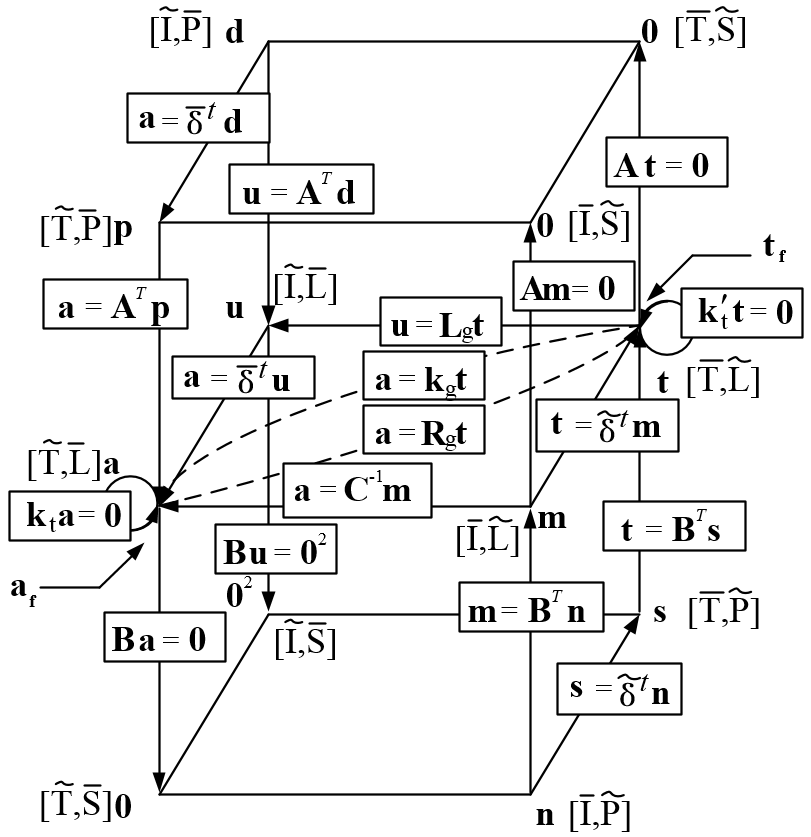}
              \caption{Matrix operators on dual cell complexes}
              \label{fig:Overall_RLC_Extended}
        \end{subfigure}
        \begin{subfigure}{.5\textwidth}
              \centering
              \includegraphics[width=.8\linewidth]{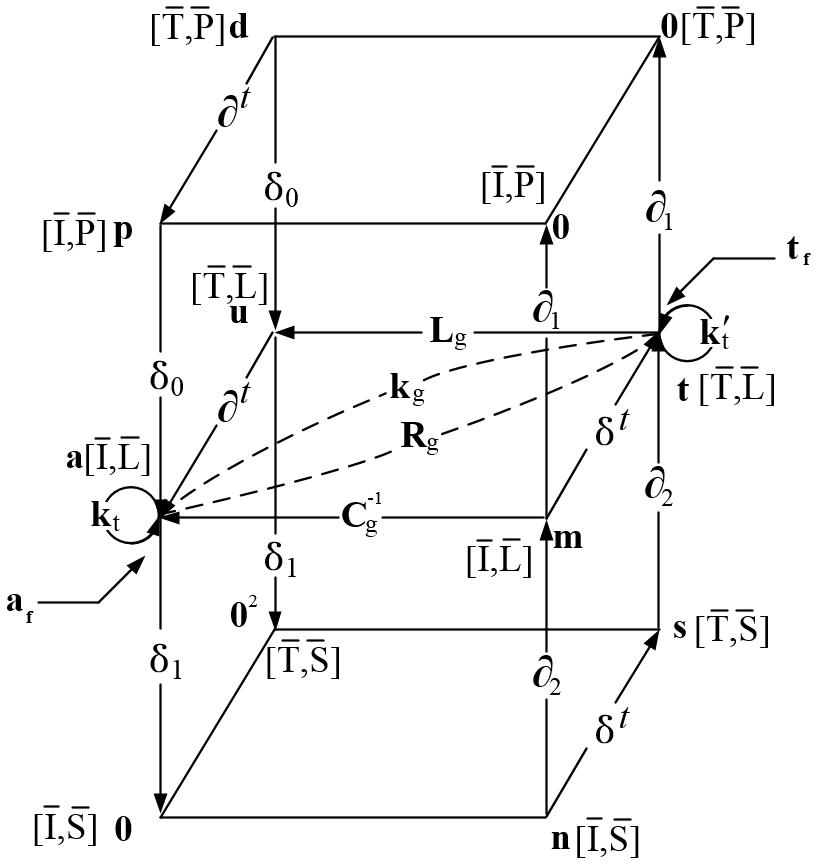}
              \caption{Topological operators on a single cell complex}
              \label{fig:Overall_RLC_Extended_T}
        \end{subfigure}
            \caption{Generalized extended Tonti diagram for generalized RLC network system}
            \label{fig:Genenralized_Overall_RLC}
    \end{figure}

With such a generalization, all the physical variables of the same space-time type are replaced by their generalized counterparts, effectively transforming model of the heterogeneous multi-domain system in an abstract  (generalized) homogeneous system.  The behavior of this system is governed by the generalized Tonti diagram shown in Figure \ref{fig:Genenralized_Overall_RLC}. As before, the generalized primal and dual cochains are related by (generalized) constitutive relations: resistance ${\bf{R}}_g$, capacitance ${\bf{C}}_g$, and inductance ${\bf{L}}_g$.  

Furthermore, since all physical quantities are generalized, the actions of ideal transformers and gyrators can be modeled simply as additional constraints on the cochains in the generalized Tonti diagram.  Traditionally, a transformer is abstracted as a linear transformation 
\begin{equation} \label{transformer_constitutive}
\left[ {\begin{array}{*{20}{c}}
{{a_1}}\\
{{t_1}}
\end{array}} \right] = \left[ {\begin{array}{*{20}{c}}
k_t&0\\
0&{{1 \mathord{\left/
 {\vphantom {1 k_t}} \right.
 \kern-\nulldelimiterspace} k_t}}
\end{array}} \right]\left[ {\begin{array}{*{20}{c}}
{{a_2}}\\
{{t_2}}
\end{array}} \right]
\end{equation}
where $k_t$ is the transformer's modulus measuring the ratio between two (generalized) voltages $a_1$ and $a_2$, as well as the reciprocal ratio between the generalized currents $t_1$ and $t_2$ in order to enforce energy balance $a_1 t_1 = a_2 t_2$.  It is easy to see that the same relationships may be enforced by a pair of linear constraints  
\begin{equation} \label{transformer_constitutive_g_v}
\left\{ \begin{array}{l}
\left[ {\begin{array}{*{20}{c}}
1&{ - {k_t}}
\end{array}} \right] \cdot {\left[ {\begin{array}{*{20}{c}}
{{a_1}}&{{a_2}}
\end{array}} \right]^T} = 0\\
\left[ {\begin{array}{*{20}{c}}
1&{ - k_t^{ - 1}}
\end{array}} \right] \cdot {\left[ {\begin{array}{*{20}{c}}
{{t_1}}&{{t_2}}
\end{array}} \right]^T} = 0
\end{array} \right.
\end{equation}
Generalizing, every ideal transformer can be represented by a pair of linear constraints 
$${{\bf{k}}_t} {{\bf{a}}} = 0; \qquad 
{{\bf{k}}_t^\prime} {{\bf{t}}} = 0
$$
on cochains of generalized voltages ${\bf{a}}$ and currents ${\bf{t}}$.
These constraints are indicated on the generalized Tonti diagram in Figure  \ref{fig:Genenralized_Overall_RLC} by two cycles.

Similarly, the effect of an abstract gyrator is usually described by a linear transformation
\begin{equation} \label{gyrator_constitutive_m}
\left[ {\begin{array}{*{20}{c}}
{{a_1}}\\
{{a_2}}
\end{array}} \right] = \left[ {\begin{array}{*{20}{c}}
0&{{k_g}}\\
{{k_g}}&0
\end{array}} \right]\left[ {\begin{array}{*{20}{c}}
{{t_1}}\\
{{t_2}}
\end{array}} \right]
\end{equation}
where the modulus $k_g$ relates the dual quantities in two interacting domains:  generalized voltage $a_1$  of the first domain is proportional to the generalized current $t_2$ of the second domain, and vice versa, again satisfying the ideal energy balance law.  Equivalently, a generalized gyrator may be represented by  a linear transformation  ${{\bf{k}}_g}$ that relates  the cochains of generalized voltages and currents, as indicated by a dotted arrow in the Tonti diagram in Figure \ref{fig:Genenralized_Overall_RLC}.

\subsection {System state equations of multi-domain systems}

With all physical variables generalized, the heterogeneous multi-domain system now becomes a homogeneous multi-domain system in terms of generalized physical variables. Instead of multiple 2-cochain complexes associated with different types of physical variables, the algebraic topological model of the multi-domain system  is now a set of 2-cochain complexes associated with the same (generalized) type of physical variables that are defined over a single cell complex and are constrained by abstract transformers and gyrators.

\begin{figure}[!htb]
        \centering
        \begin{subfigure}[b]{0.24\textwidth}
              \centering
              \includegraphics[width=\linewidth]{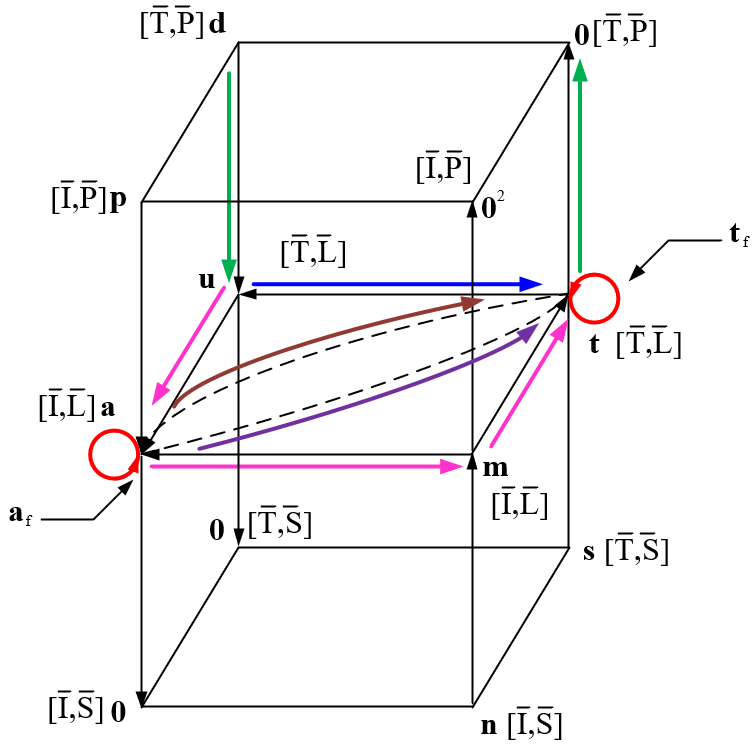}
              \caption{State variable: ${{\bf{d}}^0}$ }
              \label{fig:state_variable_psi0_generalized}
        \end{subfigure}
        \begin{subfigure}[b]{0.225\textwidth}
              \centering
              \includegraphics[width=\linewidth]{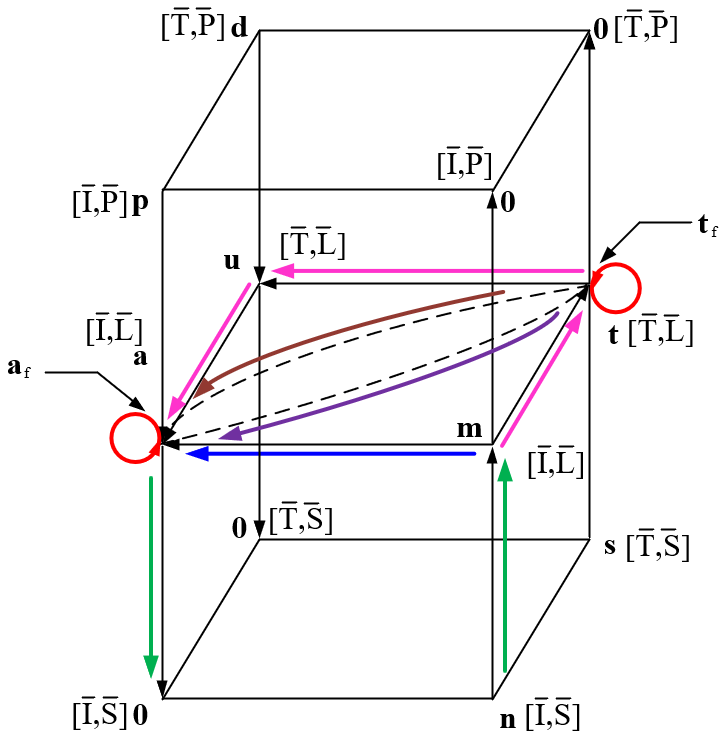}
              \caption{State variable:  ${{\bf{n}}^2}$ }
              \label{fig:state_variable_q0_generalized}
        \end{subfigure} 
        \begin{subfigure}[b]{0.24\textwidth}
          \centering
          \includegraphics[width=\linewidth]{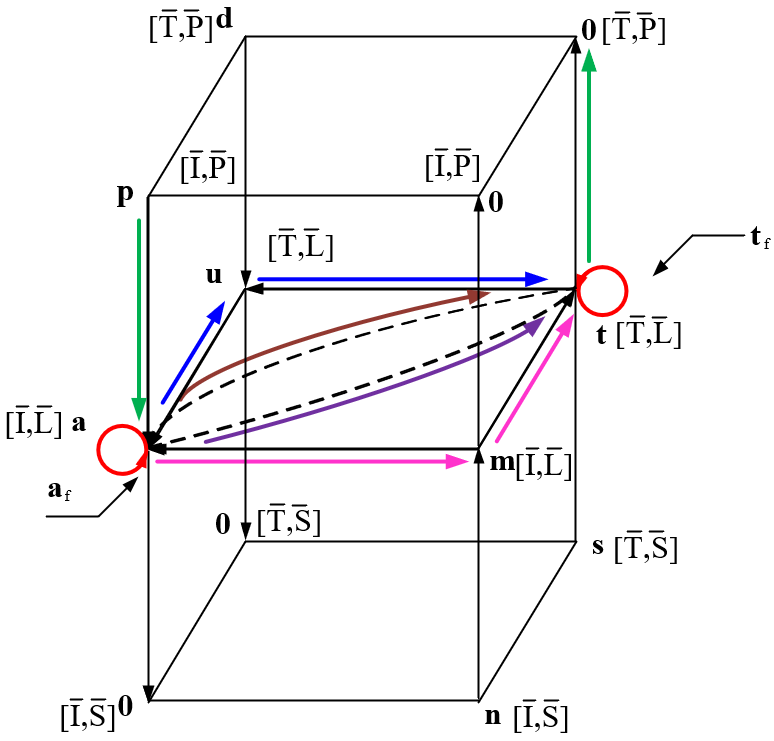}
          \caption{State variable: ${{\bf{p}}^0}$}
          \label{fig:state_variable_e0_generalized}
        \end{subfigure}
        \begin{subfigure}[b]{0.225\textwidth}
          \centering
          \includegraphics[width=\linewidth]{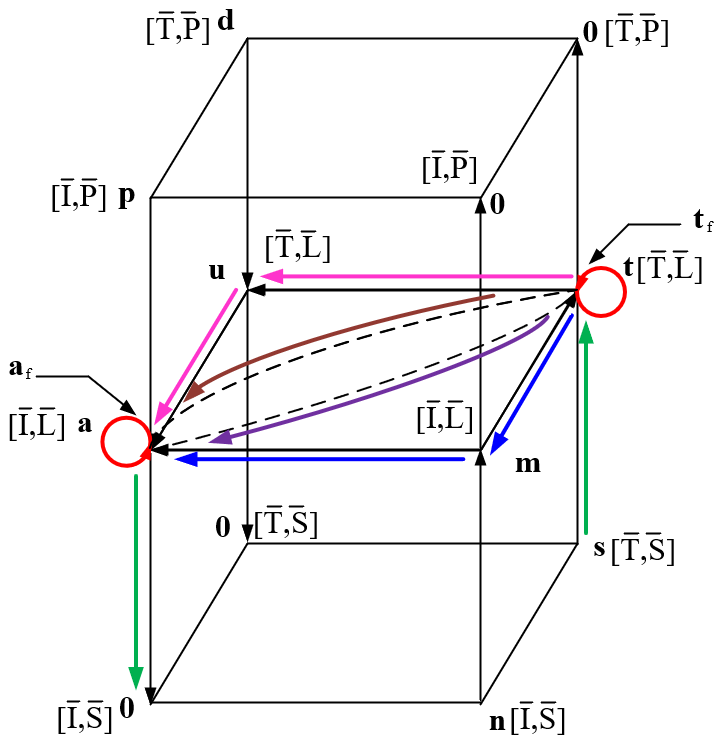}
          \caption{State variable: ${{\bf{s}}^2}$}
          \label{fig:state_variable_i0_generalized}
	   \end{subfigure} \par\medskip
        \begin{subfigure}[b]{0.24\textwidth}
              \centering
              \includegraphics[width=\linewidth]{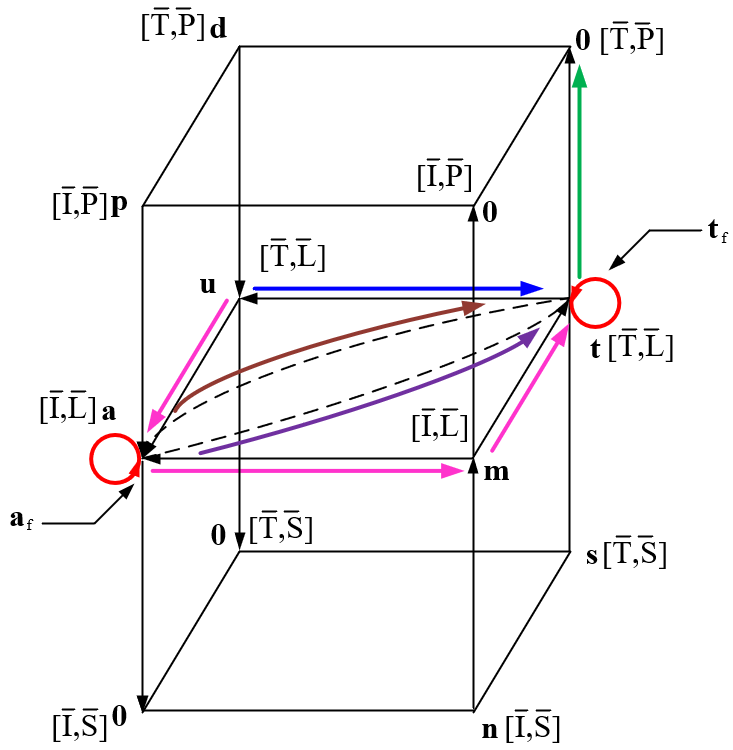}
              \caption{State variable: ${{\bf{u}}^1}$}
              \label{fig:state_variable_j1_generalized}
        \end{subfigure}
        \begin{subfigure}[b]{0.225\textwidth}
              \centering
              \includegraphics[width=\linewidth]{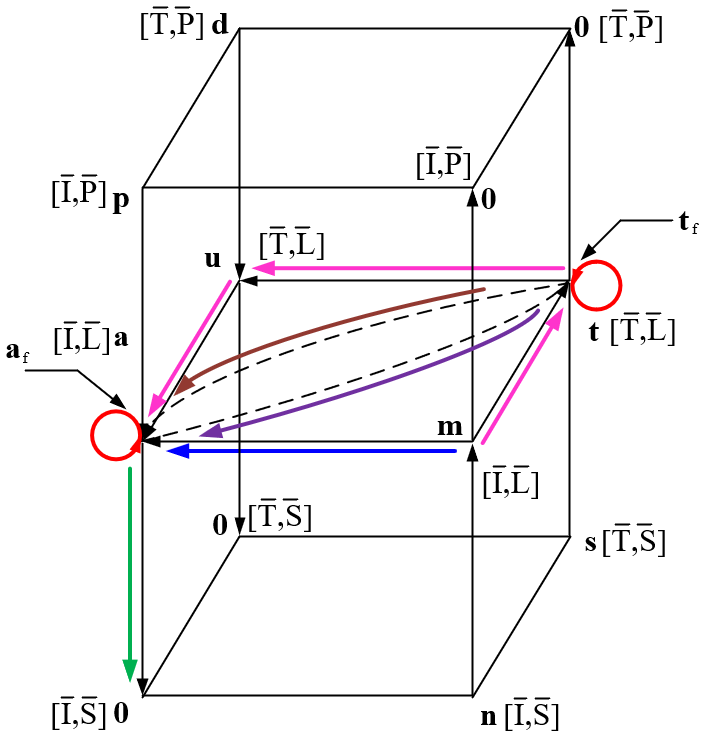}
              \caption{State variable: ${{\bf{m}}^1}$}
              \label{fig:state_variable_phi1_generalized}
        \end{subfigure} 
        \begin{subfigure}[b]{0.24\textwidth}
              \centering
              \includegraphics[width=\linewidth]{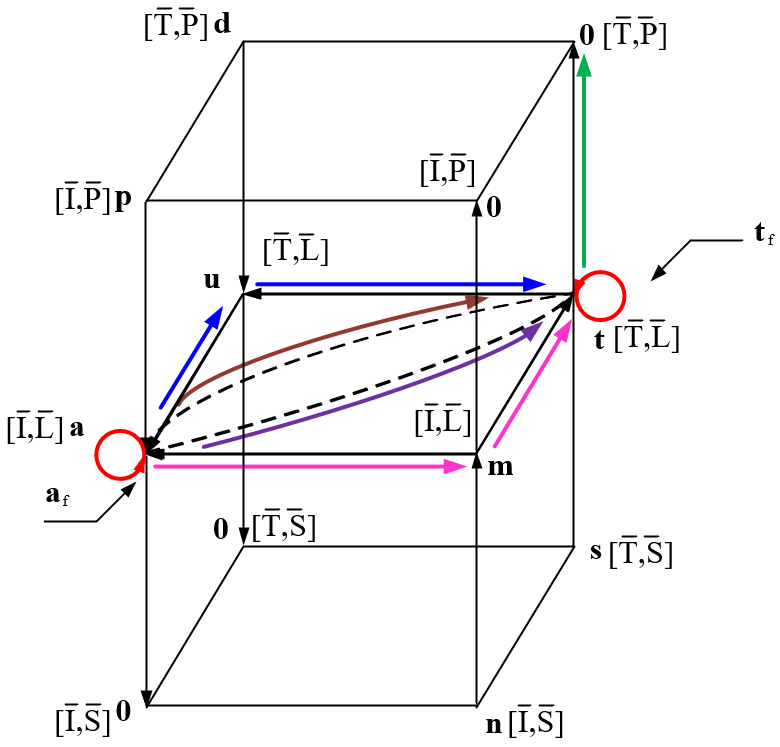}
              \caption{State variable: ${{\bf{a}}^1}$}
              \label{fig:state_variable_Q1_generalized}
        \end{subfigure}
        \begin{subfigure}[b]{0.225\textwidth}
              \centering
              \includegraphics[width=\linewidth]{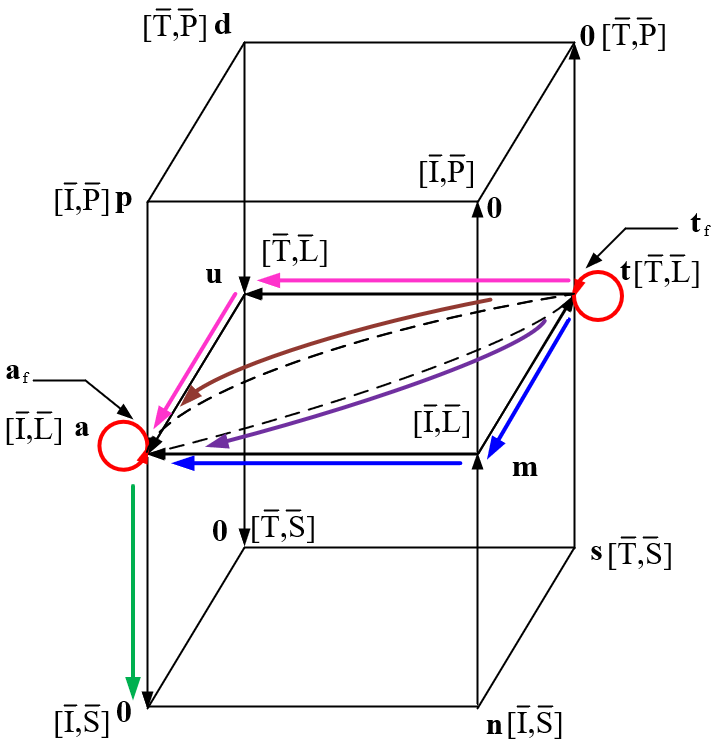}
              \caption{State variable: ${{\bf{t}}^1}$}
              \label{fig:state_variable_v1_generalized}
        \end{subfigure}
            \caption{State equation generation paths on the generalized extended Tonti diagram}
            \label{fig:paths and generalized state equations}
        \end{figure}

Eight different methods of generating the
state equations are indicated by paths in the generalized Tonti diagram shown in Figure \ref{fig:paths and generalized state equations}. Just as with the single domain systems, each path is a sequence of the arrows indicating composition of the corresponding physical laws. In contrast to the single domain diagram, the middle horizontal section of the generalized Tonti diagram allows an additional  alternative
path relating the primal 1-cochain of generalized voltages ${{\bf{a}}^1}$ and the dual 1-cochain of generalized currents ${{\bf{t}}^1}$ by  gyrators  as well as  two alternative cyclic paths (shown in red), which respectively constrain the generalized voltages and currents of transformers.

For example, if we use the paths in Figure \ref{fig:state_variable_psi0_generalized} to generate the state equations, then the 0-cochain ${{\bf{d}}^0}$ is selected as the state variable. The system state equation can be generated by composition of seven physical laws (two topological and five constitutive) starting with a 0-cochain ${{\bf{d}}^0}$.  First, coboundary operator in space $\delta_0$ applied to generalized potential magnetic fluxes ${{\bf{d }}^0}$ in order to generate generalized magnetic fluxes ${{\bf{u}}^1}$.  Now the path splits into two: the blue arrow corresponds to constitutive law ${\bf{L}}_g^{ - 1}$ that relates generalized magnetic fluxes to generalized currents of generalized inductors;  the pink arrow takes the generalized magnetic fluxes to generate generalized voltages ${{\bf{a}}^1}$ via the boundary operator in time $\partial _1^t$.  From here the path splits into four:  (1) the purple arrow corresponds to the constitutive law ${\bf{R}}_g^{ - 1}$ that relates generalized voltages to generalized currents of generalized resistors; (2) the brown arrow corresponds to the constitutive law ${\bf{k}}_g^{ - 1}$ that relates generalized voltages to generalized currents of gyrators; (3) the left red cyclic arrow corresponds to constitutive law  ${{\bf{k}}_t}$ that constrains the generalized voltages of generalized transformers; 
(4) the pink arrow takes the generalized voltages to generate generalized electric charge of generalized capacitors by using constitutive law ${\bf{C}}_g$, followed by taking the generalized electric charge to generate generalized currents of generalized capacitors via the coboundary operator in time $\delta _0^t$. Note that these four paths and the right red cyclic path\footnote{Note that the generalized currents of transformers ${\bf{t}}_T^1$ cannot be directly obtained from the generalized voltages of transformers ${\bf{a}}_T^1$, so we treat them as additional unknown variables in the system state equation.} corresponding to five constitutive laws merge into a single 1-cochain of generalized through variables ${{\bf{t}}^1}$, which is then transformed one more time by the upward green arrow corresponding to KCL ${\partial _1}{{\bf{t}}^1}= {\bf{0}}$. Taking into account the generalized sources, collecting the terms with known generalized sources and moving them to the right
hand side, above procedure results in the state equation Eq.\ref{node_potential_state_eq_mixed}, with Eq.\ref{Constraint_Eq_voltage} and Eq.\ref{Constraint_Eq_current} being the constraint equations generated from two cyclic red paths.

\resizebox{.9\linewidth}{!}{
  \begin{minipage}{\linewidth}
\begin{equation} \label{node_potential_state_eq_mixed}
\begin{aligned}
&  {{\partial _{1}}}\left( {\underbrace {\delta _0^t {{\bf{C}}_g} \partial _1^t {{ {{\delta _0}} }}{{\bf{d }}^0}}_{\substack{\text{generalized currents} \\ \text{of generalized C}}} +  \underbrace { {\bf{R}}_g^{ - 1} \partial _1^t {{ {{\delta _0}} }}{{\bf{d }}^0}}_{\substack{\text{generalized currents} \\ \text{of generalized R}}} + \underbrace { {\bf{L}}_g^{ - 1} {{ {{\delta _0}} }}{{\bf{d }}^0}}_{\substack{\text{generalized currents} \\ \text{of generalized L}}} + \underbrace { {\bf{k}}_g^{ - 1} \partial _1^t { {{\delta _0}} }{{\bf{d }}^0}}_{\substack{\text{generalized currents} \\ \text{of gyrators}}} +  \underbrace {{{\bf{t}}_T^1}}_{\substack{\text{generalized currents} \\ \text{of transformers}}}} \right)  \\&=  {{\partial _{1}}} \left[ {\underbrace {{\bf{t}}_{\bf{f}}^1}_{\substack{\text{generalized currents} \\ \text{sources}}} - \underbrace {\left( {\delta_0^t {{\bf{C}}_g} + {\bf{R}}_g^{ - 1} }+ {\bf{k}}_g^{ - 1}+ {\bf{L}}_g^{ - 1}  \delta_0^t\right){\bf{a}}_{\bf{f}}^1}_{\substack{\text{equivalent generalized current}\\ \text{sources generated from} \\ \text{generalized voltage sources}}}} \right] 
\end{aligned}
\end{equation}
  \end{minipage}
}

\begin{equation} \label{Constraint_Eq_voltage}
{{\bf{k}}_t}\left( {\partial _1^t{\delta _0}{{\bf{d}}^0} - {\bf{a}}_{\rm{f}}^1} \right) = {\bf{0}}
\end{equation} 

\begin{equation} \label{Constraint_Eq_current}
{\bf{k}}_t^\prime\left( {{{\bf{t}}^1} - {\bf{t}}_{\rm{f}}^1} \right) = {\bf{0}}
\end{equation}


Assuming that the number of state variables is \textit{N} and the number of transformers is \textit{M}, then the system of equations Eq.\ref{node_potential_state_eq_mixed} has \textit{N+2M} unknowns. The two transformer's constraints  generate  \textit{2M} constraint equations, while the other paths generate \textit{N} state equations. As expected, the number of unknowns equals to the total number of state and constraint equations.

Other methods for generating the system state equation follow the different paths in Figure \ref{fig:state_variable_q0_generalized} $\sim$ Figure \ref{fig:state_variable_v1_generalized}. For example, in Figure \ref{fig:state_variable_q0_generalized}, the process starts with the dual 2-cochain of generalized mesh electric charges ${{\bf{n}}^2}$ selected as the state variable and amounts to another composition of the seven physical laws indicated by the corresponding paths. The blue, purple, pink and brown path corresponds to the four generalized constitutive laws (generalized capacitance, resistance, inductance and gyrator), relating the generalized currents to generalized voltages. The two green arrows correspond to the $\partial_2$ operator transforming ${{\bf{n}}^2}$ to generalized electric charges ${{\bf{m}}^1}$ and application of KVL (${\delta _1}{{\bf{a}}^1} = {\bf{0}}$). The two red cyclic arrows correspond to the constitutive equations of transformers. Putting it all together and taking into account the generalized sources, the composition procedures results in Eq.\ref{mesh_current_state_eq_mixed} 
, with Eq.\ref{Constraint_Eq_current_q0} and Eq.\ref{Constraint_Eq_voltage_q0} being the constraint equations generated from two cyclic red paths.

\resizebox{.9\linewidth}{!}{
  \begin{minipage}{\linewidth}
\begin{equation} \label{mesh_current_state_eq_mixed}
\begin{aligned}
&{\delta _{1}}\left( {\underbrace {\partial _1^t{{{\bf{L}}_g}} \delta _0^t\partial _2{{\bf{n}}^2}}_{\substack{\text{generalized voltages} \\ \text{of generalized L}}} + \underbrace {{{{\bf{R}}_g}}\delta _0^t\partial _2{{\bf{n}}^2}}_{\substack{\text{generalized voltages} \\ \text{of generalized R}}} + \underbrace {{{\bf{C}}_g^{ - 1}}\partial _2{{\bf{n}}^2}}_{\substack{\text{generalized voltages} \\ \text{of generalized C}}} + \underbrace {{{\bf{k}}_g} \delta _0^t\partial _2{{\bf{n}}^2}}_{\substack{\text{generalized voltages} \\ \text{of gyrators}}} + \underbrace {{{\bf{a}}_T^1}}_{\substack{\text{generalized voltages} \\ \text{of transformers}}}} \right)  \\&={\delta _{1}}\left[ {\underbrace { - {\bf{a}}_{\bf{f}}^1}_{\substack{\text{generalized voltages} \\ \text{sources}}} + \underbrace {\left( {\partial _1^t{{{\bf{L}}_g}} + {{{\bf{R}}_g}}}+ {{{\bf{k}}_g}}+{\bf{C}}_g^{ - 1} \partial_1^t \right){\bf{t}}_{\bf{f}}^1}_{\substack{\text{equivalent generalized voltage} \\ \text{sources generated from}\\ \text{generalized current sources}}}} \right]
\end{aligned}
\end{equation}
  \end{minipage}
}

\begin{equation} \label{Constraint_Eq_current_q0}
{\bf{k}}_t^\prime\left( {\delta _0^t{\partial _2}{{\bf{n}}^2} - {\bf{t}}_{\rm{f}}^1} \right) = {\bf{0}}
\end{equation}

\begin{equation} \label{Constraint_Eq_voltage_q0}
{{\bf{k}}_{\rm{t}}}\left( {{{\bf{a}}^1} - {\bf{a}}_{\rm{f}}^1} \right) = {\bf{0}}
\end{equation}

As with any dynamic system,  interpreting boundary $\partial_1^t$ and coboundary $\delta_0^t$ operations as differentiation in time syntactically transforms Eq. \ref{node_potential_state_eq_mixed} 
$\sim$ Eq.\ref{Constraint_Eq_current} and  Eq.\ref{mesh_current_state_eq_mixed} $\sim$ Eq.\ref{Constraint_Eq_voltage_q0} to a more familiar form:

\resizebox{.9\linewidth}{!}{
  \begin{minipage}{\linewidth}
\begin{equation} \label{node_potential_state_AE_simpler_ode_mixed} 
\begin{array}{*{20}{l}}
{{\partial _1}\left( {\underbrace {{{\bf{C}}_g}{\delta _0}{{{\bf{\ddot d}}}^0}}_{\substack{\text{generalized currents} \\ \text{of generalized C}}} + \underbrace {{\bf{R}}_g^{ - 1}{\delta _0}{{{\bf{\dot d}}}^0}}_{\substack{\text{generalized currents} \\ \text{of generalized R}}} + \underbrace {{\bf{L}}_g^{ - 1}{\delta _0}{{\bf{d}}^0}}_{\substack{\text{generalized currents} \\ \text{of generalized L}}} + \underbrace {{\bf{k}}_g^{ - 1}{\delta _0}{{{\bf{\dot d}}}^0}}_{\substack{\text{generalized currents} \\ \text{of gyrators}}} + \underbrace {{\bf{t}}_T^1}_{\substack{\text{generalized currents} \\ \text{of transformers}}}} \right)}\\
{ = {\partial _1}\left[ {\underbrace {{\bf{t}}_{\bf{f}}^1}_{\substack{\text{generalized current} \\ \text{sources}}} - \underbrace {\left( {{{\bf{C}}_g}{\bf{\dot a}}_{\bf{f}}^1 + {\bf{R}}_g^{ - 1}{\bf{a}}_{\bf{f}}^1 + {\bf{k}}_g^{ - 1}{\bf{a}}_{\bf{f}}^1 + {\bf{L}}_g^{ - 1}\int {{\bf{a}}_{\bf{f}}^1} dt} \right)}_{\substack{\text{equivalent generalized current} \\ \text{sources generated from}\\ \text{generalized voltage sources}}}} \right]}
\end{array}
\end{equation}		

  \end{minipage}
}

\begin{equation} \label{Constraint_Eq_voltage_diff}
{{\bf{k}}_t}\left( {{\delta _0}{{{\bf{\dot d}}}^0} - {\bf{a}}_{\rm{f}}^1} \right) = {\bf{0}}
\end{equation} 

\begin{equation} \label{Constraint_Eq_current_diff}
{\bf{k}}_t^\prime\left( {{{\bf{t}}^1} - {\bf{t}}_{\rm{f}}^1} \right) = {\bf{0}}
\end{equation}

\resizebox{.9\linewidth}{!}{
  \begin{minipage}{\linewidth}
\begin{equation} \label{mesh_current_state_AE_simpler_ode_mixed} 
\begin{array}{l}
{\delta _1}\left( {\underbrace {{{\bf{L}}_g}{\partial _2}{{{\bf{\ddot n}}}^2}}_{\substack{\text{generalized voltages} \\ \text{of generalized L}}} + \underbrace {{{\bf{R}}_g}{\partial _2}{{{\bf{\dot n}}}^2}}_{\substack{\text{generalized voltages} \\ \text{of generalized R}}} + \underbrace {{\bf{C}}_g^{ - 1}{\partial _2}{{\bf{n}}^2}}_{\substack{\text{generalized voltages} \\ \text{of generalized C}}} + \underbrace {{{\bf{k}}_g}{\partial _2}{{{\bf{\dot n}}}^2}}_{\substack{\text{generalized voltages} \\ \text{of gyrators}}} + \underbrace {{\bf{a}}_T^1}_{\substack{\text{generalized voltages} \\ \text{of transformers}}}} \right)\\
 = {\delta _1}\left( {\underbrace { - {\bf{a}}_{\bf{f}}^1}_{\substack{\text{generalized voltages} \\ \text{sources}}} + \underbrace {{{\bf{L}}_g}{\bf{\dot t}}_{\bf{f}}^1 + {{\bf{R}}_g}{\bf{t}}_{\bf{f}}^1 + {{\bf{k}}_g}{\bf{t}}_{\bf{f}}^1 + {\bf{C}}_g^{ - 1}\int {{\bf{t}}_{\bf{f}}^1} dt}_{\substack{\text{equivalent generalized voltage} \\ \text{sources generated from}\\ \text{generalized current sources}}}} \right)
\end{array}
\end{equation}
  \end{minipage}
}

\begin{equation} \label{Constraint_Eq_current_q0_diff}
{\bf{k}}_t^\prime\left( {{\partial _2}{{{\bf{\dot n}}}^2} - {\bf{t}}_{\rm{f}}^1} \right) = {\bf{0}}
\end{equation}

\begin{equation} \label{Constraint_Eq_voltage_q0_diff}
{{\bf{k}}_{\rm{t}}}\left( {{{\bf{a}}^1} - {\bf{a}}_{\rm{f}}^1} \right) = {\bf{0}}
\end{equation}

\begin{exmp} 
 \begin{figure}[!htb]
        \centering
        \begin{subfigure}[b]{\textwidth}
          \centering
          \includegraphics[width=.55\linewidth]{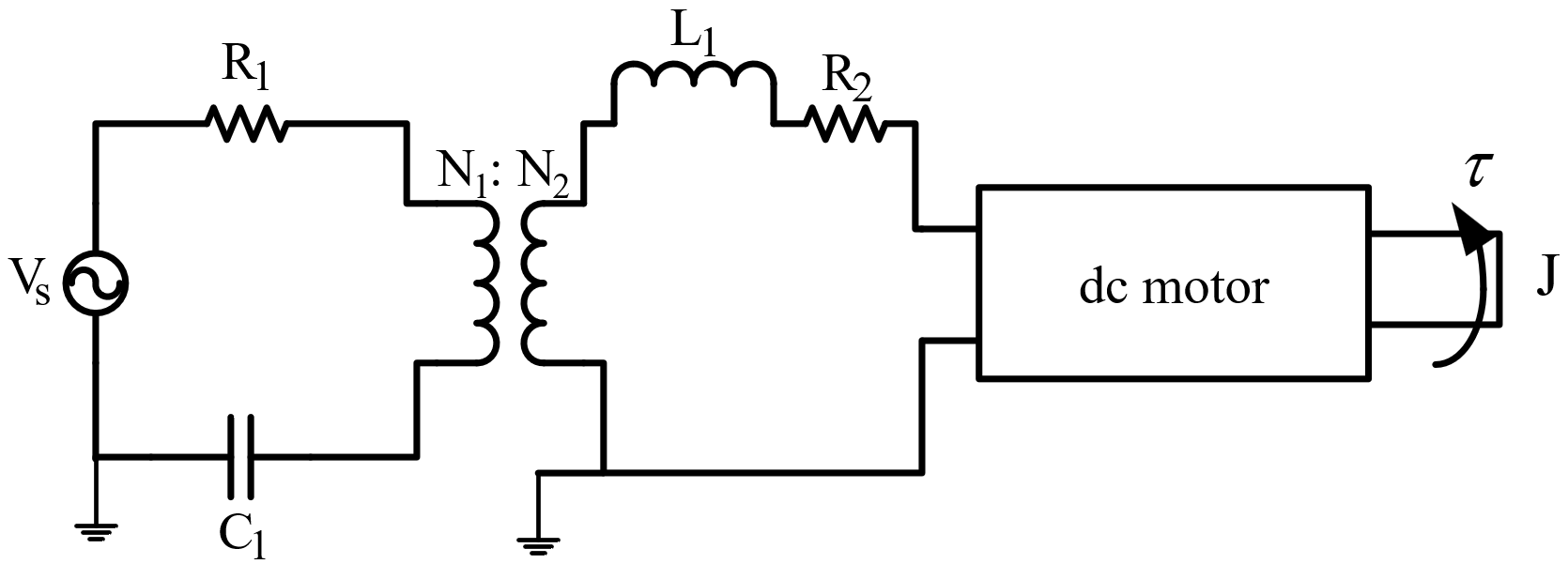}
          \caption{A multi-domain electro-mechanical system}
        \label{fig:mix_motor_RLC}
        \end{subfigure} \par\medskip

        \centering
        \begin{subfigure}[b]{\textwidth}
          \centering
		\includegraphics[width=.55\linewidth]{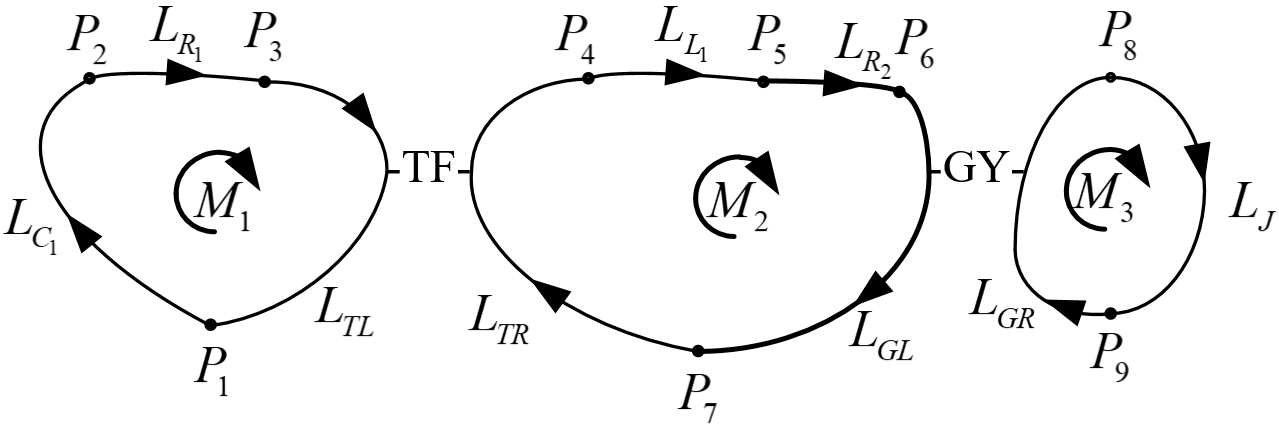}
        \caption{Topological structure}
        \label{fig:AT_model_mix_motor_RLC}
        \end{subfigure}
        \label{fig:AT_model_mix_motor_RLC_Topo}
        \caption{An electro-mechanical system and its topological structure}        
    \end{figure}

We will use an example of multi-domain electrical-mechanical system in Figure \ref{fig:mix_motor_RLC} to illustrate the derivation of Eq.\ref{node_potential_state_AE_simpler_ode_mixed} and Eq.\ref{mesh_current_state_AE_simpler_ode_mixed}
. The shown electrical-mechanical system contains two resistors $R_1$, $R_2$, one capacitor $C_1$, one inductor $L_1$, one voltage source $V_s$, one moment of inertia $J$, one external torque $\tau$, one electrical transformer and one DC motor.
The transformer between two electrical domains is an ideal electrical transformer, where the ratio of voltage drops (currents) equals the ratio (inverse ratio) of the winding numbers ${N_1}/{N_2}$; the gyrator between the electrical and the mechanical domain is an ideal DC motor, where the ratio (inverse ratio) of the voltage drop (current) and the rotational velocity (torque) is a constant number $k_g$. Topologically, the system is a 2-cell complex shown in Figure \ref{fig:AT_model_mix_motor_RLC}, and consisted of nine 0-cells (${P_1}$,${P_2}$,${P_3}$,${P_4}$,${P_5}$,${P_6}$,${P_7}$,${P_8}$,${P_9}$), nine 1-cells (${L_{{R_1}}}$,${L_{{C_1}}}$,${L_{TL}}$,${L_{TR}}$, ${L_{{R_2}}}$,${L_{{L_1}}}$,${L_{GL}}$,${L_{GR}}$,${L_{J}}$) and three 2-cells (${M_1}$,${M_2}$,${M_3}$). We use symbol -TF- to represent the abstract transformer and a symbol -GY- to represent the abstract gyrator. These two symbols identify the cells where the energy transaction may occur.

The algebraic topological model of the system contians: primal 0-cochain generalized potential magnetic fluxes (${{\bf{d}}^0} = {d_1} \cdot {P_1} + {d_2} \cdot {P_2} + {d_3} \cdot {P_3} + {d_4} \cdot {P_4} + {d_5} \cdot {P_5} + {d_6} \cdot {P_6} + {d_7} \cdot {P_7} + {d_8} \cdot {P_8} + {d_9} \cdot {P_9}$), primal 0-cochain generalized potentials (${{\bf{p}}^0} = {p_1} \cdot {P_1} + {p_2} \cdot {P_2} + {p_3} \cdot {P_3} + {p_4} \cdot {P_4} + {p_5} \cdot {P_5} + {p_6} \cdot {P_6} + {p_7} \cdot {P_7} + {p_8} \cdot {P_8} + {p_9} \cdot {P_9}$), primal 1-cochain generalized magnetic fluxes (${{\bf{u}}^1} = {u_1} \cdot {L_{{R_1}}} + {u_2} \cdot {L_{{C_1}}} + {u_3} \cdot {L_{TL}} + {u_4} \cdot {L_{TR}} + {u_5} \cdot {L_{{R_2}}} + {u_6} \cdot {L_{{L_1}}} + {u_7} \cdot {L_{GL}} + {u_8} \cdot {L_{GR}} + {u_9} \cdot {L_J}$), primal 1-cochain generalized voltages (${{\bf{a}}^1} = {a_1} \cdot {L_{{R_1}}} + {a_2} \cdot {L_{{C_1}}} + {a_{TL}} \cdot {L_{TL}} + {a_{TR}} \cdot {L_{TR}} + {a_5} \cdot {L_{{R_2}}} + {a_6} \cdot {L_{{L_1}}} + {a_{GL}} \cdot {L_{GL}} + {a_{GR}} \cdot {L_{GR}} + {a_9} \cdot {L_J}$), dual 1-cochain generalized currents (${{\bf{t}}^1} = {t_1} \cdot {L_{{R_1}}} + {t_2} \cdot {L_{{C_1}}} + {t_{TL}} \cdot {L_{TL}} + {t_{TR}} \cdot {L_{TR}} + {t_5} \cdot {L_{{R_2}}} + {t_6} \cdot {L_{{L_1}}} + {t_{GL}} \cdot {L_{GL}} + {t_{GR}} \cdot {L_{GR}} + {t_9} \cdot {L_J}$), dual 1-cochain generalized electric charges (${{\bf{m}}^1} = {m_1} \cdot {L_{{R_1}}} + {m_2} \cdot {L_{{C_1}}} + {m_{TL}} \cdot {L_{TL}} + {m_{TR}} \cdot {L_{TR}} + {m_5} \cdot {L_{{R_2}}} + {m_6} \cdot {L_{{L_1}}} + {m_{GL}} \cdot {L_{GL}} + {m_{GR}} \cdot {L_{GR}} + {m_9} \cdot {L_J}$), dual 2-cochain generalized mesh electric charges ${{\bf{n}}^2} = {n_1} \cdot {M_1} + {n_2} \cdot {M_2} + {n_3} \cdot {M_3}$, dual 2-cochain generalized mesh currents ${{\bf{s}}^2} = {s_1} \cdot {M_1} + {s_2} \cdot {M_2} + {s_3} \cdot {M_3}$ and four cochains that are always ${\bf{0}}$: 2-cochain of generalized mesh magnetic fluxes, 2-cochain of generalized mesh voltages, 0-cochain of generalized node currents and 0-cochain of generalized node electric charges. In order to obtain unique solution of the state equations, we consider 0-cells $P_1$, $P_7$ and $P_9$ as the reference node. Following the paths shown in Figure \ref{fig:state_variable_psi0_generalized},  generates Eq.\ref{node_potential_state_AE_simpler_ode_mixed} $\sim$ Eq.\ref{Constraint_Eq_current_diff}, with individual terms as follows:

\begin{equation} \label{restrictedboundary10_mixed}
{\partial _1} = \left[ {\begin{array}{*{20}{c}}
{ - 1}&0&{ + 1}&0&0&0&0&0&0\\
{ + 1}&{ - 1}&0&0&0&0&0&0&0\\
0&{ + 1}&{ - 1}&0&0&0&0&0&0\\
0&0&0&{ + 1}&{ - 1}&0&0&0&0\\
0&0&0&0&{ + 1}&{ - 1}&0&0&0\\
0&0&0&0&0&{ + 1}&{ - 1}&0&0\\
0&0&0&{ - 1}&0&0&{ + 1}&0&0\\
0&0&0&0&0&0&0&{ + 1}&{ - 1}\\
0&0&0&0&0&0&0&{ - 1}&{ + 1}
\end{array}} \right]
\end{equation}

\begin{equation}\label{transpose} 
{\delta _0} = \partial _1^T	 
\end{equation} 

\begin{equation}\label{mixed_global_C}
{{\bf{C}}_g} = \left[ {\begin{array}{*{20}{c}}
{{C_{{g_1}}}}&0&0&0&0&0&0&0&0\\
0&0&0&0&0&0&0&0&0\\
0&0&0&0&0&0&0&0&0\\
0&0&0&0&0&0&0&0&0\\
0&0&0&0&0&0&0&0&0\\
0&0&0&0&0&0&0&0&0\\
0&0&0&0&0&0&0&0&0\\
0&0&0&0&0&0&0&0&0\\
0&0&0&0&0&0&0&0&0
\end{array}} \right]
\end{equation}

\begin{equation}\label{mixed_global_inverseR}
{\bf{R}}_g^{ - 1} = \left[ {\begin{array}{*{20}{c}}
0&0&0&0&0&0&0&0&0\\
0&{R_{{g_1}}^{ - 1}}&0&0&0&0&0&0&0\\
0&0&0&0&0&0&0&0&0\\
0&0&0&0&0&0&0&0&0\\
0&0&0&0&0&0&0&0&0\\
0&0&0&0&0&{R_{{g_2}}^{ - 1}}&0&0&0\\
0&0&0&0&0&0&0&0&0\\
0&0&0&0&0&0&0&0&0\\
0&0&0&0&0&0&0&0&0
\end{array}} \right]
\end{equation}

\begin{equation}\label{mixed_global_inverseL}
{\bf{L}}_g^{ - 1} = \left[ {\begin{array}{*{20}{c}}
0&0&0&0&0&0&0&0&0\\
0&0&0&0&0&0&0&0&0\\
0&0&0&0&0&0&0&0&0\\
0&0&0&0&0&0&0&0&0\\
0&0&0&0&{L_{{g_1}}^{ - 1}}&0&0&0&0\\
0&0&0&0&0&0&0&0&0\\
0&0&0&0&0&0&0&0&0\\
0&0&0&0&0&0&0&0&0\\
0&0&0&0&0&0&0&0&{L_{{g_2}}^{ - 1}}
\end{array}} \right]
\end{equation}

\begin{equation}\label{mixed_gyrator_inverseK}
\begin{array}{*{20}{l}}
{{\bf{k}}_g^{ - 1} = \left[ {\begin{array}{*{20}{c}}
0&0&0&0&0&0&0&0&0\\
0&0&0&0&0&0&0&0&0\\
0&0&0&0&0&0&0&0&0\\
0&0&0&0&0&0&0&0&0\\
0&0&0&0&0&0&0&0&0\\
0&0&0&0&0&0&0&0&0\\
0&0&0&0&0&0&0&{k_g^{ - 1}}&0\\
0&0&0&0&0&0&{k_g^{ - 1}}&0&0\\
0&0&0&0&0&0&0&0&0
\end{array}} \right]}\\
\;
\end{array}
\end{equation}
	
\begin{equation}\label{mixed_transformer_kt}	
{{\bf{k}}_t} = \left[ {\begin{array}{*{20}{c}}
0&0&1&{{{ - {N_1}} \mathord{\left/
 {\vphantom {{ - {N_1}} {{N_2}}}} \right.
 \kern-\nulldelimiterspace} {{N_2}}}}&0&0&0&0&0
\end{array}} \right]
\end{equation}

\begin{equation}\label{mixed_transformer_ktprime}	
{{{\bf{k'}}}_t} = \left[ {\begin{array}{*{20}{c}}
0&0&1&{{{ - {N_2}} \mathord{\left/
 {\vphantom {{ - {N_2}} {{N_1}}}} \right.
 \kern-\nulldelimiterspace} {{N_1}}}}&0&0&0&0&0
\end{array}} \right]
\end{equation}

\begin{equation}\label{mixed_transformer_current}	
{\bf{t}}_T^1 = {\left[ {\begin{array}{*{20}{c}}
0&0&{{t_{TL}}}&{{t_{TR}}}&0&0&0&0&0
\end{array}} \right]^T}
\end{equation}

\begin{equation}\label{mixed__voltage_source}
{\bf{a}}_{\rm{f}}^1 = {\left[ {\begin{array}{*{20}{c}}
0&{ - {a_{{{\rm{f}}_1}}}}&0&0&0&0&0&0&{ - {a_{{{\rm{f}}_2}}}}
\end{array}} \right]^T}
\end{equation}

\begin{equation}\label{mixed__current_source}
{\bf{t}}_{\rm{f}}^1 = {\left[ {\bf{0}} \right]_{9 \times 1}}
\end{equation}

Substitute Eq.\ref{restrictedboundary10_mixed} $\sim$ Eq.\ref{mixed__current_source} to Eq.\ref{node_potential_state_AE_simpler_ode_mixed} $\sim$ Eq.\ref{Constraint_Eq_current_diff}, the generated system state equation is as follows: 

\begin{equation} \label{mixed_state_equation_node_potential}
\left\{ \begin{array}{l}
\left[ {\begin{array}{*{20}{c}}
{{C_{{g_1}}}}&{ - {C_{{g_1}}}}&0&0&0&0&0&0&0\\
{ - {C_{{g_1}}}}&{{C_{{g_1}}}}&0&0&0&0&0&0&0\\
0&0&0&0&0&0&0&0&0\\
0&0&0&0&0&0&0&0&0\\
0&0&0&0&0&0&0&0&0\\
0&0&0&0&0&0&0&0&0\\
0&0&0&0&0&0&0&0&0\\
0&0&0&0&0&0&0&0&0\\
0&0&0&0&0&0&0&0&0
\end{array}} \right]{{{\bf{\ddot d}}}^0} + \left[ {\begin{array}{*{20}{c}}
0&0&0&0&0&0&0&0&0\\
0&{R_{{g_1}}^{ - 1}}&{ - R_{{g_1}}^{ - 1}}&0&0&0&0&0&0\\
0&{ - R_{{g_1}}^{ - 1}}&{R_{{g_1}}^{ - 1}}&0&0&0&0&0&0\\
0&0&0&0&0&0&0&0&0\\
0&0&0&0&0&0&0&0&0\\
0&0&0&0&0&{R_{{g_2}}^{ - 1}}&{ - R_{{g_2}}^{ - 1}}&{ - k_g^{ - 1}}&{k_g^{ - 1}}\\
0&0&0&0&0&{ - R_{{g_2}}^{ - 1}}&{R_{{g_2}}^{ - 1}}&{k_g^{ - 1}}&{ - k_g^{ - 1}}\\
0&0&0&0&0&{ - k_g^{ - 1}}&{k_g^{ - 1}}&0&0\\
0&0&0&0&0&{k_g^{ - 1}}&{ - k_g^{ - 1}}&0&0
\end{array}} \right]{{{\bf{\dot d}}}^0}\\
 + \left[ {\begin{array}{*{20}{c}}
0&0&0&0&0&0&0&0&0\\
0&0&0&0&0&0&0&0&0\\
0&0&0&0&0&0&0&0&0\\
0&0&0&{L_{{g_1}}^{ - 1}}&{ - L_{{g_1}}^{ - 1}}&0&0&0&0\\
0&0&0&{ - L_{{g_1}}^{ - 1}}&{L_{{g_1}}^{ - 1}}&0&0&0&0\\
0&0&0&0&0&0&0&0&0\\
0&0&0&0&0&0&0&0&0\\
0&0&0&0&0&0&0&{L_{{g_2}}^{ - 1}}&{ - L_{{g_2}}^{ - 1}}\\
0&0&0&0&0&0&0&{ - L_{{g_2}}^{ - 1}}&{L_{{g_2}}^{ - 1}}
\end{array}} \right]{{\bf{d}}^0} + \left[ {\begin{array}{*{20}{c}}
{{t_{TL}}}\\
0\\
{ - {t_{TL}}}\\
{{t_{TR}}}\\
0\\
0\\
{ - {t_{TR}}}\\
0\\
0
\end{array}} \right] = \left[ {\begin{array}{*{20}{c}}
0\\
{ - R_{{g_1}}^{ - 1}{a_{{{\rm{f}}_{\rm{1}}}}}}\\
{R_{{g_1}}^{ - 1}{a_{{{\rm{f}}_{\rm{1}}}}}}\\
0\\
0\\
0\\
0\\
{ - L_{{g_2}}^{ - 1}\int {{a_{{{\rm{f}}_2}}}dt} }\\
{L_{{g_2}}^{ - 1}\int {{a_{{{\rm{f}}_2}}}dt} }
\end{array}} \right]\\
{{\left( {{\bf{\dot d}}_{(3)}^0 - {\bf{\dot d}}_{(1)}^0} \right)} \mathord{\left/
 {\vphantom {{\left( {{\bf{\dot d}}_{(3)}^0 - {\bf{\dot d}}_{(1)}^0} \right)} {\left( {{\bf{\dot d}}_{(7)}^0 - {\bf{\dot d}}_{(4)}^0} \right) = {{{N_1}} \mathord{\left/
 {\vphantom {{{N_1}} {{N_2}}}} \right.
 \kern-\nulldelimiterspace} {{N_2}}}}}} \right.
 \kern-\nulldelimiterspace} {\left( {{\bf{\dot d}}_{(7)}^0 - {\bf{\dot d}}_{(4)}^0} \right) = {{{N_1}} \mathord{\left/
 {\vphantom {{{N_1}} {{N_2}}}} \right.
 \kern-\nulldelimiterspace} {{N_2}}}}}\\
{{{t_{TL}}} \mathord{\left/
 {\vphantom {{{t_{TL}}} {{t_{TR}} = {{{N_2}} \mathord{\left/
 {\vphantom {{{N_2}} {{N_1}}}} \right.
 \kern-\nulldelimiterspace} {{N_1}}}}}} \right.
 \kern-\nulldelimiterspace} {{t_{TR}} = {{{N_2}} \mathord{\left/
 {\vphantom {{{N_2}} {{N_1}}}} \right.
 \kern-\nulldelimiterspace} {{N_1}}}}}
\end{array} \right.
\end{equation}

Similarly following the paths  in Figure \ref{fig:state_variable_q0_generalized}, generates Eq.\ref{mesh_current_state_AE_simpler_ode_mixed} $\sim$ Eq.\ref{Constraint_Eq_voltage_q0_diff}, with individual terms as follows:

	\begin{equation} \label{coboundary12_mixed}
	{\delta _{1}} = \left[ {\begin{array}{*{20}{c}}
	{ + 1}&{ + 1}&{ + 1}&0&0&0&0&0&0\\
	0&0&0&{ + 1}&{ + 1}&{ + 1}&{ + 1}&0&0\\
	0&0&0&0&0&0&0&{ + 1}&{ + 1}
	\end{array}} \right] 
	\end{equation}
	
\begin{equation}\label{transpose} 
 \partial _2 ={\delta _1} ^T
\end{equation} 

\begin{equation}\label{mixed_global_L}	
{\delta _1}{{\bf{L}}_g}\partial _2 = {\delta _1}\left[ {\begin{array}{*{20}{c}}
0&0&0&0&0&0&0&0&0\\
0&0&0&0&0&0&0&0&0\\
0&0&0&0&0&0&0&0&0\\
0&0&0&0&0&0&0&0&0\\
0&0&0&0&{{L_{{g_1}}}}&0&0&0&0\\
0&0&0&0&0&0&0&0&0\\
0&0&0&0&0&0&0&0&0\\
0&0&0&0&0&0&0&0&0\\
0&0&0&0&0&0&0&0&{{L_{{g_2}}}}
\end{array}} \right]\partial _2 = \left[ {\begin{array}{*{20}{c}}
0&0&0\\
0&{{L_{{g_1}}}}&0\\
0&0&{{L_{{g_2}}}}
\end{array}} \right]
\end{equation}	

\begin{equation}\label{mixed_global_R}	
{\delta _1}{{\bf{R}}_g}\partial _2 = {\delta _1}\left[ {\begin{array}{*{20}{c}}
0&0&0&0&0&0&0&0&0\\
0&{{R_{{g_1}}}}&0&0&0&0&0&0&0\\
0&0&0&0&0&0&0&0&0\\
0&0&0&0&0&0&0&0&0\\
0&0&0&0&0&0&0&0&0\\
0&0&0&0&0&{{R_{{g_2}}}}&0&0&0\\
0&0&0&0&0&0&0&0&0\\
0&0&0&0&0&0&0&0&0\\
0&0&0&0&0&0&0&0&0
\end{array}} \right]\partial _2 = \left[ {\begin{array}{*{20}{c}}
{{R_{{g_1}}}}&0&0\\
0&{{R_{{g_2}}}}&0\\
0&0&0
\end{array}} \right]
\end{equation}	

\begin{equation}\label{mixed_global_inverseC}	
{\delta _1}{\bf{C}}_g^{ - 1}\partial _2 = {\delta _1}\left[ {\begin{array}{*{20}{c}}
{C_{{g_1}}^{ - 1}}&0&0&0&0&0&0&0&0\\
0&0&0&0&0&0&0&0&0\\
0&0&0&0&0&0&0&0&0\\
0&0&0&0&0&0&0&0&0\\
0&0&0&0&0&0&0&0&0\\
0&0&0&0&0&0&0&0&0\\
0&0&0&0&0&0&0&0&0\\
0&0&0&0&0&0&0&0&0\\
0&0&0&0&0&0&0&0&0
\end{array}} \right]\partial _2 = \left[ {\begin{array}{*{20}{c}}
{C_{{g_1}}^{ - 1}}&0&0\\
0&0&0\\
0&0&0
\end{array}} \right]
\end{equation}	

\begin{equation}\label{mixed_global_gyrator_K}	
{\delta _1}{{\bf{k}}_g}\partial _2 = {\delta _1}\left[ {\begin{array}{*{20}{c}}
0&0&0&0&0&0&0&0&0\\
0&0&0&0&0&0&0&0&0\\
0&0&0&0&0&0&0&0&0\\
0&0&0&0&0&0&0&0&0\\
0&0&0&0&0&0&0&0&0\\
0&0&0&0&0&0&0&0&0\\
0&0&0&0&0&0&0&{{k_g}}&0\\
0&0&0&0&0&0&{{k_g}}&0&0\\
0&0&0&0&0&0&0&0&0
\end{array}} \right]\partial _2 = \left[ {\begin{array}{*{20}{c}}
0&0&0\\
0&0&{{k_g}}\\
0&{{k_g}}&0
\end{array}} \right]	
\end{equation}

	\begin{equation}\label{mixed_global_transformer_voltage}	
{\delta _1}{\bf{a}}_T^1 = {\delta _1}{\left[ {\begin{array}{*{20}{c}}
0&0&{{{\rm{a}}_{TL}}}&{{{\rm{a}}_{TR}}}&0&0&0&0&0
\end{array}} \right]^T} = \left[ {\begin{array}{*{20}{c}}
{{{\rm{a}}_{TL}}}\\
{{{\rm{a}}_{TR}}}\\
0
\end{array}} \right]
\end{equation}

Substitute Eq.\ref{mixed_transformer_kt} $\sim$ Eq.\ref{mixed__current_source}, Eq.\ref{coboundary12_mixed} $\sim$ Eq.\ref{mixed_global_transformer_voltage} to Eq.\ref{mesh_current_state_AE_simpler_ode_mixed} $\sim$ Eq.\ref{Constraint_Eq_voltage_q0_diff}, the generated system state equation is as follows: 

\begin{equation} \label{mixed_state_equation_mesh_current}
\left\{ \begin{array}{l}
\left[ {\begin{array}{*{20}{c}}
0&0&0\\
0&{{L_{{g_1}}}}&0\\
0&0&{{L_{{g_2}}}}
\end{array}} \right]{{{\bf{\ddot n}}}^2} + \left[ {\begin{array}{*{20}{c}}
{{R_{{g_1}}}}&0&0\\
0&{{R_{{g_2}}}}&{{k_g}}\\
0&{{k_g}}&0
\end{array}} \right]{{{\bf{\dot n}}}^2} + \left[ {\begin{array}{*{20}{c}}
{C_{{g_1}}^{ - 1}}&0&0\\
0&0&0\\
0&0&0
\end{array}} \right]{{\bf{n}}^2} + \left[ {\begin{array}{*{20}{c}}
{{{\rm{a}}_{TL}}}\\
{{{\rm{a}}_{TR}}}\\
0
\end{array}} \right] = \left[ {\begin{array}{*{20}{c}}
{{a_{{{\rm{f}}_1}}}}\\
0\\
{{a_{{{\rm{f}}_2}}}}
\end{array}} \right]\\
{{{{\rm{a}}_{TL}}} \mathord{\left/
 {\vphantom {{{{\rm{a}}_{TL}}} {{{\rm{a}}_{TL}} = {{{N_1}} \mathord{\left/
 {\vphantom {{{N_1}} {{N_2}}}} \right.
 \kern-\nulldelimiterspace} {{N_2}}}}}} \right.
 \kern-\nulldelimiterspace} {{{\rm{a}}_{TL}} = {{{N_1}} \mathord{\left/
 {\vphantom {{{N_1}} {{N_2}}}} \right.
 \kern-\nulldelimiterspace} {{N_2}}}}}\\
{{{\bf{\dot n}}_{(1)}^2} \mathord{\left/
 {\vphantom {{{\bf{\dot n}}_{(1)}^2} {{\bf{\dot n}}_{(2)}^2 = {{{N_2}} \mathord{\left/
 {\vphantom {{{N_2}} {{N_1}}}} \right.
 \kern-\nulldelimiterspace} {{N_1}}}}}} \right.
 \kern-\nulldelimiterspace} {{\bf{\dot n}}_{(2)}^2 = {{{N_2}} \mathord{\left/
 {\vphantom {{{N_2}} {{N_1}}}} \right.
 \kern-\nulldelimiterspace} {{N_1}}}}}
\end{array} \right.
	\end{equation}

\end{exmp}


\section{Conclusion}

\subsection {Summary and significance}

Formal semantics of engineering models is a principal ingredient of rigorous and algorithmic foundations for model-based systems engineering.  Standardizing on such a semantics is also a key to solving widespread interchangeability and interoperability challenges that arise with rapid proliferation of model-based engineering language and tools.  
In this paper, we proposed a formal semantics for a large and important class of lumped-parameter systems that are widely used for systems engineering, physical modeling,  and design activities.  The proposed semantics relies only on standard tools from algebraic topology and known results in classification of physical theories and systems.  The semantics is  effectively `representation free' in that it is independent of specific implementation assumptions, coordinates,  linguistic constructs, or numerical simulation schemes.
 
We showed that (extended and generalized) Tonti diagrams provide a canonical method for representing behaviors of lumped parameter systems computationally.  Based on known classification of physical theories, behavior of any lumped parameter system may be described either as a collection of interacting single-domain Tonti diagrams or as a single generalized Tonti diagram with energy transduction represented by additional constraints. In other words, a Tonti diagram can be viewed as a representation scheme and a data structure for representing most known physical behaviors.  We have also seen that this representation 
supports algorithmic generation of all possible forms of the governing state equations as paths in the (collection of) Tonti diagrams.  
The topological and constitutive operators appear as labels on the edges of the diagram and may be interpreted either symbolically,  giving differential equations,  or numerically (e.g. as finite difference operators), corresponding to executable approximations of such models.

\subsection {Interchangeability and interoperability}

Achieving greater interoperability of lumped parameter systems tools and languages was the main motivation for developing the proposed canonical semantic model.  
The existence of common semantics implies that all compliant models become fully interchangeable irrespective of specific syntax or modeling concepts adapted in a particular simulation tool. The representation of this semantics using the corresponding Tonti diagrams can serve as a formal neutral format for exchange of all such models.  
Furthermore, this representation supports exchange not only of the system models but also of the adapted simulation approaches in individual systems,  now represented as paths in the corresponding Tonti diagrams. 
In practical terms,  the existence of the common semantic model also eliminates the need for customized point-to-point translators or new super-languages attempting to subsume and unify currently existing tools.  
New compatible languages and modeling tools, for example in SysML,  could be developed rigorously and rapidly,  as long as they adhere to the adapted semantic model. 
 
While model  translation and exchange is a widely adapted and practiced in  industry, it is a rather limited form of interoperability.
When distinct tools and languages are based on a common reference semantics,  they should be able to interoperate directly in a truly object-oriented fashion,  where distinct portions of the system models are represented and simulated in different tools.  This would allow better protection of proprietary models and fine-tuning selection of tools that are most appropriate for specific modeling tasks.

\subsection {Open issues and promising directions}

The semantic model proposed in this paper provides a foundation and a starting point for developing a more general semantic framework for modeling behavior of physical systems.   In particular, the strongly-typed classification of physical theories and their representation by Tonti diagrams tacitly assumes that the constructed models are subject to energy balance laws. 
The  majority of lumped-parameter system models satisfy this assumption.   However, many modeling systems support numerous additional operations and constructs, such as signal flows and arbitrary mathematical transformations, that are not necessarily based on first principles.  Whether and how such constructs should be included in the common semantics model is not entirely clear.  In principle,  they could be treated as special cases of energetic processes (as, for example, is advocated in bond graph literature \cite{karnopp1990system}), or they may require introduction of additional types of constraints to the underlying topological model.

In this paper we focused on lumped parameter systems that are intrinsically two-dimensional and devoid of geometry.  However, both classification of physical theories and Tonti diagrams span the entire spectrum of full-dimensional physical models and behaviors.   This implies that the proposed formal semantic model can be extended to all spatially distributed models whose behavior is commonly described by integral and partial differential equations.   In fact, developing a semantics model for an single-domain physical model whose behavior is described by a known Tonti diagram is a straightforward task that would follow the development in Section 4 of this paper.  Different paths on such a diagram would correspond to different methods to generate governing equations;  replacing symbolic operators by numerical approximations yields rich variety of standard numerical simulation methods \cite{Mattiussi,felippa2004introduction,ferretti2015cell}.
However,  in contrast to lumped parameter systems that are all isomorphic, spatially distributed models are heterogeneous in their dimension, type of variables, and topological structures.  Interaction, composition, and transformation of such Tonti diagrams and the systems they represent remain active open  research issues.   

Treating the common semantic model and its representation by Tonti diagrams as the first class objects, that explicitly represent physical behaviors, opens up a number of promising and exciting opportunities in computational design and model-based engineering.  Algorithmic construction, editing, composition, and transformation of such models would support a broad range of design engineering activities, from  concept generation to detailed system modeling.  Because many such engineering activities are performed using SysML, it may be reasonable to expect that Tonti diagrams should appear as one of the standard diagrams in SysML in the near future.   Such a diagram would provide immediate support for including physical behaviors and their simulation into a broad range of model-based systems engineering activities.

\section*{Acknowledgements}
The authors would like to thank Conrad Bock for numerous discussions and encouragement.   This research was supported in part by the National Institute of Standards and Technology (NIST) under cooperative agreement 70NANB14H248, by NSF grants CMMI-1344205, CMMI-1361862, and CMMI-1547189,  and by 
Defense Advanced Research Projects Agency's FUNdamental Design Program.
The responsibility for errors and omissions lies solely with the
authors.

\bibliographystyle{unsrt}
\bibliography{asme2e,refs-chard}

\end{document}